\documentclass[a4paper,11pt]{article}
\usepackage{jheppub}

\usepackage{amsthm}
\usepackage{marginnote}
\usepackage{enumerate}
\usepackage{subfigure}
\usepackage{tabularx}

\newcolumntype{Y}{>{\centering\arraybackslash}X}

  \usepackage{tkz-euclide}
  \usetikzlibrary{patterns}
  \usetikzlibrary{calc}
  \usetikzlibrary{positioning}
  \usetikzlibrary{decorations.pathmorphing}
  \tikzstyle{point}=[draw,circle,fill=black,scale=.3] 
  \tikzset{baseline={([yshift=-.5ex]current bounding box.center)}, every picture}

\makeatletter
\newcommand*{\defeq}{\mathrel{\rlap{%
  \raisebox{0.3ex}{$\m@th\cdot$}}%
  \raisebox{-0.3ex}{$\m@th\cdot$}}%
=}
\makeatother

  \newcommand{\be}{\begin{equation}}
  \newcommand{\ee}{\end{equation}}
  \newcommand{\del}{\partial}

  \newcommand{\NN}{\mathbb{N}}
  \newcommand{\ZZ}{\mathbb{Z}}
  \newcommand{\QQ}{\mathbb{Q}}
  \newcommand{\RR}{\mathbb{R}}
  \newcommand{\CC}{\mathbb{C}}
  
  \newcommand{\HH}{\mathbb{H}}
  \newcommand{\DD}{\mathbb{D}}
  \newcommand{\RP}{\mathbb{RP}}

  \DeclareMathOperator{\SL}{SL}
  \DeclareMathOperator{\GL}{GL}
  \DeclareMathOperator{\uSL}{\widetilde{SL}}
  \DeclareMathOperator{\SO}{SO}
  \DeclareMathOperator{\PSU}{PSU}
  \DeclareMathOperator{\PSL}{PSL}
  \DeclareMathOperator{\Vect}{Vect}
  
  \DeclareMathOperator{\Diff}{Diff}
  \DeclareMathOperator{\Hol}{Hol}
  \DeclareMathOperator{\Stab}{Stab}
  \DeclareMathOperator{\Isom}{Isom}
  \DeclareMathOperator{\Sch}{Sch}

  \newcommand{\OO}{\mathcal O}
  \newcommand{\vir}{\mathfrak{vir}}
  \newcommand{\M}{\mathcal{M}}
  \newcommand{\Tr}{\mathrm{Tr}\,}
  \renewcommand{\Im}{\mathrm{Im}}
  \newcommand{\Al}{\mathcal{A}_{\ell}}

  \newcommand{\mat}[4]{\begin{pmatrix} #1 & #2 \\ #3 & #4 \end{pmatrix}}
  \newcommand{\vek}[2]{\begin{pmatrix} #1 \\ #2 \end{pmatrix}}

\title{\boldmath{$\mathrm{SL}(2,\mathbb{R})$} Gauge Theory, Hyperbolic Geometry and Virasoro Coadjoint Orbits}

  \author[a]{Matthias Blau}
  \author[b]{and Donald Ray Youmans}

  \affiliation[a]{Albert Einstein Center for Fundamental Physics,
    Institute for Theoretical Physics,
  University of Bern, Sidlerstrasse 5, CH-3012 Bern, Switzerland}
  \affiliation[b]{Institute for Mathematics, Ruprecht-Karls-Universit\"at Heidelberg, Im Neuenheimer Feld 205, DE-69120 Heidelberg, Germany}

  \emailAdd{blau@itp.unibe.ch}
  \emailAdd{youmans@mathi.uni-heidelberg.de}

  \abstract{

It has long been known that the moduli space of hyperbolic metrics
on the disc can be identified with the Virasoro coadjoint orbit
$\mathrm{Diff}^+(S^1) / \mathrm{SL}(2,\mathbb{R})$. The interest in this relationship has
recently been revived in the study of two-dimensional JT gravity
and it raises the natural question if all Virasoro
orbits $\mathcal{O}$ arise as moduli spaces of hyperbolic metrics. In this article,
we give an affirmative answer to this question using 
$\mathrm{SL}(2,\mathbb{R})$ gauge theory on a cylinder $S$: to any $L\in\mathcal{O}$ we assign a
flat $\mathrm{SL}(2,\mathbb{R})$ gauge field $A_L =
(g_L)^{-1} dg_L$, and we explain 
how the global properties and singularities of the hyperbolic geometry are 
encoded in the monodromies and winding numbers of $g_L$, 
and how they depend on the Virasoro orbit. 
In particular, we show that the 
somewhat mysterious geometries associated with Virasoro orbits with
no constant representative $L$ arise from large gauge transformations
acting on standard (constant $L$ ) funnel or cuspidal geometries,
shedding some light on their potential physical 
significance: e.g.\ they describe new
topological sectors of two-dimensional gravity, characterised by
twisted boundary conditions. Using a gauge theoretic gluing construction,
we also obtain a complete
dictionary between Virasoro coadjoint orbits and moduli spaces of
hyperbolic metrics with specified boundary projective structure. 

}

  \global\parskip=4pt

  \makeatletter
  \gdef\@fpheader{}
  \makeatother

\begin{document}

  \maketitle

\section{Introduction}

It has long been known that there are intriguing relations between the theory of Virasoro coadjoint orbits 
\cite{Kirillov,Segal,Witten,Balog_Feher_Palla} and models of two-dimensional
gravity, going back at least to \cite{AS1,RR1}. More recently, this relationship
has reemerged and played a prominent role in the context of the widely-studied
subject of quantum Jackiw-Teitelboim (JT) gravity (see e.g.\ \cite{MertensTuriaci:Review} 
for a review). 
In particular, the boundary dynamics of JT gravity is governed by a Schwarzian
action $\int dt\; \Sch(f)(t)$, with 
      \begin{equation}
	\Sch(f) = \frac{f'''}{f'} - \frac32 \left( \frac{f''}{f'} \right)^2 
      \end{equation}
capturing the broken conformal (reparametrisation) invariance 
at the boundary \cite{AP,Jensen,MSY,EMV,GNW}. The relation
with Virasoro coadjoint orbits then arises since the Schwarzian is precisely 
the corresponding cocycle as in \eqref{eq:iLf} below. 

While there are many ways to understand and explain the emergence 
of the Schwarzian theory \cite{Mertens:Origins,MertensTuriaci:Review}, 
from a purely classical gravitational or geometric point of view
the most elementary way to understand the emergence of a Virasoro
cadjoint action in this context is as an action of ``large'' diffeomorphisms on
hyperbolic metrics (constant scalar curvature $R=-2$). 
This is particularly
easy to see in the Fefferman-Graham (FG) gauge, which is well adapted to the
study of asymptotic symmetries. The exact solution 
on a cylinder $S$, say, with FG coordinates $(\rho,\varphi)$ and fixed 
asymptotic boundary condition at $\rho\to\infty$ (so from the JT perspective
we are looking at thermal Eulidean geometries) is determined by a 
single (and arbitrary) periodic function $L(\varphi)$, and is given explicitly
by 
\begin{equation}
ds^2(L) = d\rho^2 +\left( e^\rho - L(\varphi) e^{-\rho}\right)^2 d\varphi^2 
\label{eq:FG_metric_intro}
\end{equation}
Large diffeomorphisms preserving the FG gauge (and inducing a non-trivial
diffeomorphism $f(\varphi)$ on the boundary circle $S^1$ - hence ``large'') act via 
$ds^2(L)\to ds^2(L^f)$, where 
\begin{equation}
\label{eq:iLf}
L^f(\varphi) = f'(\varphi)^2 L(f(\varphi)) + \frac12\Sch(f)(\varphi)
\end{equation}
is precisely the Virasoro coadjoint action of $f(\varphi)$ on $L(\varphi)$
(thought of as a quadratic differential). Thus the (moduli) space of hyperbolic 
metrics on $S$ in the FG gauge can be identified with the space of $L(\varphi)$, i.e.\ with 
the (smooth) dual $\vir^*$ of the Virasoro algebra, 
\begin{equation}
\label{mfgvir}
\M_{FG}(S)\cong \{L(\varphi)\} \cong \vir^*
\end{equation}
and can be decomposed into Virasoro coadjoint orbits. 

Remarkably, a very similar structure arises in a slightly different
context, namely solutions of Lorentzian 2+1 dimensional AdS gravity
with Brown-Henneaux boundary conditions, which are labelled by pairs
of such functions $L^\pm(x^\pm)$. See \cite{SJY1,SJY2} for an analysis of
these Ba\~nados solutions \cite{Banados} 
from the point of view of Virasoro coadjoint orbits,
in a spirit very similar to the one that we will adopt in this
paper.

As we will review
in Appendix \ref{app:vir}, following 
\cite{Witten,Balog_Feher_Palla,Oblak}
these orbits can be conveniently studied by analysing the solutions 
of the associated Hill's equation 
\be
\label{eq:ihill}
\psi''(\varphi) + L(\varphi)\psi(\varphi) = 0
\ee
The outcome of the classification is that caodjoint 
orbits are labelled by a pair $(\sigma,n_0)$ 
where $\sigma$ labels a conjugacy class in $\PSL(2,\RR)$ (and these
can be partitioned into degenerate, elliptic, hyperbolic and parabolic 
conjugacy classes respectively), and an integer winding number $n_0\in\NN_0$.

The identification \eqref{mfgvir} suggests
an intriguing relation between certain moduli spaces of 
hyperbolic metrics and Virasoro coadjoint orbits, 
\be
\label{MO}
\{\text{Moduli Spaces of Hyperbolic Metrics}\} \quad\Leftrightarrow\quad 
\{\text{Virasoro Coadjoint Orbits}\}
\ee
One particular instance of this correspondence is well known and well
understood: for the special constant value $L_0=1/4$ one finds (after a
constant shift of $\rho$) the standard
hyperbolic metric on the (infinite) disc, 
\be
ds^2(L_0=1/4) = d\rho^2 + \sinh^2(\rho) d\varphi^2
\ee
(equivalent to the standard Poincar\'e
metric on the unit disc $\DD$), and the corresponding coadjoint orbit
\be
\label{eq:td}
\OO_{L_0=1/4} \equiv \OO_{\sigma=0,n_0=1}\cong \Diff^+(S^1)/\SL(2,\RR)
\ee
can be identified with the moduli space of smooth hyperbolic metrics on the
disc. Indeed this space is precisely 
the Teichm\"uller space of the disc $\DD$ - see e.g.\ \cite{VV}. By 
\cite{NV,HR} it can also be regarded as the smooth part of what is known
as the universal Teichm\"uller space. In 
the JT context, this space is known as the ``vacuum orbit'', and parametrises
the Nambu -- Goldstone bosons that arise from the spontaneous breaking of diffeomorphism
invariance to the $\SL(2,\RR)$ isometry group of the ground state metric on the
disc. Another special value is $L_0=0$ which, upon quantisation of the
corresponding orbit, is related to 
the two-dimensional Polyakov action for gravity \cite{AS1,RR1} (for
generalisations see \cite{GNW}). 

In the JT context, constant values 
of $L_0\neq 1/4$ are (e.g.\ from a three-dimensional point of view) 
understood to be related to topological defects in JT gravity 
\cite{MertensTuriaci:Defects}. From a geometric point of view these can 
readily be seen to correspond to 
\begin{itemize}
\item an $n$-fold branched covering of $\DD$ for $L_0 = n^2/4$
\item a punctured disc $\DD^*$ with a conical singularity at the origin for $L_0>0, L_0 \neq
n^2/4$
\item an annular or funnel-like geometry for $L_0 < 0$
\item a punctured disc $\DD^*$ with a cusp singularity at the origin for $L_0=0$
\end{itemize}
(this will be reviewed and explained in Section \ref{sec:examples}). 
Other orbits, those without constant 
representatives, and the geometries they represent are not well understood, 
even classically, and are considered to be ``highly quantum'' \cite{MertensTuriaci:Review}
from the point of view of the JT path integral (since they do not contain 
any saddle points of the Schwarzian action).

The purpose of this paper is to study this tentative correspondence \eqref{MO} between 
moduli spaces of hyperbolic metrics and Virasoro coadjoint orbits in 
more detail. 
In particular, our aim is to understand
\begin{itemize}
\item the geometry at a general point $L_0^f$ in the orbit $\OO_{L_0}$ of $L_0$ 
\item what the geometries in a given caodjoint orbit have in common
respectively what distinguishes them 
\item what distinguishes the geometries in different coadjoint orbits from each
other
\item the exotic geometries described by orbits without constant representatives
\end{itemize}
In order to accomplish this, we systematically adopt an $\SL(2,\RR)$ gauge
theory perspective which is useful for connecting the gravitational
and geometric aspects (hyperbolic metrics or hyperbolic structures) to the 
classification of Virasoro coadjoint orbits, via the theory of Hill's equation.
In Section \ref{sec:gauge_theory}, we review the basic results of 
hyperbolic gravity mentioned above in this $\SL(2,\RR)$ gauge theory language 
(well known from the $\SL(2,\RR)$ BF gauge theory description of JT gravity 
\cite{FK,IT})
of a flat $\SL(2,\RR)$ connection $A(L)$ encoding the vielbeins and spin connection
of the hyperbolic metric $ds^2(L)$. 
In a suitable gauge theoretic analogue of the FG gauge, this gauge field takes the 
form \eqref{eq:A}
  \begin{equation}
    A(L) = 
    \mat{\frac{d\rho}{2}}{-L(\varphi)e^{-\rho}d\varphi}{e^{\rho}d\varphi}{-\frac{d\rho}{2}}
\label{eq:Aint}
  \end{equation}
We also provide a purely gauge theoretic
derivation of the Virasoro coadjoint action on hyperbolic metrics by determining 
the gauge transformations that leave the form of $A(L)$ invariant 
(Section \ref{sec:Fred}).

Locally (and away from possible singularities) any hyperbolic metric
is of course isometric to the Poincar\'e upper half plane $\HH$ or Poincar\'e 
disc $\DD$ equipped with its standard metric.
Moreover it is known that any isolated singularity of a hyperbolic metric is 
either conical or a cusp \cite{Nitsche}, and that in a neighbourhood of such a
singularity the geometry  
always takes the standard form provided by the constant $L=L_0$ geometries
listed above \cite{FengShiXu}. In the case at hand, such a 
\textit{uniformisation map} 
\be
z_L: S\to \HH 
\label{eq:z_intro}
\ee
from the FG cylinder $S$ to the upper half plane $\HH$
should have the property that 
  in terms of the local coordinates $z_L(\rho,\varphi)$ the metric
  \eqref{eq:FG_metric_intro} takes the form of
  the standard Poincar\'e  upper half plane metric $ds^2_\HH =
dz\;d\bar{z}/\Im(z)^2$, i.e. 
  \begin{equation}
    ds^2(L) = \frac{dz_L(\rho,\varphi) d\bar z_L(\rho,\varphi)}{\Im\left(
    z_L(\rho,\varphi)\right)^2} 
    \equiv z_L^*ds^2_{\HH}
  \end{equation}
For the simple 
constant $L=L_0$ geometries mentioned above, such a 
uniformisation map which makes this 
explicit is well-known or in any case readily found. 
The general construction of $z_L$ which makes this explicit for any metric $ds^2(L)$, 
not just for constant $L=L_0$, is a priori a non-trivial
task. 

We will explain in Section \ref{sec:uniformization_map} 
how to obtain such a uniformisation map $z_L$ directly 
from the $\SL(2,\RR)$ gauge theory perspective.
Indeed, we will show that $z_L$ can 
be constructed from the local expression
$A(L)= g_L^{-1}dg_L$ for the corresponding flat $\SL(2,\RR)$-connection $A(L)$,
where the $\SL(2,\RR)$ matrix
$g_L(\rho,\varphi)$ can in turn be constructed explicitly from the solutions of the 
associated Hill's equation \eqref{eq:ihill}.
Given $g_L$, we can then define $z_L$ by 
\eqref{eq:z_via_g}
\be
A(L) = g_L^{-1} dg_L\quad\Rightarrow\quad z_L(\rho,\varphi) = g_L(\rho,\varphi) \cdot i
\label{eq:zint}
\ee
where $g_L\cdot i$ denotes the fractional linear (M\"obius) action of
$\mathrm{(P)}\SL(2,\RR)$ on the point $i\in \HH$.
We then  determine the local and 
global properties of $z_L$. In particular we explain 
\begin{itemize}
\item how the holonomy of $A(L)$ determines the 
monodromy of $z_L$ and how the 
conjugacy class $\sigma$ of this monodromy (appearing in the classification of Virasoro
orbits by pairs $(\sigma,n_0)$) determines the  global properties of the geometry $ds^2(L)$ 
\item how the winding number $n_0$, the other member of the pair $(\sigma,n_0)$,  
is related to covering geometries and large gauge transformations
\item and how to use the uniformisation map to construct the 
bulk extension $\tilde{f}$ of a given boundary diffeomorphism
$f$ preserving the FG gauge and implementing the transformation $ds^2(L)\mapsto
ds^2(L^f)$, 
  \begin{equation}
    ds^2(L^f) = \tilde f^* ds^2(L)
    \label{eq:gluing_intro}
  \end{equation}
\end{itemize}
While our main interest in the following will be in the uniformisation map
$z_L$ and its local and global properties, the construction of $\tilde f$ is of
some interest in its own right. For example, 
exact expressions for the diffeomorphism $\tilde f$ were previously
obtained in the literature for $L_0=0$ \cite{GNW}
and for the Teichm\"uller orbit $L_0 = 1/4$ \cite{Choi_Larsen}.
Our construction systematically generalises these results to all
Virasoro orbits. Explicit expressions for those orbits that admit a
constant representative $L_0$ are constructed in
Appendix \ref{app:explicit_calc}. 

With the aid of the uniformisation map and an understanding of its
global properties, we are then able to obtain a detailed understanding of the
hyperbolic geometries $ds^2(L)$ asociated with Virasoro coadjoint orbits 
in Section \ref{sec:geometry}. While this
reproduces the known results for the metrics $ds^2(L_0)$ for constant
$L_0$ mentioned above, we also discuss these cases in some detail
in order to be able to contrast and compare these standard geometries with
the more exotic geometries (exotic funnels and cusps) that one
obtains from the orbits without constant representatives. In particular,
we will see that the exotic geometries can be seen 
\begin{itemize}
\item either as ``twisted''
counterparts of standard funnels and cusps, the twist being encoded
in the winding number of the uniformisation map at the asymptotic
boundary, 
\item or as deformations of the standard hyperbolic disc geometry 
which break the isometry group down 
to a hyperbolic or parabolic subgroup of $\SL(2,\RR)$
(and coverings thereof).
\end{itemize}

This distinctive boundary behaviour leads us to then study the
proposed correspondence \eqref{MO} from the point of view of boundary
conditions (which should in any case be part of the definition of
a moduli space) in Section \ref{sec:moduli_spaces}. The usual way
to think about this in the context of JT gravity is in terms of
``wiggly boundaries'' and their Schwarzian extrinsic curvature
(see e.g.\ \cite{MSY,SSS}), in which $L(\varphi)$ determines the subleading 
behaviour of the extrinsic curvature of the (regularised) boundary. 

  While quite natural from the JT gravity point of view, 
geometrically at first this sounds somewhat unusual. 
In order to better understand this, we 
reformulate the correspondence \eqref{MO} more carefully in terms
of geometric structures (as defined e.g.\ in \cite{Goldman,GoldmanHiggs}),
specifically hyperbolic structures on $S$ and projective structures
on $S^1$. In particular, as we will recall, hyperbolic
structures on $S$ induce projective structures on
the asymptotic boundary $\del_\infty S  \cong 
S^1$ of $S$, and  projective structures on $S^1$ (modulo the action of
diffeomorphisms) have the same classification as Virasoro coadjoint
orbits. Thus specifiying the extrinsic curvature via $L(\varphi)$
is equivalent to specifying a boundary projective structure, and is 
indeed the natural boundary condition for hyperbolic structures.

\begin{figure}[htb]
    \centering
    \hspace{4cm}
    \begin{tikzpicture}
	    \tkzDefPoint(0,0){C}

	    \tkzDefPoint(0,.5){A1}
	    \tkzDefPoint(0,1.7){A2}
	    \tkzDefPoint(0,2){A3}
	    \tkzDefPoint(0,2.3){A4}

	    \tkzDrawCircle[thick,black,dashed,pattern=north east lines,pattern color=gray](C,A3)
	    \tkzDrawCircle[thick,black,fill=white](C,A2)
	    \tkzDrawCircle[thick,black](C,A4)
	    \tkzDrawCircle[thick,black](C,A1)

	    \tkzLabelPoint(0,-0.6){\small $A(L_0)$}
	    \tkzLabelPoint(135:3.5){\small $A(L_1)$}
	    \tkzLabelPoint(10:4.5){\small $A(L_1) = A(L_0)^{h_{L_0L_1}}$}

	    \tkzDefPoint(9:2.7){X}
	    \tkzDefPoint(10:1.85){Y}

	    \tkzDefPoint(132:2.7){P}
	    \tkzDefPoint(130:2.2){Q}

	    \tkzDrawSegment[-latex,thick](X,Y)
	    \tkzDrawSegment[-latex,thick](P,Q)

	  \end{tikzpicture}
	  \caption{Construction of the gauge field $A$ on $S$.}
	  \label{fig:intro_gluing_A}
  \end{figure}
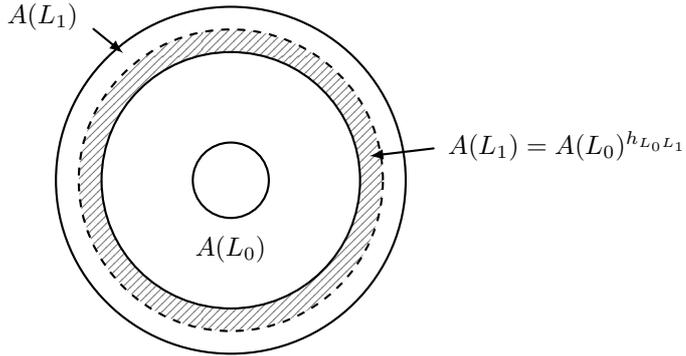

It is therefore natural to define the moduli space of
hyperbolic structures $\M_{\sigma,n_0}(S)$ on $S$
of type $(\sigma,n_0)$ to be the moduli space of hyperbolic structures on $S$
which induce a projective structure of type $(\sigma,n_0)$ on the asymptotic 
boundary. 
  In practice (Section \ref{sub:hs}), a point in 
$\M_{\sigma,n_0}$ is defined by defining an appropriate gauge field $A$ on $S$ which encodes the hyperbolic structure in terms of an appropriate hyperbolic metric.
Deep in the bulk, this gauge field takes the form $A(L_0)$ and describes the model geometry obtained by $L_0$. 
Close to the boundary the gauge field takes the form $A(L_1)$ and describes the geometry obtained from $L_1$ which has to be chosen to satisfy the boundary condition given by the projective structure of type $(\sigma,n_0)$. 
In an intermediate region, $A$ is patched up via a gauge transformation $h_{L_0L_1}$ which takes $A(L_0)$ to $A(L_1)$, see Figure \ref{fig:intro_gluing_A}.
For $A(L_0)$ and $A(L_1)$ to be gauge equivalent, it is crucial that the
holonomies (monodromies) agree: $M_{L_0} = M_{L_1}$.
This determines preferred models for the bulk geometries, namely those corresponding to constant values of $L_0$ and it follows immediately that the only singularities that can appear in the bulk are those of the prototypical geometries described by $ds^2(L_0)$.

We can now give a precise formulation of the proposed correspondence
  \eqref{MO} (Section \ref{sub:modp}):

\paragraph{Main Statement:}

There is a canonical isomorphism 
\begin{equation}
\label{eq:ms1}
\M_{\sigma,n_0}(S) \cong \OO_{\sigma,n_0}
\end{equation}
between the moduli space 
$\M_{\sigma,n_0}(S)$ of hyperbolic structures of type $(\sigma,n_0)$ 
and the Virasoro 
coadjoint orbit $\OO_{\sigma,n_0}$ obtained by associating to any $L_{\sigma,n_0} 
\in\OO_{\sigma,n_0}$ the hyperbolic metric $ds^2(L_{\sigma,n_0})$.

To the best of our knowledge, this precise version of the correspondence
\eqref{MO}
has not been observed in the literature before.
It generalises the well-known correspondence \eqref{eq:td} for the moduli space of 
hyperbolic metrics of the disc $\DD$ to the punctured disc $\DD^*$ (and its
coverings) allowing a conical or cuspidal singularity at the puncture and to
annular geometries (funnels) with standard and exotic boundary conditions.
The details of this correspondence for the individual
Virasoro orbits are summarised in Table \ref{tab:main_statement} in Section 
\ref{sub:modp}. To give a flavour and illustration of the results, some of the 
entries of that table are reproduced in Table \ref{tab:imain_statement} below.
\begin{table}[htb]
  \centering
  \begin{tabularx}{\textwidth}{c|Y|Y|c}
    $\M_{\sigma,n_0}$ & topology & singularity & Virasoro orbit \\ \hline
    $\M_{0,n}$ & $n$-fold branched covers $\DD^*_{0,n}$ of punctured disc $\DD^*$ & conical & $\OO_{0,n}\cong \Diff^{+}(S^1) /
    \PSL^{(n)}(2,\RR)$ \\ \hline
    $\M_{\alpha,0}$ & cone $\DD^*_{\alpha}$ with opening angle $2\pi\alpha$, $\alpha
    \in (0,1)$ & conical &
    $\OO_{\alpha,0}\cong \Diff^{+}(S^1) / S^1$ \\ \hline
    $\M_{\pm,n}$ & exotic cusps $\DD^*_{\pm,1}$ (and their $n$-fold branched covers) & cuspidal
     &
    $\OO_{\pm,n}\cong \Diff^{+}(S^1) / \RR \times \ZZ_n$ 
  \end{tabularx}
 \caption{Examples of the Main Statement (excerpt from Table 
  \ref{tab:main_statement})}
  \label{tab:imain_statement}
\end{table}

In the final Section \ref{sec:outlook}, we return to the physics
questions that were part of the original motivation for this work and 
comment on the possible 
significance and interpretation of the exotic cuspidal and
funnel geometries that we describe from various points of
view in this paper. In particular, we emphasise that they are 
completely understood from a gauge theory perspective, arising
as large gauge transformations of the standard geometries, and 
are thus analogous to the (Lorentzian) ``kink geometries'' of \cite{SchallerStrobl}.
We 
also stress their role as giving rise to new topological sectors
of the theory characterised by boundary conditions
with a non-zero winding number.

  \section{Hyperbolic Gravity and \texorpdfstring{$\SL(2,\RR)$}{SL(2,R)} Gauge Theory}\label{sec:gauge_theory}

  We recall the basics of hyperbolic gravity described as a
  $\SL(2,\RR)$ gauge theory.
  We will describe the prototypical examples and establish a first link with Virasoro coadjoint orbits. 

  \subsection{2d Gravity and \texorpdfstring{$\SL(2,\RR)$}{SL(2,R)} Gauge Theory in the Fefferman-Graham Gauge}\label{sub:cartan}

  Let $S$ be a two-dimensional surface that is topologically 
  a cylinder $\RR\times S^1$. While  
  we will generally refer to $S$ as a cylinder, 
  depending on the metric one puts on it it may be more appropriate to think of 
  $S$ as e.g.\ a (punctured) disc, or an annulus - 
  see Section \ref{sec:examples} for simple examples.
  In order to study hyperbolic metrics on the cylinder $S$, 
  locally modelled on $\SL(2,\RR) / \SO(2)$,
  we consider a flat $\SL(2,\RR)$ connection $A$ on $S$ of the form
  \begin{equation}
    A = \frac12\mat{e^1}{e^2+\omega}{e^2-\omega}{-e^1}
  \end{equation}
  This Cartan connection $A$ encodes both the $\SO(2)$-connection $\omega$ and the vielbeins
  (coframe) $e^a = e^a_{\mu}dx^{\mu}$ defining the metric via the line element
  $ds^2 = \delta_{ab}e^ae^b$. 
  In particular, the $\SO(2)$ spin connection takes values in the
  $\mathfrak{so}(2) \subset \mathfrak{sl}(2,\RR)$ subalgebra of
  $\mathfrak{sl}(2,\RR)$.  This parametrisation is such that 
  \begin{itemize}
    \item 
      the action of coframe $\SO(2)$-rotations on $(e^a,\omega)$ is realised by 
      $\SO(2) \subset \SL(2,\RR)$ gauge transformations of $A$: indeed, for 
      \be
      g(x) =
      \mat{\phantom{-}\cos\Lambda(x)/2}{\sin\Lambda(x)/2}{-\sin\Lambda(x)/2}{\cos\Lambda(x)/2}
      \ee
      one has that 
      $A\to A^g =g^{-1} A g + g^{-1} dg$ 
      implies 
      \be
      A\to A^g \quad\Rightarrow\quad
      \vek{e^1}{e^2}\to
      \mat{\cos\Lambda}{-\sin\Lambda}{\sin\Lambda}{\phantom{-}\cos\Lambda}\vek{e^1}{e^2}
      \quad,\quad
      \omega \to \omega + d\Lambda
      \ee
    \item the flatness
      condition $F_A=0$ for the connection 
      $A$ is precisely equivalent to the statement that the metric has 
      constant negative curvature $-2$: 
      \begin{equation}
	F_A = dA + \frac12 [A,A] = 0\quad\Leftrightarrow\quad
	T^a = de^a + \varepsilon^a_{\ b}\omega \wedge e^b = 0 \quad , \quad d\omega = - e^1 \wedge e^2 
      \end{equation}
      The first condition is the same as the no-torsion condition on 
      $\omega$, and determines the spin connection $\omega = \omega(e)$ in terms of the vielbein.
      With this, the second condition is equivalent to constant negative scalar curvature $R=-2$ of the 
      metric.  
  \end{itemize}
  In the following, we will refer to such metrics on $S$ 
  with constant negative curvature as \emph{hyperbolic metrics}.

  In order to describe these metrics more explicitly, we now introduce coordinates
  $(\rho,\varphi)$, where $\varphi\sim\varphi + 2\pi$ is a standard coordinate
  on $S^1$ and $\rho$ is some radial or axial coordinate. 
  It will turn out to be extremely convenient and useful
  to study hyperbolic metrics on $S$ in the Gaussian or Fefferman-Graham (FG)
  coordinates adapted to the metric. They are characterised
  by the fact that in these coordinates the line element takes the form
  \begin{equation}
    ds^2 = d\rho^2 + g_{\varphi\varphi}(\rho,\varphi) d\varphi^2 
    \quad\Leftrightarrow\quad 
    g_{\rho\rho} = 1\quad,\quad g_{\rho\varphi} = 0
    \label{eq:FG_cond}
  \end{equation}
  In these coordinates, $S$ is naturally foliated by hypersurfaces (circles) 
  of constant $\rho$, with geodesic normal curves---the coordinate lines of
  $\rho$. As we will see, for hyperbolic metrics 
  such FG coordinates will always exist
  asymptotically, for sufficiently large $\rho$, but will not 
  necessarily be defined globally (coordinates rarely are). For most purposes
  this is not a problem (and is amply compensated by the convenience of working in FG
  coordinates).  We will readdress this shortcoming on various occasions in 
  the following - in particular in Section 
  \ref{sec:moduli_spaces}, where we provide a more coordinate-independent
  perspective.

  The above FG conditions on the metric 
  translate to two conditions on the vielbein, namely
  \begin{equation}
    (e^1_{\rho})^2 + (e^2_{\rho})^2 = 1 \quad , \quad e^1_{\varphi}e^1_{\rho} + e^2_{\varphi}e^2_{\rho} \hskip3pt = 0
  \end{equation}
  Moreover, by a coframe rotation, we can set $e^1_{\rho} = 1$.
  It follows that the first condition implies $e^2_{\rho} = 0$, and the second condition yields $e^1_{\varphi} = 0$. Thus we define the gauge theoretic counterpart of the FG gauge by the $\SL(2,\RR)$ 
  gauge conditions 
  \begin{equation}
    e^1_{\rho} = 1 \quad , \quad e^1_{\varphi} = 0 \quad , \quad e^2_{\rho} = 0
    \quad\Leftrightarrow\quad 
    e^1 = d\rho \quad , \quad e^2 = e^2_{\varphi}(\rho,\varphi) d\varphi 
    \label{eq:gfg}
  \end{equation}
  In this gauge, the connection $A$ takes the form
  \begin{equation}
    A = \frac12 \mat{d\rho}{e^2_{\varphi}d\varphi + \omega}{e^2_{\varphi}d\varphi - \omega}{-d\rho}
  \end{equation}
  where $e^2_\varphi = e^2_{\varphi}(\rho,\varphi)$ and $\omega =
  \omega(\rho,\varphi)$ are two initially arbitrary and independent functions and
  one-forms respectively.
  In order to describe hyperbolic geometries, we now impose the flatness condition
  $F_A=0$ on this connection $A$. 
  First of all, the no-torsion condition fixes the spin connection $\omega$ 
  in terms of the function $e^2_{\varphi}(\rho,\varphi)$, 
  \begin{equation}
    T^a =0 \quad\Rightarrow\quad 
    \omega = - \del_{\rho}e^2_{\varphi}(\rho,\varphi) d\varphi
  \end{equation}
  Then the curvature condition $d\omega + e^1\wedge e^2 = 0$ implies
  \begin{equation}
    d\omega + e^1\wedge e^2 = 0 
    \quad\Rightarrow\quad
    (\del_{\rho})^2 e^2_{\varphi}(\rho,\varphi) = e^2_{\varphi}(\rho,\varphi)
  \end{equation}
  whose general solution is parametrized by two periodic functions $u(\varphi)$ and $L(\varphi)$, 
  \[
    e^2_{\varphi}(\rho,\varphi) = u(\varphi) e^{\rho} - L(\varphi)e^{-\rho}
  \]
  Thus in the FG gauge a flat gauge field $A$ has the form 
  \begin{equation}
    A = \mat{\frac{d\rho}{2}}{-L(\varphi)e^{-\rho}d\varphi}{u(\varphi)e^{\rho}d\varphi}{-\frac{d\rho}{2}}
    \label{eq:flat_FG_A}
  \end{equation}
  and the hyperbolic metric it descibes is given by the line element 
  \begin{equation}
    ds^2(L) = d\rho^2 + \left( u(\varphi) e^{\rho} - L(\varphi) e^{-\rho}\right)^2 d\varphi^2 
    \label{eq:ds_w_u}
  \end{equation}
  In this FG gauge, the ideal boundary (the circle at infinity) 
  $\del_{\infty}S$ is located at $\rho \to \infty$.

  We now want to impose an asymptotic boundary condition by demanding that as
  $\rho$ tends to $\infty$ the 
  metric converges to the standard hyperbolic metric there. 
  The standard hyperbolic metric in FG-coordinates is 
  \be
  ds^2 = d\rho^2 + \sinh^2(\rho) d\varphi^2
  \label{eq:ds_std_polar}
  \ee
  The characteristic feature of this metric is that $g_{\varphi\varphi} \sim
  e^{2\rho}$ for $\rho\to\infty$, with a $\varphi$-independent constant coefficient
  of proportionality that can be scaled by a constant shift of $\rho$. 
  Keeping this freedom in the shift of $\rho$ in mind, for simplicity and for
  the time being, we fix
  this constant to be $1$. 
  Comparison with \eqref{eq:ds_w_u} then shows that we need to impose the
  condition $u(\varphi)=1$. 
  Later on, e.g.\ in the discussion of the examples in Section
  \ref{sec:examples}, we will occasionally find it convenient to use the freedom
  in the shift of $\rho$ to put the metrics into some other preferred form,
  e.g.\ as in
  \eqref{eq:ds_std_polar} which has $u = 1/2$. 
  Note that by imposing an asymptotic boundary
  condition that determines $u(\varphi)$, we are now left with only one degree of freedom, namely the choice of the
  periodic function $L(\varphi)$.

  The gauge field \eqref{eq:flat_FG_A} then becomes
  \begin{equation}
    A \equiv A(L) = 
    \mat{\frac{d\rho}{2}}{-L(\varphi)e^{-\rho}d\varphi}{e^{\rho}d\varphi}{-\frac{d\rho}{2}}
    \label{eq:A}
  \end{equation}
  which now describes a hyperbolic metric in the FG gauge, namely
  \begin{equation}
    ds^2(L) = d\rho^2 + \left( e^{\rho} - L(\varphi) e^{-\rho} \right)^2d\varphi^2
    \label{eq:FG_gauge}
  \end{equation}
  subject to the asymptotic boundary condition $g_{\varphi\varphi}\to e^{2\rho}$.
  The corresponding vielbein and spin connection are given by
  \begin{equation}
    e^1 = d\rho \quad , \quad e^2 = \left( e^{\rho} - L(\varphi) e^{-\rho}
    \right)d\varphi \quad , \quad \omega = - \left( e^{\rho} + L(\varphi) e^{-\rho} \right)d\varphi
  \end{equation}

  \subsection{\texorpdfstring{$\SL(2,\RR)$}{SL(2,R)} Holonomy of a Hyberbolic Metric}\label{sec:holonomy1}

  Given a flat $\SL(2,\RR)$-connection $A$ on the cylinder $S$, 
  the $\SL(2,\RR)$-holonomy of $A$, which should not be confused with the
  $\SO(2)$-holonomy of the spin connection, around a circle of constant $\rho$ is defined by 
  the path ordered exponential
  \begin{equation}
    \Hol_{\rho}(A) = \mathrm{P} \exp\left( \int_0^{2\pi} A \right) 
  \end{equation}
  Since the gauge field $A$ is flat, locally, it can generally be written in pure gauge form, 
  $A = g^{-1}dg$, for some group-valued field $g(\rho,\varphi)$.
  Globally, the group-valued field $g(\rho,\varphi)$ is quasi-periodic 
  in the angular coordinate $\varphi$, and the holonomy is precisely the measure of this
  quasi-periodicity, 
  \begin{equation}
    A = g^{-1}dg \quad\Rightarrow\quad  \Hol_{\rho}(A) = \mathrm{P} 
    \exp\left( \int_0^{2\pi} A \right) = g^{-1}(\rho,0) g(\rho,2\pi)
    \label{eq:holag}
  \end{equation}
  An explicit construction of such a $g(\rho,\varphi)$ (in terms of solutions to Hill's equation
  for $L(\varphi)$) will be given in Section \ref{sec:Hill_and_gauge_theory}.

  While the holonomy, as defined above, depends on various choices, its conjugacy class 
  in $\SL(2,\RR)$ (and thus in particular also its trace) does not:
  \begin{enumerate}
\item While $\Hol_\rho(A)$ depends on a choice of base point on $S^1$, its
  conjugacy class $[\Hol_\rho(A)]$ is independent of the choice of base point. 
\item Under gauge transformations that are periodic in $\varphi$, $h(\rho,\varphi+2\pi) = 
  h(\rho,\varphi)$ the holonomy changes
  by conjugation, 
  \be
  \Hol_\rho(A^h) = h^{-1}(\rho,0)\Hol_\rho(A)h(\rho,0)\;\;,
  \ee
  and thus the 
  conjugacy class $[\Hol_\rho(A)]$ is gauge invariant. 
\item Homotopic paths give rise to conjugate holonomies, and circles at different values 
  of $\rho$ are homotopic. Therefore the 
  conjugacy class $[\Hol_\rho(A)]$ is independent of $\rho$. 

  In the case at hand, this can be established very explicitly by observing that 
  the flat gauge field $A = A(L)$ \eqref{eq:A},
  \begin{equation}
    A(L) = \mat{\frac{d\rho}{2}}{-L(\varphi)e^{-\rho}d\varphi}{e^{\rho}d\varphi}{-\frac{d\rho}{2}}
  \end{equation}
  ``factorises'' in the sense that it can be written in terms of a 
  gauge field $A_0(L)$ 
  depending solely on $\varphi$ and a purely $\rho$-dependent gauge transformation $s(\rho)$
  that acts by scaling. Indeed, 
  with 
  \begin{equation}
    A_0(L) = \mat{0}{-L(\varphi)}{1}{0}d\varphi,  \qquad s(\rho) = \mat{e^{\rho/2}}{0}{0}{e^{-\rho/2}},
  \end{equation}
  one has
  \begin{equation}
    A(L) = s^{-1}(\rho) A_0(L) s(\rho) + s^{-1}(\rho) ds(\rho).
    \label{eq:A_factorized}
  \end{equation}
  Hence, the holonomy 
  \begin{equation}
    \Hol_{\rho}(A) = s^{-1}(\rho) \Hol(A_0) s(\rho), \qquad \Hol(A_0) = \mathrm{P} \exp\left( \int_0^{2\pi} A_0 \right) 
  \end{equation}
  is conjugate to the $\rho$-independent holonomy $\Hol(A_0)$ of the $\rho$-independent gauge field 
  $A_0(L)$ and one has
  \be
  [\Hol_\rho(A)] = [\Hol(A_0)]
  \ee

\item Gauge fields $A(L_1)$ and $A(L_2)$ corresponding to two 
  \textit{diffeomorphic} hyperbolic metrics $ds^2(L_{1,2})$ of the form 
  \eqref{eq:FG_gauge} turn out 
  to be gauge equivalent, and thus the conjugacy class 
  $[\Hol(A_0(L))]$ is also diffeomorphism invariant! 
  This will be established infinitesimally in Section \ref{sec:Fred}
  below, and globally in Section \ref{sec:gbulk}.

\item Related to this, and last but not least, 
  it is well known that Virasoro coadjoint orbits can be classified in terms of 
  the conjugacy class of the monodromy matrix of the associated Hill equation
  \cite{Balog_Feher_Palla} (see Appendix \ref{app:vir} for a quick review). 
  As we will explain in Section \ref{sec:Hill_and_gauge_theory}, this monodromy class coincides with the conjugacy class of $\Hol(A_0)$. 
  \end{enumerate}

  For constant $L(\varphi)=L_0$, it is straightforward to calculate this holonomy
  directly. Indeed, since in this case 
  $A_0(L_0)$ is $\varphi$-independent, the path ordered exponential is an
  ordinary matrix exponential which can easily be calculated by noting that 
  \begin{equation}
    \mat{0}{-L_0}{1}{\phantom{-}0} \mat{0}{-L_0}{1}{\phantom{-}0} = -L_0 \mat{1}{0}{0}{1} 
  \end{equation}
  Then one finds 
  \begin{equation}
    \Hol(A_0) = 
    \mat{\cos(2\pi\sqrt{L_0})}{-\sqrt{L_0}\sin(2\pi\sqrt{L_0})}
    {(1/\sqrt{L_0})\sin(2\pi\sqrt{L_0})}{\cos(2\pi\sqrt{L_0})}
    \label{eq:hola0}
  \end{equation}

  In order to determine the conjugacy class to which $\Hol(A_0)$ belongs for a given 
  $L_0$, let us recall (see Appendix \ref{app:virsl2r}) 
  that in $\SL(2,\RR)$ one has the following types of conjugacy classes. 
  We will say that $g \in \SL(2,\RR)$ is 
  \begin{enumerate}
\item \emph{degenerate} if $g$ is conjugate (and hence equal) to plus or minus the identity
  matrix, 
  \begin{equation}
    g = \pm \mat{1}{\phantom{+}0}{0}{\phantom{+}1} = \pm \mathbb{I}
  \end{equation}

\item \emph{elliptic} if $g$ is conjugate to a matrix of the form $\pm M_\alpha$, where
  $\alpha\in (0,1)$ and
  \be
  M_\alpha = 
  \mat{\cos(\pi\alpha)}{-\sin(\pi\alpha)}{\sin(\pi\alpha)}{\phantom{-}\cos(\pi\alpha)} 
  \ee

\item \emph{hyperbolic} if $g$ is conjugate to a matrix of the form $\pm M_\ell$ where
  $\ell \in \RR_+$ and
  \begin{equation}
    M_\ell = \mat{e^{-\pi\ell}}{0}{0}{e^{+\pi\ell}}
  \end{equation}

\item \emph{parabolic} if $g$ is conjugate to a matrix of the form $\pm M_\pm$ (independent signs), where
  \be
  M_\pm = \mat{\phantom{\pm}1}{\phantom{+}0}{\pm1}{\phantom{+}1}
  \ee
  \end{enumerate}

  Correspondingly, for the holonomies $\Hol(A_0)$ 
  there are now 4 cases to distinguish (and in the following we use a more informative notation
  for the different possible constant values of what we have so far just called
  $L_0$; the notation follows that for the classification of Virasoro orbits
  recalled in Appendix \ref{app:vir}):

  \begin{enumerate}

\item Degenerate Holonomy: $L_{0,n} = n^2 / 4, n \in \NN \equiv \NN_{\neq 0}$. 

  In this case one finds the degenerate holonomies
  \begin{equation}
\label{eq:hol0n}
    \Hol(A_0) = (-1)^n \mat{1}{0}{0}{1}
  \end{equation}

\item Elliptic Holonomy: $L_{\alpha,n_0} = (\alpha+n_0)^2 / 4 > 0, \alpha \in (0,1),
  n_0 \in \NN_0$

  In this case, the holonomy \eqref{eq:hola0} is elliptic and conjugate to $\pm M_\alpha$, 
  \begin{equation}
    \Hol(A_0) =
    \mat{\cos(\pi(\alpha+n_0))}{-\frac{\alpha+n_0}{2}\sin(\pi(\alpha+n_0))}{\frac{2}{\alpha+n_0}\sin(\pi(\alpha+n_0))}{\cos(\pi(\alpha+n_0))}
    \sim (-1)^{n_0} M_{\alpha}
  \end{equation}
  Note that (as we will see below) for example 
  $L_{\alpha,0}$ and $L_{\alpha,2}$ correspond to different (and indeed non-diffeomorphic) 
  metrics with the same holonomy.

\item Hyperbolic Holonomy: $L_{\ell,0} = -\ell^2/4 < 0, \ell \in \RR_+$

  In this case the holonomy \eqref{eq:hola0} is hyperbolic and conjugate to $M_\ell$, 
  \begin{equation}
    \Hol(A_0) = 
    \mat{\cosh(\pi \ell)}{\frac{\ell}{2} \sinh(\pi \ell)}{\frac{2}{\ell}\sinh(\pi \ell)}{\phantom{\frac{\ell}{2}}\cosh{(\pi \ell)}} \sim M_\ell
  \end{equation}

\item Parabolic Holonomy: $L_{+,0} = 0$

  In the limit $L_0 \to L_{+,0} = 0$ one obtains from \eqref{eq:hola0} the parabolic holonomy
  \begin{equation}
    \Hol(A_0) = \mat{1}{0}{1}{1} = M_{+}
  \end{equation}
Let us also note for (much) later use that the ``missing'' holonomy $M_-
= (M_+)^{-1}$ can be
obtained from the parabolic gauge field $A_0(L_{+,0}=0)$ by reversing the 
orientation of the circle, $\varphi \to -\varphi$, 
\begin{equation}
\label{eq:Aminus}
    A_0(L_{+,0}) = \mat{\phantom{+}0}{\phantom{+}0}{+1}{\phantom{+}0}d\varphi \to 
    \mat{\phantom{+}0}{\phantom{+}0}{-1}{\phantom{+}0}d\varphi  \quad\Rightarrow\quad \Hol(A_0) = M_+ \to M_-
\end{equation}
  \end{enumerate}

  \subsection{Holonomy and Geometry: Hyperbolic Discs, Cones, Funnels and Cusps}\label{sec:examples}

  Having determined the holonomies, in this section we take a quick look at the 
  corresponding geometries described by the metrics
  \be
  ds^2(L_0) 
  = d\rho^2 + \left( e^{\rho} - L_0 e^{-\rho} \right)^2d\varphi^2
  \label{eq:fgl0}
  \ee
  for constant $L(\varphi)=L_0$. All of these turn out to be familiar
  and well-understood hyperbolic metrics. 
  In the following we will denote the (polar) coordinates in which
  these metrics with constant $L_0$ take a standard form by
  $(\rho_0,\varphi_0)$. 
  \begin{enumerate}
\item Degenerate Holonomy: the disc and its (branched) coverings

  For $L_{0,n}=n^2/4$, the metric initially takes the form
  \begin{equation}
    ds^2(L_{0,n}) = d\rho^2 + \left( e^{\rho} - \frac{n^2}{4}e^{-\rho} \right)^2d\varphi^2,
  \end{equation}
  In order to put this into a more convenient and familiar form, in which the potential degeneracy 
  of $g_{\varphi\varphi}(\rho,\varphi)$ arises not at $\rho = \log(n/2)$ but at the origin, we shift
  $\rho = \rho_0 + \log(n/2)$. With $\varphi = \varphi_0$ one has
  \begin{equation}
    \label{eq:lnm}
    ds^2(L_{0,n}) = d\rho_0^2 + n^2 \sinh^2(\rho_0) d\varphi_0^2.
  \end{equation}
  There are now two different subcases to discuss:
  \begin{itemize}
    \item $L_{0,1}=1/4$

      For $n=1$, one immediately recognises the metric as the standard hyperbolic metric on the (infinite) 
      disc in polar coordinates, 
      \be
      \label{eq:l1m}
      ds^2(L_{0,1}) = d\rho_0^2 + \sinh^2(\rho_0) d\varphi_0^2\;\;.
      \ee
      With the peridocity of $\varphi_0$ fixed to be $2\pi$, 
      as for the standard Euclidean metric in polar coordinates the degeneracy of the metric at 
      $\rho_0=0$ is merely a coordinate singularity. $\rho_0=0$ is a single point, the origin, 
      and negative values of $\rho_0$ should be excluded.

    \item $L_{0,n}=n^2/4, n > 1$

      For $n>1$, the metric is 
\begin{equation} 
\label{eq:lnm2} ds^2(L_{0,n})
      = d\rho_0^2 + n^2\sinh^2(\rho_0) d\varphi_0^2 = d\rho_0^2
      + \sinh^2(\rho_0) d(n\varphi_0)^2 
\end{equation} 
The disc is evidently
      covered once when $\varphi_0$ varies in $[0,2\pi/n)$. Thus,
	given the fixed $2\pi$-range of $\varphi_0$, for $n>1$
	this describes an $n$-fold \textit{covering} of the disc. 
	More
	precisely, this is an $n$-fold cover of the punctured disc 
	(the disc with the origin removed), and an
	$n$-fold \emph{branched covering of the disc} with branch point the origin. 
	This will be explained in more detail in Section
	\ref{sec:degenerate_monodromy_geometry}.

    \end{itemize}

  \item Elliptic Holonomy: Cones and their coverings 

    For $L_{\alpha,n_0} = (\alpha+n_0)^2/4 > 0$, $\alpha \in (0,1), n_0\in \NN_0$
    and after shifting
    $\rho=\rho_0 + \log (\alpha +n_0)/2$, analogously 
    to the above one obtains the metric
    \be
    ds^2(L_{\alpha,n_0}) = d\rho_0^2 + (\alpha+n_0)^2 \sinh^2(\rho_0) d\varphi_0^2
    \label{eq:ds_L_alpha_0}
    \ee
    It is useful to distinguish the two cases $n_0=0$ and $n_0=n\in\NN$.

    \begin{itemize}
      \item $L_{\alpha,0} = \frac{\alpha^2}{4}$, $\alpha \in (0,1)$

	In order to understand the nature of the geometry of the metric at $\rho_0=0$, recall that 
	in standard polar coordinates $(r,\theta)$ in the Euclidean 2-plane $\RR^2$, in the line element
	\be
	ds^2 = dr^2 + \alpha^2 r^2 d\theta^2 = dr^2 + r^2 d(\alpha\theta)^2
	\ee
	the ray through the origin with angle $\alpha\theta$ is to be identified with the ray with angle
	angle $\alpha\theta + 2\pi\alpha$. For
	$0 < \alpha < 1$, this can be visualised by removing a wedge with angle 
	$2\pi (1-\alpha) \equiv 2\pi \delta$ from the disc, and identifying the sides, resulting
	in a cone with opening angle (or cone angle) $2\pi\alpha$, respectively with a deficit
	angle $2\pi\delta$.

	Adopting this terminology in the present case, we see that
	$ds^2(L_{\alpha,0})$ describes a 
	conical hyperbolic geometry, i.e.\ a hyperbolic disc with a conical singularity with 
	cone angle $2\pi\alpha$ at the origin. Again the range of $\rho_0$ should be restricted
	to non-negative values. 

      \item $L_{\alpha,n} = \frac{(\alpha + n)^2}{4}$, $\alpha \in (0,1)$, $n \in
	\NN$.

	For $n\geq 1$, one still has a conical singularity, but in that case
	the opening angle $2\pi(\alpha +n)$ of the cone is larger than $2\pi$, and 
	it is more appropriate to talk of excess angles rather than deficit angles.
	For a nice visualisation of excess angles we refer the reader to
	Elizabethan ruffs \cite{Dunajski_Gavrea}.
	Moreover, just as in the case of the disc, the geometries defined
	for $n > 1$ describe $n$-fold covers of the basic excess geometries one finds
	for $n=1$. In 
	particular, as we will see in Section \ref{sub:excess}, the excess geometry 
	with opening angle $2\pi (n+\alpha) \in (2\pi n, 2\pi(n+1))$ is an $n$-fold
	covering of the excess geometry with angle $2\pi (1+ \alpha/n) \in (2\pi,4\pi)$.

    \end{itemize}

  \item Hyperbolic Holonomy: Hyperbolic Cylinders and Funnels 

    For $L_{\ell,0} = - \ell^2/4$, the metric initially takes the form 
    \be
    ds^2(L_{\ell,0}) = d\rho^2 
    + \left( e^{\rho} + \frac{\ell^2}{4}e^{-\rho} \right)^2d\varphi^2,
    \ee
    which is everywhere non-degenerate. Nevertheless also in this case it is convenient to shift
    $\rho=\rho_0 + \log (\ell/2)$, so that the metric takes the standard (and 
    manifestly $\rho_0 \leftrightarrow -\rho_0$ symmetric) form
    \begin{equation}
      ds^2(L_{\ell,0}) = d\rho_0^2 + \ell^2 \cosh^2(\rho_0) d\varphi_0^2.
      \label{eq:ds2_hyp_prototypical}
    \end{equation}
    Note that this metric has the same asymptotic behaviour for
    $\rho_0\to\pm\infty$ as the hyperbolic disc metric \eqref{eq:l1m}
    for $\rho_0\to +\infty$. Thus
    this metric describes a (wormhole-like) hyperbolic cylinder interpolating between two 
    hyperbolic discs at $\rho_0 \rightarrow \pm \infty$. The significance of the 
    parameter $\ell$ is that the geometry has a unique 
    periodic geodesic at the throat $\rho_0 = 0$, of length $2\pi\ell$.

    Thus this geometry (accidentally) contains two asymptotic regions. Of more 
    relevance in the context of two-dimensional hyperbolic geometry and gravity is
    the so-called \textit{funnel} or \textit{trumpet}, which one obtains 
    by cutting off the space at $\rho_0=0$. Thus a funnel has one geodesic boundary 
    at $\rho_0=0$ and one asymptotic end at $\rho_0\to +\infty$.

  \item Parabolic Holonomy: Cusps 

    When $L_0=L_{+,0}=0$, the metric takes the particularly simple form 
    \begin{equation}
      ds^2(L_{+,0}) = d\rho_0^2 + e^{2\rho_0} d\varphi_0^2.
      \label{eq:ds_par_ex}
    \end{equation}
    The metric is regular for all finite values of $\rho_0$ but
    degenerates as $\rho_0\to -\infty$ and describes what is known
    as a \textit{cusp singularity} there.

    Note that the cuspidal singularity we have just obtained can be regarded and
    obtained as the limit of a conical singularity (as the opening angle
    $\alpha\rightarrow 0$) or a funnel geometry (as the length parameter
    $\ell\rightarrow 0$), as can be seen directly from the FG form \eqref{eq:fgl0}
    of the metric, 
    \be
    \lim_{L_0\to 0_\pm} ds^2(L_0) = d\rho^2 + e^{2\rho}d\varphi^2 = ds^2(L_{+,0})
    \label{eq:limitcusp}
    \ee
    \end{enumerate}

    While this concludes our quick overview of the standard metrics one obtains 
    for $L=L_0$ constant, let us close this section with some remarks and an
    outlook:
    \begin{enumerate}
  \item 

    Locally (and away from possible singularities) any hyperbolic metric
    is of course isometric to the Poincar\'e upper half plane $\HH$ or Poincar\'e 
    disc $\DD$
    equipped with its standard metric. For the simple 
    constant $L=L_0$ metrics we have 
    just discussed, such a \emph{uniformisation map} which makes this 
    explicit is well-known or in any case readily found (see Section
    \ref{sec:geometry} for some more details and figures):
    \begin{itemize}
      \item E.g.\ for the metric 
	$ds^2(L_{0,1})$, the coordinate transformation 
	\begin{equation}
	  w(\rho_0,\varphi_0) = \tanh\left( \frac{\rho_0}{2} \right) e^{i 
	  \varphi_0}
	  \label{eq:trafo_exc1}
	\end{equation}
	maps the metric to the standard Poincar\'e metric on the unit disc $\DD
	= \{ w \in \CC \mid \lvert w \rvert < 1 \}$, 
	\begin{equation}
	  ds^2(L_{0,1}) = ds^2_{\DD} = \frac{4 dw d\bar w}{(1 - \lvert w \rvert^{2})^2}
	  \label{eq:Pdisc}
	\end{equation}
      \item
	For $L_{0,n}$ or $L_{\alpha,0}$, the same objective is accomplished by setting
	\begin{equation}
	  w_n(\rho_0,\varphi_0) = \tanh\left( \frac{\rho_0}{2} \right)
	  e^{i n \varphi_0} \label{eq:trafo_exc}
	  \quad,\quad 
	  w_{\alpha}(\rho_0,\varphi_0) = \tanh\left( \frac{\rho_0}{2} \right)
	  e^{i\alpha \varphi_0} 
	\end{equation} 
	respectively. 
	In these cases, the global geometry is encoded in the nontrivial 
	\textit{winding} $n$ of the
	coordinate $w_n$ for $n>1$, or the non-trivial \textit{monodromy} (lack of
	periodicity) of the coordinate
	$w_\alpha$, 
	\begin{equation}
	  w_\alpha(\rho_0,\varphi_0 + 2\pi) = e^{2\pi i \alpha} w_\alpha(\rho_0,\varphi_0)
	\end{equation} 
	leading to the conical identification of the geometry. 
      \item
	Likewise, for the hyperbolic geometries $L_{\ell,0}$,
	the coordinate transformation 
	\begin{equation}
	  z_{\ell}(\rho_0,\varphi_0) = e^{\ell \varphi_0} \frac{e^{\rho_0} +
	i}{e^{\rho_0} - i} 
	\label{eq:trafo_hyp}
      \end{equation}
      maps the metric to the standard Poincar\'e metric 
      \begin{equation}
	ds^2_{\HH} = \frac{dz\;d\bar{z}}{\Im(z)^2}
      \end{equation}
      on the complex upper half plane $\HH$.
      In this case, the global features are again captured by the non-trivial
      monodromy of the coordinate transformtion. 
    \item
      Finally, for the parabolic geometry $L_{+,0}$, the same is accomplished by setting
      \be
      z_0(\rho_0,\varphi_0) = \frac{1}{2\pi}(\varphi_0 + i e^{-\rho_0}) 
      \label{eq:trafo_par}
      \ee
      In this description, the cuspidal singularity at $z=i\infty$ again arises from 
      the mondromy,  i.e.\ from the periodic identification $z \sim z+1 $ dictated by the periodicity of $\varphi_0$.
  \end{itemize}
\item
  We will explain in Section
  \ref{sec:uniformization_map} how to obtain this uniformisation map for
\textit{any} (not necessarily constant) $L(\varphi)$ 
  from the $\SL(2,\RR)$ gauge theory perspective. 
  The
  global structure of the geometry is then encoded in the monodromies and windings 
  of the uniformising coordinate, and we explain these from the gauge
  theoretic point of view in Sections \ref{sec:M} and \ref{sec:F}.
  This will then allow us to obtain a
  detailed understanding of the geometries $ds^2(L)$ for any
  $L(\varphi)$ in Section \ref{sec:geometry}.
\item
  In all cases (except for the disc $L_{0,1}$) it is possible, and
  occasionally convenient, to perform a further coordinate transformation 
  to single-valued coordinates without winding or monodromy (see Section 
  \ref{sec:geometry} for details), identifying the FG-cylinder
  $(S,ds^2(L_0))$ for a constant $L_0$ with either of the following
  hyperbolic (constant negative curvature) geometries on domains in
  $\CC$:
  \begin{itemize}
    \item the Poincar\'e disc metric $(\DD,ds^2_{\DD})$ \eqref{eq:Pdisc}
      or its (branched) covering $(\DD^{(n)},ds^2_{\DD^{(n)}})$ \eqref{eq:dsdn}
    \item the punctured disc $(\DD^*_{\alpha}, ds^2_{\DD^*_{\alpha}})$
      with metric
      \eqref{eq:ds_alpha_punctured_disc2}, with 
      a conical singularity with opening angle $2\pi\alpha$ at the origin 
    \item the annulus $(\mathcal A_{\ell},ds^2_{\mathcal A_{\ell}})$
      \eqref{eq:annulus2} in the complex plane, 
      with a metric  \eqref{eq:ds_annulus2} 
      with a periodic geodesic of length $\ell$
    \item the punctured disc $(\DD^*_0,ds^2_{\DD^*_0})$ with metric
      \eqref{eq:ds_par_punctured_disc2} and a 
      cuspidal singularity at the origin 
  \end{itemize}

  \end{enumerate}

  \subsection{Diffeomorphism Symmetries and Virasoro Coadjoint Orbits}\label{sub:diffbc}

  For the following, it will be important to know which diffeomorphisms leave the FG gauge invariant, 
  i.e.\ how much of the diffeomorphism gauge freedom of the metric is fixed by the FG gauge. We will look 
  at this question infnitesimally. If $\xi = \xi^{\rho}(\rho,\varphi)\del_{\rho} +
  \xi^{\varphi}(\rho,\varphi)\del_{\varphi}$ is any vector field on $S$,
  then an infinitesimal variation of the metric is given by its Lie
  derivative. If we want to preserve the FG gauge \eqref{eq:FG_cond},
  we thus need to impose the conditions 
  \begin{equation}
    \mathcal L_{\xi} g_{\rho\rho} = 0, \qquad \mathcal L_{\xi} g_{\rho\varphi} = 0
    \label{eq:FG_cond_metric}
  \end{equation}
  on $\xi$.
  An explicit calculation shows that any $\xi$ subject to the conditions above must be of the form
  \begin{equation}
    \xi^{\rho}(\rho,\varphi) = \sigma(\varphi), \qquad
    \xi^{\varphi}(\rho,\varphi) = v(\varphi) - \sigma'(\varphi) \int\frac{d\rho}{g_{\varphi\varphi}(\rho,\varphi)}.
    \label{eq:xi_FG}
  \end{equation}
  We now look at this result in two different situations, for diffeomorphisms that 
  are asymptotically trivial (the identity on the boundary), and for
  diffeomorphisms that are asymptotically non-trivial but that 
  preserve in addition to the FG gauge the asymptotic 
  boundary condition $u(\varphi)=1$. In the context of asymptotic symmetries, 
  the former are regarded as true gauge symmetries while the 
  latter play the role of global symmetries of the theory:

  \begin{enumerate} 
\item Asymptotically trivial diffeomorphisms

  Any diffeomorphism that integrates $\xi$ and is asymptotically
  trivial must be such that $\xi$ vanishes on $\del_\infty S$. If $\xi$
  vanishes on $\del_\infty S$ (or indeed on any of the circles
  $\del_\rho S$ of constant $\rho$ for finite $\rho$), then one must
  have $\sigma(\varphi) = v(\varphi) = 0$ and therefore $\xi(\rho,\varphi)=0$
  everywhere. But then any diffeomorphism integrating $\xi$ is trivial
  and we can conclude that the FG gauge fixes all diffeomorphism which
  restrict to the identity on $\del_{\infty}S$.

\item Infinitesimal diffeomorphisms that preserve the boundary condition

  Let us now consider 
  an infinitesimal diffeomorphism of $S$, generated by 
  a vector field $\xi$, that preserves not only the FG gauge, i.e.\ that satisfies
  \eqref{eq:FG_cond_metric}, but that also preserves the boundary condition 
  which fixes $u(\varphi)=1$ in \eqref{eq:ds_w_u}. 
  Such a diffeomorphism can then only change the function
  $L(\varphi)$, i.e.\ 
  \begin{equation}
    L(\varphi) \to L(\varphi)  + (\delta_\xi L)(\varphi)  
  \end{equation}
  for some $(\delta_\xi L)(\varphi)$.
  Correspondingly the two conditions 
  \eqref{eq:FG_cond_metric} need to be supplemented by 
  \begin{equation}
    \mathcal L_{\xi} g_{\varphi\varphi} \stackrel{!}{=}
    -2\Big( \delta_{\xi}L(\varphi) \Big) \Big( 1 - e^{-2\rho}L(\varphi) \Big)
    \label{eq:Killing}
  \end{equation}
  As we have seen in \eqref{eq:xi_FG}, a vector field $\xi$ satisfying
  the two conditions in \eqref{eq:FG_cond_metric} is parametrized by two periodic functions
  $\sigma(\varphi)$ and $v(\varphi)$, cf. \eqref{eq:xi_FG}.  It is
  straightforward to see that the third condition \eqref{eq:Killing}
  expresses $\sigma(\varphi)$ in terms of $v(\varphi)$ as $\sigma(\varphi)
  = - v'(\varphi)$. Then $\xi = \xi_v(L)$ takes the form 
  \begin{equation}
    \xi_v(L) = -v'(\varphi) \del_{\rho} + \left( v(\varphi) -\frac{v''(\varphi)}{2}\frac{1}{e^{2\rho} - L(\varphi)} \right)\del_{\varphi}
    \label{eq:xi}
  \end{equation}
  and is parametrized by a unique vector field $v=v(\varphi)\del_\varphi$ on the ideal boundary
  $\del_{\infty}S \cong S^1$.
  \end{enumerate}
  The crucial observation for current purposes is now 
  that the induced variation of $L(\varphi)$ generated by $\xi_v(L)$ is
  \begin{equation}
    \delta_{\xi_v(L)}L(\varphi) = 2v'(\varphi)L(\varphi) + v(\varphi)L'(\varphi)
+ \frac{v'''(\varphi)}{2} \equiv \delta_v L(\varphi)
    \label{eq:dxil}
  \end{equation}
  which is \emph{precisely} the infinitesimal coadjoint action
  \eqref{eq:infcoad} of the Virasoro algebra (for more details see Appendix 
  \ref{app:vir} and references given there),  
  with $L(\varphi)$ regarded as an element of the (smooth) dual $\vir^*$ of the
  Virasoro algebra.
  This is the first
  indication of a surprising relation between moduli spaces of hyperbolic metrics and 
  Virasoro coadjoint orbits which we will explore in much more detail in 
  the following. 

  As a first step, let us show that the vector fields $\xi_v(L)$ form a Lie algebra
  isomorphic to the (Witt) Lie algebra $\Vect(S^1)$ of vector fields on the circle 
  $S^1$ (so that the above coadjoint action \eqref{eq:dxil} can be regarded as a 
  projective representation of this algebra of vector fields). To that end, 
  note first of all that $\xi_v(L)$ is linear in $v$, 
  \begin{equation}
    \xi_{v_1 + v_2}(L) = \xi_{v_1}(L) + \xi_{v_2}(L)
  \end{equation}
  Since the $\xi_v(L)$ are field-dependent (in the case at
  hand, they depend on the metric via their
  dependence on $L$), they do not form a closed Lie algebra under the ordinary Lie bracket
  $[\xi_v(L),\xi_w(L)]$. 
  Rather, one has to subtract from this Lie bracket the explicit 
  action of $\xi_v(L)$ on the $L$ appearing in $\xi_w(L)$
  and vice-versa. In this way one is led to the Barnich-Troessaert (BT) bracket 
  \cite{Barnich_Troessaert_I} which is defined by subtracting those contributions, 
  namely
  \begin{equation}
    [\![\xi_v(L),\xi_w(L) ]\!] \defeq [\xi_v(L),\xi_w(L)] 
    - \delta_{\xi_v(L)}\xi_w(L) + \delta_{\xi_w(L)}\xi_v(L)
    \label{eq:BT_bracket}
  \end{equation}
  with 
  \begin{equation}
    \delta_{\xi_v(L)} \xi_w(L) \defeq - \frac{w''}{2}\frac{\delta_v L(\varphi)}{\left( e^{2\rho} - L(\varphi) \right)^2}\del_{\varphi}
  \end{equation}
  With respect to this bracket one has 
  \be
  [\![\xi_v(L),\xi_w(L) ]\!] = \xi_{[v,w]}(L)
  \ee
  where $[v,w]$ denotes the ordinary Lie bracket (or commutator) of vector fields in $\Vect(S^1)$. 
  Thus we see that the set of these vector fields $\xi_v(L)$ equipped with the BT bracket 
  is isomorphic to the (Witt) Lie algebra $\Vect(S^1)$, 
  \be
  \left(\{\xi_v(L)\},[\![.,.]\!]\right) \cong \Vect(S^1) 
  \label{eq:btv1}
  \ee

  Here are some further remarks and observations about the vector fields
  $\xi_v(L)$:

  \begin{itemize}
    \item Gauge Theory Derivation of $\xi_v(L)$ and the Virasoro Coadjoint Action 

      In Section \ref{sec:Fred} below, we will provide a gauge theoretic derivation 
      of the above results by determining the infinitesimal $\SL(2,\RR)$ gauge
      transformations that leave the form \eqref{eq:A} of the gauge field $A(L)$ in the FG gauge 
      invariant.

    \item Bulk Extensions of Boundary Diffeomorphisms

      The vector field $\xi_v(L)$ \eqref{eq:xi} 
      can be regarded as providing an extension of the boundary vector field 
      $v(\varphi)\del_\varphi$ to the bulk spacetime. 
      In Section \ref{sec:bulk_extension} and 
      Appendix \ref{app:explicit_calc}, we will study
      in quite some detail the corresponding non-infinitesimal
      problem, i.e.\ that of extending a given 
      boundary diffeomorphism $f\in \Diff^+(S^1)$ to a bulk diffeomorphism
$\tilde{f}$ of the hyperbolic 
      metric $ds^2(L)$ preserving the FG gauge. 

    \item Stabilisers and Killing Vectors

      Note also that by \eqref{eq:Killing} infinitesimal coadjoint transformations
      that stabilise $L(\varphi)$ (i.e.\ leave $L(\varphi)$ invariant)
      correspond to Killing vectors of the metric.
      This correspondence will be further explored in Section \ref{sec:Isom}.

  \end{itemize}

  \subsection{Virasoro Coadjoint Orbits from Infinitesimal Gauge Transformations}\label{sec:Fred}

  As a gauge theoretic analogue of the previous construction, we will
  now determine those residual (infinitesimal) gauge transformations 
  \be
  \delta_X A = d_AX \equiv dX + [A,X] 
  \ee
  that leave the form \eqref{eq:A} of the gauge field $A(L)$ invariant, i.e.\ that act as 
  \be
  \delta_X A  \stackrel{!}{=} 
  \mat{0}{-(\delta_X L(\varphi)) e^{-\rho}d\varphi}{0}{0}
  \ee
  for some $\delta_X L(\varphi)$. A straightforward calculation shows that such
an $X=X_v(L)$ is 
  parametrised by a single periodic function $v(\varphi)$ on $S^1$ and takes the form
  \begin{equation}
    X_v(L) = \mat{-\frac{v'(\varphi)}{2}}{-\frac12 e^{-\rho} 
  \left( v''(\varphi) + 2 L(\varphi) v(\varphi) \right)}{v(\varphi) e^{\rho}}{\frac{v'(\varphi)}{2}}.
  \label{eq:Xvl}
\end{equation}
Crucially, the variation of $L(\varphi)$ induced by this infinitesimal gauge transformation, 
\begin{equation}
  \delta_{X_v(L)}  L(\varphi) = 2v'(\varphi)L(\varphi) + v(\varphi)L(\varphi)' + \frac{v'''(\varphi)}{2},
  \label{eq:delta_L}
\end{equation}
is identical to the diffeomorphism variation $\delta_{\xi_v(L)} L$
\eqref{eq:dxil}, and thus also acts on $L(\varphi)$ as the infinitesimal
coadjoint action of the Virasoro group corresponding to the vector
field $v= v(\varphi)\del_{\varphi}$.

Note that the $X_v(L)$ do not form a Lie algebra under usual matrix commutator brackets,
\begin{equation}
  [X_v(L),X_w(L)] - X_{[v,w]}= \mat{0}{-e^{-\rho}(v\delta_w L(\varphi) - w\delta_v L(\varphi))}{0}{0}
\end{equation}
This is to be expected since the $X_v(L)$ are a field-dependent generators of gauge transformations
(in the case at hand, they depend on the connection $A(L)$ via their dependence on $L$).
Indeed, the right hand side is exactly the contribution of the
infinitesimal action of the gauge transformation on the connection
via its explicit action on $L(\varphi)$. By analogy with the above
BT construction for vector fields \cite{Barnich_Troessaert_I}, the
correct commutator is defined by subtracting those contributions,
i.e.\ one sets
\begin{equation}
  [\![ X_v(L),X_w(L) ]\!] \defeq [X_v(L),X_w(L)] - \delta_{X_v(L)}X_w(L) + \delta_{X_w(L)}X_v(L)
\end{equation}
with
\begin{equation}
  \delta_{X_v(L)}X_w(L) = \mat{0}{-e^{-\rho}w\delta_v L(\varphi)}{0}{0}
\end{equation}
Then one indeed finds 
\begin{equation}
  [\![ X_v(L),X_w(L) ]\!] = X_{[v,w]}(L)
\end{equation}
so that, in precise analogy with \eqref{eq:btv1}, 
the set of generators $X_v(L)$ equipped with the BT bracket  
is isomorphic to the (Witt) Lie algebra $\Vect(S^1)$, 
\begin{equation} 
\left(\{X_v(L)\},[\![.,.]\!]\right) \cong \Vect(S^1)
\label{eq:btv2}
\ee
In the light of this, it is of considerable interest to understand the relation
between the generators $\xi_v(L)$ \eqref{eq:xi} of infinitesimal diffeomorphisms
and the corresponding generators $X_v(L)$ \eqref{eq:Xvl} of gauge transformations. 
This relation can be understood as a consequence of the charming fact (well known e.g.\
in the context of Chern-Simons gravity) that for a flat connection $A$, $F_A=0$, 
an infinitesimal diffeomorphism can be realised as an infinitesimal field-dependent
gauge transformation. Namely, simply using the Cartan formula for the Lie derivative
and following one's nose one has, with $\Lambda_{\xi} = \iota_{\xi}A$, 
\begin{equation}
  \begin{split} 
    \mathcal L_{\xi} A &= (d\iota_{\xi} + \iota_{\xi}d)A \\
    &= d\Lambda_{\xi} + \iota_{\xi}F_A - \iota_{\xi} \frac12[A,A] \\
    &= d\Lambda_{\xi} + [A,A_{\xi}] + \iota_{\xi}F_A \\
    &= d_A \Lambda_{\xi}
  \end{split}
\end{equation}
This is, as claimed, an infinitesimal gauge transformation, with generator
$\Lambda_{\xi}$. If one 
wishes to prerserve a certain gauge condition on $A$, then the above transformation may have 
to be accompanied by another infinitesimal gauge transformation required to return to the 
chosen gauge.

In the case at hand, one finds 
by direct computation, that the infinitesimal diffeomorphism generator $\xi_v(L)$ and the gauge 
transformation generator $X_v(L)$ are indeed precisely related in the above manner, 
up to an infinitesimal coframe $\SO(2)$-rotation (required to maintain the $\SO(2)$ gauge condition $e^1=d\rho$).
Indeed, one has 
\begin{equation}
\label{eq:XLR}
  X_v(L) = \Lambda_{\xi_v(L)} + \mat{0}{-\theta_v(\rho,\varphi)}{\theta_v(\rho,\varphi)}{\phantom{-}0}
\end{equation}
where the coframe rotation angle is given by 
\begin{equation}
  \theta_v(\rho,\varphi) = \frac{v''(\varphi)}{2} \frac{1}{e^{\rho} - L(\varphi)e^{-\rho}}
\end{equation}
Finally, note that the above general argument that on a flat connection 
infinitesimal diffeomorphisms are realised via infinitesimal gauge transformations
also shows that the conjugacy class of the holonomy $\Hol(A(L))$ is invariant
under infinitesimal diffeomorphisms. In Section
\ref{sec:gbulk} we will establish
this more generally for non-infinitesimal diffeomorphisms $f$, i.e.\ we will see that
\be
[\Hol(A(L^f))]= [\Hol(A(L)]
\ee

\subsection{Isometries and Virasoro Stabilisers}\label{sec:Isom}

As noted before, equation \eqref{eq:Killing} shows that Killing vector fields of the metric 
\[
  ds^2(L) = d\rho^2 + \left( e^{\rho} - L(\varphi)e^{-\rho} \right)^2d\varphi^2
\] 
must satisfy $\delta_{\xi} L = 0$ and are hence in one-to-one correspondence with vector fields $v$ on the circle which stabilise $L(\varphi)$ under the Virasoro coadjoint action, i.e.\ which are such that
\begin{equation}
  \delta_v L(\varphi) = 2v'(\varphi)L(\varphi) + v(\varphi) L'(\varphi) + \frac{v'''(\varphi)}{2} = 0
  \label{eq:delta_v_L_Isom}
\end{equation}
It turns out that for any $L(\varphi)$ there is always at least one non-trivial
periodic solution to \eqref{eq:delta_v_L_Isom}. 
In particular, for any constant $L(\varphi) = L_0$, there is always the obvious solution 
$v_0(\varphi) = 1$.
By \eqref{eq:xi}, this solution generates the constant  Killing vector field 
\begin{equation}
  \label{eq:xi0}
  \xi_{v_0} = \del_{\varphi}.
\end{equation}
Indeed, the metric $ds^2(L_0)$ for constant $L_0$ 
is manifestly invariant under shifts of $\varphi$.
For constant $L_0 \neq n^2/4$ with $n \in \NN$, $v_{0}(\varphi) = 1$ is (up
to normalisation) the unique periodic solution of \eqref{eq:delta_v_L_Isom}
\cite{Witten} and hence the only Killing vector field. An analogous statement 
then also applies to any other metric in the Virasoro diffeomorphism orbit
of such an $L_0$. 

For any non-constant $L$ (whether or not in the orbit of a constant $L_0$) 
one can also follow the discussion in Appendix
\ref{app:stab}
to construct a periodic solution of \eqref{eq:delta_v_L_Isom} 
(in terms of bilinears of solutions to Hill's equation). 
With \eqref{eq:xi} this periodic solution can then be lifted to a Killing 
vector field of the metric $ds^2(L)$. This general construction will turn out to
be useful in our discussion of exotic funnels and cusps in Section
\ref{sec:geometry}. 

For the degenerate cases, $L_{0,n} = n^2/4$, there exist two more periodic solutions,
namely  $v_1(\varphi) = \cos(n\varphi)$ and $v_2(\varphi) = \sin(n\varphi)$.
Equation \eqref{eq:xi} then yields the following two additional Killing vector fields
\begin{equation}
  \label{eq:xi12}
  \begin{split}
    \xi_{v_1} &= \phantom{-}n\sin(n\varphi) \del_{\rho} +
\cos(n\varphi)\coth(\rho)\del_{\varphi} \\
    \xi_{v_2} &= -n\cos(n\varphi) \del_{\rho} +
\sin(n\varphi)\coth(\rho)\del_{\varphi}
  \end{split}
\end{equation}
Since $\{v_0,v_1,v_2\}$ (seen as vector fields on $S^1$) form an
$\mathfrak{sl}(2,\RR)$ algebra with commutation relations
\be
[v_0,v_1] = -nv_2, \qquad [v_0,v_2]= n v_1, \qquad [v_1,v_2] = n v_0,
\ee
so do $\{\xi_{v_0}, \xi_{v_1}, \xi_{v_2}\}$ under the Barnich-Troessaert bracket
\eqref{eq:BT_bracket}. Since the $v_i$ are in the stabiliser of $L$, the BT bracket 
agrees with the ordinary Lie bracket, and therefore one has 
\be
[\![\xi_{v_{i}},\xi_{v_j}]\!] = [\xi_{v_i},\xi_{v_j}] = \xi_{[v_i,v_j]}.
\ee
For $n > 1$, the stabiliser/isometry group generated by $\{\xi_{v_0}, \xi_{v_1},
\xi_{v_2}\}$ is an $n$-fold cover $\SL^{(n)}(2,\RR)$ of $\SL(2,\RR)$
\cite{Witten}. This fact, somewhat obscure from the point of view of the classification of
Virasoro orbits, 
of course fits in perfectly with our observation made in connection with 
\eqref{eq:lnm2} in Section \ref{sec:examples} that the corresponding 
geometries are $n$-fold (branched) covers of the disc,  and therefore have $n$-fold covers of the 
isometry group $\SL(2,\RR)$ of the disc as their isometry groups. 

There is one subtlety, however, regarding this last assertion, that we
need to address. Namely, as we saw in Section \ref{sec:examples} (and will
explain in more detail in Section 
\ref{sec:degenerate_monodromy_geometry}), for $n>1$ the origin at $\rho\to 0$ 
is a singular (branch) point
of the geometry, with cone opening angle $2\pi n$ (whereas it is of course only
a coordinate singularity for $n=1$). Given such a special 
point,  
\begin{itemize}
  \item one can either look at all isometries of the metric, including those that
    move that point: this gives rise to the $\SL^{(n)}(2,\RR)$ isometry group 
    obtained above;
  \item or one can restrict to those isometries that leave that special point
    invariant: in this case the isometry group is reduced to the $\SO(2)^{(n)}\subset
    \SL^{(n)}(2,\RR)$ subgroup of rotations around the origin, 
    generated by $\zeta_{v_0}=\del_\varphi$. 
\end{itemize}
We believe that it is a matter of taste and context which point of view one
adopts.

\section{Uniformisation of Hyperbolic Metrics 
  from \texorpdfstring{$\SL(2,\RR)$}{SL(2,R)} Gauge Theory}\label{sec:uniformisation}

    We construct a \emph{uniformisation map}, a local
    isometry $z_L\colon (S,ds^2(L)) \cong (\HH, ds^2_{\HH})$ from a gauge
    theory perspective. 
    We then study its monodromy and the implications on global aspects of the geometry of
  $(S,ds^2(L))$.

  \subsection{Hill's Equation and \texorpdfstring{$\SL(2,\RR)$}{SL(2,R)} Gauge
Theory}
\label{sec:Hill_and_gauge_theory}

For a more global understanding of the relation among $\SL(2,\RR)$ gauge theory, hyperbolic geometry
and coadjoint orbits explored in the previous section it turns out to be extremely convenient to make
use of the solution to 
an auxiliary problem, namely the properties of solutions to Hill's equation
\begin{equation}
  \psi''(\varphi) + L(\varphi) \psi(\varphi) = 0
  \label{eq:Hill_eqn2}
\end{equation} 
Suppose that $\psi_1(\varphi),\psi_2(\varphi)$ are two linear independent
solutions of \eqref{eq:Hill_eqn2} with normalised Wronskian,
\begin{equation}
  \label{eq:wronski2}
  \psi_1(\varphi)\psi_2'(\varphi) - \psi_1'(\varphi)\psi_2(\varphi) = 1
\end{equation} 
and denote by $\Psi_L = (\psi_1,\psi_2)^t$
the corresponding solution vector. 
The main properties of the solutions of this equation are reviewed in Appendix 
\ref{app:vir}, in particular in relation to the monodromy matrix 
$M_{\Psi_L}\in \SL(2,\RR)$ defined by \eqref{eq:psimono}
\begin{equation}
  \Psi_L(\varphi + 2\pi) = M_{\Psi_L} \Psi_L(\varphi)
\end{equation}
Another useful and ubiquitous object is the \textit{prepotential} \eqref{eq:F}
\be
F_{\Psi_L} = \frac{\psi_2}{\psi_1} 
\ee
which satisfies, in particular \eqref{eq:LSchF}
\begin{equation}
  \frac12 \Sch(F_{\Psi_L})(\varphi) = L(\varphi)
\end{equation}
(hence the name prepotential). 
From the current gauge-theoretic point of view, however, the most useful and informative
object to study turns out to be the corresponding Wronskian matrix 
$W_{\Psi_L}$ defined by
\be
W_{\Psi_L}(\varphi) =
\left(\Psi_L(\varphi)
\Psi'_L(\varphi)\right)
= \mat{\psi_1(\varphi)}{\psi_1'(\varphi)}{\psi_2(\varphi)}{\psi_2'(\varphi)}
\in \SL(2,\RR) 
\label{eq:Wdef}
\ee
The Wronskian matrix has the following important properties:
\begin{enumerate}
    \item \label{prop:W_change_of_basis} If $\hat{\Psi}_L$ is any other Wronskian-normalised solution vector to Hill's equation, 
      then there is a constant $\SL(2,\RR)$ matrix $\S$ such that $\hat{\Psi}_L = S\Psi_L$.
      Under such a change of basis, $W_{\Psi_L}$ transforms as 
      \be
      \hat{\Psi}_L = S\Psi_L \quad\Rightarrow\quad W_{\hat{\Psi}_L} = SW_{\Psi_L} 
      \label{eq:W_SPsi}
      \ee
    \item If $M_{\Psi_L}$ is the monodromy matrix of $\Psi_L$, then 
      $M_{\Psi_L}$ is also the monodromy matrix of the Wronskian matrix, 
      \be
      W_{\Psi_L}(\varphi + 2\pi) = M_{\Psi_L} W_{\Psi_L}(\varphi) 
      \label{eq:monodromy_W}
      \ee
\item The monodromy matrix $M_{\Psi_L}$ is invariant under 
the $\Diff^+(S^1)$ action $L\to L^f,\Psi_L\to \Psi_L^f$
\eqref{eq:Diff_action_on_Hill1} on $L$ and $\Psi_L$ \eqref{eq:mfm}, 
and therefore also the corresponding Wronskian matrices have the 
same monodromy, 
      \be
M_{\Psi_{L^f}} = M_{\Psi_L} \quad\Rightarrow\quad
      W_{\Psi_{L^f}}(\varphi + 2\pi) = M_{\Psi_L} W_{\Psi_{L^f}}(\varphi) 
\label{eq:mfm2}
\ee
    \item 
      For later use, we also note that one can associate to 
      the Wronskian 
      a winding number, counting how many integer times it 
      winds around the compact $\SO(2)\subset \SL(2,\RR)$ subgroup, 
      and that this is equal to the 
      winding number of the
      solution vector $\Psi_L$ regarded as a map to 
      $\RR^2 \setminus \{(0,0)\}$ 
      (see Appendix \ref{app:F_winding}). 
      \end{enumerate}
      With this in hand, we can now establish the connection to the gauge theoretic 
      considerations of Section \ref{sub:cartan} and, in particular, Section 
      \ref{sec:holonomy1}. 
      Namely, recall that the central
      role was played by the flat $\SL(2,\RR)$-connections 
      \be
      A(L) =
      \mat{\frac{d\rho}{2}}{-L(\varphi)e^{-\rho}d\varphi}{e^{\rho}d\varphi}{-\frac{d\rho}{2}}
      = \frac12\mat{e^1}{e^2 + \omega}{e^2 - \omega}{-e^1}
      \ee
      encoding the hyperbolic geometry through the vielbein $e^a$ and spin
      connection $\omega$ of the metric \eqref{eq:FG_gauge}.
      Recall also from \eqref{eq:A_factorized}
      that the gauge field $A(L)$ factorises in the sense that 
      $A(L) = s^{-1}(\rho) A_0(L) s(\rho) + s^{-1}(\rho) ds(\rho)$
      where 
      \begin{equation}
	A_0(L) = \mat{0}{-L(\varphi)}{1}{0}d\varphi \quad ,  
	\quad s(\rho) = \mat{e^{\rho/2}}{0}{0}{e^{-\rho/2}}
      \end{equation}
      Furthermore, since $A(L)$ is flat, it must be possible to write $A(L)$ in
      pure gauge form, $A(L) = g_L^{-1} dg_L$, for some group valued field 
      $g_L(\rho,\varphi)$ with possible non-trivial
      monodromy along the angular coordinate $\varphi$, and likewise for $A_0(L)$. 

      The required
      group valued field is provided to us by the Wronskian matrix. Indeed, a straightforward
      calculation shows that 
      \be
      \label{eq:a0gdg}
      \left(W_{\Psi_L}\right)^{-1} dW_{\Psi_L} = A_0(L) 
      \ee
      and therefore one also has 
      \be
      \left(W_{\Psi_L}s(\rho)\right)^{-1} d(W_{\Psi_L}s(\rho))  = A(L) 
      \label{eq:agdg}
      \ee
      Thus the required group-valued field is given by 
      \be
      g_{\Psi_L}(\rho,\varphi) = W_{\Psi_L}(\varphi) s(\rho)
      \label{eq:def_g}
      \ee
      with monodromy
      \begin{equation}
	g_{\Psi_L}(\rho,\varphi+2\pi) = M_{\Psi_L}g_{\Psi_L}(\rho,\varphi)
	\label{eq:monodromy_g}
      \end{equation}
      as follows directly from \eqref{eq:monodromy_W}.
      Explicitly, one has 
      \begin{equation}
	g_{\Psi_L}(\rho,\varphi) = 
	\mat{e^{\rho/2}\psi_1(\varphi)}{e^{-\rho/2}\psi_1'(\varphi)}
	{e^{\rho/2}\psi_2(\varphi)}{e^{-\rho/2}\psi_2'(\varphi)} \;\;.
	\label{eq:gpsi}
      \end{equation}
      This is the key result which will allow us to explicitly construct the
      uniformisation map for all $L(\varphi)$ in Section \ref{sec:uniformization_map},
      and to subsequently 
      obtain a detailed 
      understanding of the hyperbolic geometries $ds^2(L)$ for any $L(\varphi)$
      (Section \ref{sec:geometry}). 

      This construction of $g_{\Psi_L}$ in terms of solutions to Hill's equation
      now also allows us to establish
      a direct link between the holonomy
      \begin{equation}
	\Hol_{\rho}(A(L)) = g_{\Psi_L}^{-1}(\rho,0) g_{\Psi_L}(\rho,2\pi)
	\label{eq:Hol}
      \end{equation}
      of the connection $A(L)$, and the monodromy $M_{\Psi_L}$ of the solutions of the 
      corresponding Hill's equation. 
      Namely, from 
      \be
      g_{\Psi_L}(\rho,2\pi) = M_{\Psi_L}  g_{\Psi_L}(\rho,0) 
      \ee
      one sees that the holonomy and the monodromy are conjugate to each other, 
      \be
      \Hol_{\rho}(A(L))= g_{\Psi_L}^{-1}(\rho,0) M_{\Psi_L} g_{\Psi_L}(\rho,0) 
      \ee
      and hence define the same conjugacy class
      \begin{equation}
	[M_{\Psi_L}] = [\Hol_{\rho}(A)(L)] = [\Hol(A_0)(L)]
      \end{equation}
      The last equality follows directly from \eqref{eq:def_g} and
      \eqref{eq:Hol}.

      To complete this discussion, one can now also easily prove that 
      connections $A(L)$ and $A(L^f)$ associated to hyperbolic geometries in the
same Virasoro orbit are gauge equivalent. We will postpone the simple proof of
this assertion to Section \ref{sec:gbulk}, where we will explore such  
issues in somewhat more detail and generality. 
In any case, anticipating this result 
it follows that the conjugacy class
      $[\Hol(A(L))]$ is the same for $L$ and $L^f$ and thus 
uniquely associated to the entire cadjoint orbit $\OO_L$
      through $L$, and agrees with the conjugacy class $[M_{\Psi_L}]$  of the monodromy
      associated to $\OO_L$.

      Finally a word on  notation: the matrix $g_{\Psi_L}$ 
      only depends weakly and in an uninteresting way 
      on the choice of solution vector, namely via left-multiplication by the constant
      $\SL(2,\RR)$-matrix $S$, and the key identities  \eqref{eq:a0gdg} and \eqref{eq:agdg} are
      invariant under this transformation.
      Therefore, to unburden the notation somewhat, we will in the following
frequently 
      just write $g_L(\rho,\varphi)$, so that one has
      \begin{equation}
	g_{L}(\rho,\varphi) = 
	\mat{e^{\rho/2}\psi_1(\varphi)}{e^{-\rho/2}\psi_1'(\varphi)}
	{e^{\rho/2}\psi_2(\varphi)}{e^{-\rho/2}\psi_2'(\varphi)} 
	\label{eq:g}
      \end{equation}
      and $A(L)= (g_L)^{-1} dg_L$ etc. In Section \ref{sec:uniformization_map} below 
      we will use the maps $g_L$ to construct a uniformisation map for the hyperbolic metrics
      $ds^2(L)$.

      \subsection{Uniformisation Map from Flat 
	\texorpdfstring{$\SL(2,\RR)$}{SL(2,R)} Connections}\label{sec:uniformization_map}

	We would like to study the geometry defined by the metric associated to an
	arbitrary point $L(\varphi) = L_0^{f}(\varphi)$ of a given Virasoro coadjoint orbit $\OO_{L_0}$.
	To do so, it will be convenient to define a \emph{uniformisation map} which relates
	the metric $ds^2(L)$ on the cylinder $S$ to the standard Poincar\'e metric 
	\be
	ds^2_{\HH}= \frac{dzd\bar{z}\;}{\Im(z)^2}
	\label{eq:Poincare}
	\ee
	on the upper half plane 
	\be
	\HH = \{z\in \CC\colon \; \Im(z) >0\}
	\ee
	This map is conveniently defined in terms of the $\SL(2,\RR)$-valued field $g_L(\rho,\varphi)$ 
	\eqref{eq:g} 
	defined and studied in the previous subsection. 

	The existence of such an uniformisation map follows from the following important
	observation: Denote by $\theta_{\SL(2,\RR)}$ the Maurer-Cartan form on $\SL(2,\RR)$.
	Then 
	\begin{equation}
	  A(L) = g_L^{-1}dg_L = g_L^*\theta_{\SL(2,\RR)}
	\end{equation}
	Crucially, since $A(L)$ encodes the vielbein (and spin connection) of the metric 
	\begin{equation}
	  ds^2(L) = d\rho^2 + \left( e^{\rho} - L(\varphi) e^{-\rho} \right)^2
	  d\varphi^2
	\end{equation}
	while $\theta_{\SL(2,\RR)}$ encodes the vielbein (and spin connection) of the 
	standard hyperbolic
	metric on $\HH$, the metrics $ds^2(L)$ and $ds^2_{\HH}$, and 
	hence the geometries $(S,ds^2(L))$ and $(\HH,ds^2_{\HH})$, are (locally) diffeomorphic.

	As an aside we note that this local diffeomorphism is related
	to what is known as the \emph{developing map} in the language
	of Cartan geometry.  While a detailed account of Cartan
	geometry is beyond the scope of (and not needed for) this
	article, we would like to point the interested reader to
	\cite{Wise} and references therein for a gentle introduction
	to the topic in a context similar to ours.

	In the context of this article, we will use the more common
	(in the hyperbolic context) term \emph{uniformisation map}
	for this local diffeomorphism, and in the following we will
	give the explicit construction of this uniformisation map,
	and then study in detail its properties in the remainder
	of this Section.

	First note that \eqref{eq:g} defines a map  $g_L: S \to \SL(2,\RR)$.
	Projecting the image of this map onto the quotient space $\SL(2,\RR)/\SO(2)$
	and identifying the latter with $\HH$ gives the desired map $S \to \HH$.
	The isomorphism $\SL(2,\RR)/\SO(2) \cong \HH$ can be realised explicitly by
	\begin{equation}
	  [M] \mapsto M \cdot i \quad , \quad [M] \in \SL(2,\RR)/\SO(2)
	\end{equation}
	where $M \cdot i$ denotes the action of the $\SL(2,\RR)$ matrix $M$ 
        on $\HH$ via fractional linear transformations, here conveniently but 
        somewhat unconventionally (for reasons explained below) defined by 
	\begin{equation}
	  M = \mat{a}{b}{c}{d} \colon z \mapsto M\cdot z = \frac{c + d z}{a + b z} \quad , \quad z \in \HH
	  \label{eq:frac_lin_on_z}
	\end{equation}
These $\mathrm{(P)}\SL(2,\RR)$-transformations form the group of
orientation-preserving isometries of the Poincar\'e metric $ds^2_\HH$ on $\HH$. 
For later use we note that the Poincar\'e metric is also invariant
under the orientation-reversing transformation 
\be
P\colon z \mapsto -\bar{z} \quad\Leftrightarrow\quad x + iy \mapsto -x + iy
\label{eq:P}
\ee
which is a reflection of the complex upper half plane on the imaginary axis. 

	In summary, we define the uniformisation map by
	\begin{equation}
	  z_L \colon (\rho,\varphi) \mapsto z_L(\rho,\varphi) = g_L(\rho,\varphi)\cdot i
	  \label{eq:z_via_g}
	\end{equation}
	Explicitly, using \eqref{eq:g} we find
	\begin{equation}
	  z_L(\rho,\varphi) = \frac{e^{\rho/2} \psi_2(\varphi) + i
	  e^{-\rho/2}\psi_2'(\varphi)}{e^{\rho/2}
	  \psi_1(\varphi) + i e^{-\rho/2}\psi_1'(\varphi)} =  \frac{e^{\rho} \psi_2(\varphi) + i
	\psi_2'(\varphi)}{e^{\rho}
      \psi_1(\varphi) + i \psi_1'(\varphi)}
      \label{eq:z}
    \end{equation}

    We end this section with a remark on the definition of the fractional linear transformation \eqref{eq:frac_lin_on_z}.
    At first glance, \eqref{eq:frac_lin_on_z} looks unfamiliar and indeed one
    could have defined $z_L(\rho,\varphi)$ in terms of the standard $\SL(2,\RR)$-action on $\HH$
    \begin{equation}
      \mat{a}{b}{c}{d} \colon z \mapsto \frac{a z + b}{c z + d}\quad , \quad z \in \HH
      \label{eq:std_frac_lin_on_z}
    \end{equation}

    However, with our definition of the Wronskian \eqref{eq:wronski2}, we find the choice \eqref{eq:frac_lin_on_z} more
    natural, as the following example shows: Consider the Hill potential $L_{+,0} = 0$.
    Then a Wronskian-normalised basis is given by $\psi_1(\varphi)
    =\sqrt{2\pi}$ and $\psi_2(\varphi) = \frac{\varphi}{\sqrt{2\pi}}$.
    The uniformisation map then yields the upper half plane coordinate
    \begin{equation}
      z_{L_{+,0}}(\rho,\varphi) = \frac{1}{2\pi}\left( \varphi + i e^{-\rho}
      \right).
    \end{equation}
    If we had used \eqref{eq:std_frac_lin_on_z}, we would have
    obtained $1/z_{L_{+,0}}(\rho,\varphi)$ instead. 
Note also (again for later
use) that in this case the orientation-reversing isometry $P$ \eqref{eq:P}
simply acts as an orientation-reversal of the circle parametrised by $\varphi$, 
    \begin{equation}
     (Pz_{L_{+,0}})(\rho,\varphi) = \frac{1}{2\pi}\left( -\varphi + i e^{-\rho}
      \right) = z_{L_{+,0}}(\rho,-\varphi) 
    \end{equation}

    \subsection{Local Properties of the Uniformisation Map}\label{sec:properties_z}

    In order to better understand the uniformisation map \eqref{eq:z}, we will list
    its most important properties below.
    Many of these properties follow directly from the definition \eqref{eq:z_via_g}
    or the explicit formula \eqref{eq:z} for $z_L(\rho,\varphi)$.
    Nevertheless, it is instructive to verify them explicitly.

    \begin{enumerate}
      \item $z_L$ takes values in the upper half plane $\HH$

	It is clear from the definition \eqref{eq:z_via_g}, that $z_L(\rho,\varphi)$
	takes values in $\HH$.
	This can also be seen explicitly from \eqref{eq:z}, as
	\begin{equation}
	  \Im(z_L(\rho,\varphi)) =  \frac{e^{\rho}\big(\psi_1(\varphi)\psi_2'(\varphi) -
	\psi_1'(\varphi)\psi_2(\varphi)\big)}{e^{2\rho} \psi_1^2(\varphi) +
      \psi_1'^2(\varphi)} =  \frac{e^{\rho}}{e^{2\rho} \psi_1^2(\varphi) +
    \psi_1'^2(\varphi)} > 0
  \end{equation}
  where we used the normalisation of the Wronskian in the last step.

\item $z_L$ uniformises the metric $ds^2(L)$

  For any given $L(\varphi)$, the uniformisation map $z_L$ 
pulls back the standard
  hyperbolic metric on $\HH$ to the metric $ds^2(L)$ in the FG gauge on $S$, i.e.\ 
\be
\label{eq:zlhl}
z_L^*ds^2_{\HH} =ds^2(L)
\ee
or, more explicitly, 
  \begin{equation}
    \frac{dz_L(\rho,\varphi)\; d\bar z_L(\rho,\varphi)}{\Im(z_L(\rho,\varphi))^2}= d\rho^2 + \left( e^{\rho} - L(\varphi) e^{-\rho} \right)^2 d\varphi^2 
    \label{eq:uniformisation_metric}
  \end{equation}
  This follows on general grounds from the discussion at the beginning of
  Section \ref{sec:uniformization_map}.
  Here, we verify it explicitly.
  First note that 
  \begin{equation}
    dz_L(\rho,\varphi) = \frac{-i e^{\rho} d\rho}{(e^{\rho}\psi_1(\varphi) + i \psi_1'(\varphi))^2} + \frac{\left( e^{2\rho} - L(\varphi) \right)d\varphi}{(e^{\rho}\psi_1(\varphi) + i \psi_1'(\varphi))^2}
  \end{equation}
  It follows that 
  \begin{equation}
    \begin{split} 
      \frac{dz_L(\rho,\varphi)\; d\bar z_L(\rho,\varphi)}{\Im(z_L(\rho,\varphi))^2} &= \frac{e^{2\rho}d\rho^2 + (e^{2\rho} - L(\varphi))^2d\varphi^2}{\left( \frac{1}{2i} \right)^2\lvert e^{\rho}\psi_1(\varphi) + i \psi_1'(\varphi) \rvert^4\left( \frac{e^{\rho}\psi_2(\varphi) + i \psi_2'(\varphi)}{e^{\rho}\psi_1(\varphi) + i \psi_1'(\varphi)} - \frac{e^{\rho}\psi_2(\varphi) - i \psi_2'(\varphi)}{e^{\rho}\psi_1(\varphi) - i \psi_1'(\varphi)} \right)^2} \\
      &= \frac{e^{2\rho}\left( d\rho^2 + (e^{\rho} - L(\varphi)e^{-\rho})^2d\varphi^2\right)}{ \left(\Im (e^{\rho}\psi_2(\varphi) + i \psi_2'(\varphi))(e^{\rho}\psi_1(\varphi) - i \psi_1'(\varphi)) \right)^2} \\
      &= d\rho^2 + \left( e^{\rho} - L(\varphi) e^{-\rho} \right)^2 d\varphi^2
    \end{split}
  \end{equation}
  as claimed.

\item Uniqueness of $z_L$ up to fractional linear $\SL(2,\RR)$-transformations

  Recall that under a change of basis $\hat \Psi_L = S\Psi_L$, for some
  constant $S \in \SL(2,\RR)$, one has $\hat g_L = S g_L$ as follows from \eqref{eq:W_SPsi} and the definition of $g_L$ \eqref{eq:def_g}.
  As a consequence of the definition \eqref{eq:z_via_g}, the uniformisation map
  is defined only modulo $\SL(2,\RR)$ fractional linear transformations, i.e.\ 
  up to orientation-preserving isometries of the upper half plane metric
$ds^2_{\HH}$, 
  \begin{equation}
    \hat z_L(\rho,\varphi) = S \cdot z_L(\rho,\varphi)
    \label{eq:z_change_of_basis}
  \end{equation}
  as one would expect.
  Here, $\hat z_L$ is defined with respect to $\hat\Psi_L = S\Psi$, while $z_L$ is defined with respect to $\Psi$.

\item Singular locus of $z_L$

  As we have mentioned several times before, the uniformisation map gives a
  \emph{local} diffeomorphism between the cylinder and the upper half
  plane.
  This can be seen explicitly seen by studying its Jacobian
  \begin{equation}
    J = \mat{\frac{\del z_L}{\del \rho}}{\frac{\del z_L}{\del
    \varphi}}{\frac{\del \bar z_L}{\del \rho}}{\frac{\del \bar z_L}{\del \varphi}} = \mat{\frac{-i}{(e^{\rho}\psi_1(\varphi) + i \psi_1'(\varphi))^2}}{\frac{e^{2\rho} - L(\varphi)}{(e^{\rho}\psi_1(\varphi) + i \psi_1'(\varphi))^2}}{\frac{i}{(e^{\rho}\psi_1(\varphi) - i \psi_1'(\varphi))^2}}{\frac{e^{2\rho} - L(\varphi)}{(e^{\rho}\psi_1(\varphi) - i \psi_1'(\varphi))^2}}
  \end{equation}
  Since
  \begin{equation}
    \det J = - 2i \cdot \frac{e^{2\rho} - L(\varphi)}{\lvert e^{\rho} \psi_1(\varphi) + i \psi_1'(\varphi) \rvert^4}
    \label{eq:det_J}
  \end{equation}
  clearly, the coordinate transformation is ill-defined at points where \mbox{$e^{2\rho} = L(\varphi)$}.
  In Appendix \ref{app:explicit_calc},
  we show that if we restrict to
  geometries defined by Virasoro orbits which admit a constant representative
  $L_0$, there exist global coordinates in which the metric $ds^2(L)$ takes the
  form of prototypical geometries discussed in Section \ref{sec:examples}.
  This exposes most of the apparent singularities of the metric 
  \eqref{eq:uniformisation_metric}
  at the locus $e^{2\rho}
  = L(\varphi)$ as mere coordinate singularities. 

\item Monodromy of $z_L$ 

  The image of the FG-cylinder $S$
  under $z_L$
  is subject to certain identifications, as we now show.  
  Recall from \eqref{eq:monodromy_g} that
  \begin{equation}
    g_L(\rho,\varphi + 2\pi) = M_{\Psi_L} g_L(\rho,\varphi) 
  \end{equation}
  where $M_{\Psi_L}$ is the monodromy of the solution vector $\Psi_L$ of the Hill
  problem associated to $L$.
  From \eqref{eq:z_via_g} follows immediately that
  \begin{equation}
    z_L(\rho,\varphi+2\pi) = g_L(\rho,\varphi+2\pi)\cdot i = M_{\Psi_L}
    g_L(\rho,\varphi) \cdot i = M_{\Psi_L} \cdot z_L(\rho,\varphi)
    \label{eq:glue_z}
  \end{equation}
  where the action of $M_{\Psi_L}$ on $z_L(\rho,\varphi)$ is given by fractional linear transformations defined in \eqref{eq:frac_lin_on_z}. 
  We will discuss the consequences of this in general terms in Section
  \ref{sec:M} and subsequently in more detail in Section
  \ref{sec:geometry}. 
  In particular, we will see how singularities of the metrics
  $ds^2(L)$ arise from the non-singular metric $ds^2_{\HH}$ of the
  upper half plane.

\item Asymptotic behaviour

  Away from the zeros of $\psi_{1,2}$, the uniformisation map has the following asymptotic behaviour in the limit $\rho \to \infty$
  \begin{equation}
    \lim_{\rho \to \infty} z_{L}(\rho,\varphi) \mathop{\sim}\limits_{\rho \to \infty} F_L(\varphi) + i e^{-\rho} F_L'(\varphi) + \OO(e^{-2\rho})
    \label{eq:z_asymptotics}
  \end{equation}
  where
  \begin{equation}
    F_{L}(\varphi) \defeq \frac{\psi_2(\varphi)}{\psi_1(\varphi)}
  \end{equation}
  In particular, the boundary value of $z_L$ is given by
  \begin{equation}
    \lim_{\rho \to \infty} z_{L}(\rho,\varphi) = F_L(\varphi)
    \label{eq:imz0}
  \end{equation}
  In fact, \eqref{eq:imz0} holds for \emph{all} values of $\varphi$, i.e.\ even along zeros of $\psi_{1,2}$.

  Since $F_L(\varphi) \in \RR$, the asymptotic behaviour of $z_L$ shows that the ideal
  boundary $\del_{\infty}S$ of the FG-cylinder $S$ is mapped to the
  ideal boundary of $\HH$ sitting at ${\Im}(z_L) = 0$.

  The function $F_L$ is precisely the prepotential \eqref{eq:F} of the Hill
  potential, and plays an important role in the classification of
  Virasoro orbits, cf.\ Appendix \ref{app:F}. Here we learn that this 
  prepotential also determines the asymptotics of the uniformisation map, and we
  will revisit this important observation in the guise of a \textit{projective structure} on the boundary 
  in Section \ref{sub:proj}. 

\end{enumerate}

\subsection{Bulk \texorpdfstring{$\Diff^+(S^1)$}{Diff(S1)} 
Action I: via the Uniformisation Map} 
\label{sec:bulk_extension}

In Section \ref{sub:diffbc} we had seen that for every boundary vector field
$v(\varphi)$ there is an infinitesimal bulk
diffeomorphisms $\xi_v$ \eqref{eq:xi} 
which is such that (a) it preserves
the bulk FG gauge, and (b) it acts on the metric $ds^2(L)$ 
via the infinitesimal Virasoro coadjoint action $L\to L+\delta_v L$ \eqref{eq:dxil}.

We will now explain how to construct the corresponding bulk extension
$\tilde{f}$ of a finite boundary diffeomorphism $f\in \Diff^+(S^1)$, i.e.\
a diffeomorphism $\tilde{f}$ which acts on the hyperbolic metric $ds^2(L_0)$
(for some reference point $L_0$) 
preserving the FG gauge and mapping it (under pullback) to the hyperbolic
metrics $ds^2(L_0^f)$,
\be
\label{eq:ftllf}
\tilde{f}^*ds^2(L_0) = ds^2(L_0^f)
\ee
Here we will give a concrete prescription for $\tilde{f}$ 
in terms of the uniformisation maps 
constructed above, which can be used for explicit calculations (Appendix 
\ref{app:explicit_calc}). A more gauge theoretic
perspective on the existence and construction of such an $\tilde{f}$ is then
provided in Section \ref{sec:gbulk}.

Recall that the uniformisation map $z_L$
\eqref{eq:z} allows us to locally identify $(S,ds^2(L))$ with $(\HH,ds^2_{\HH})$.
Given a Virasoro orbit $\OO$, let us fix a reference point $L_0(\varphi) \in \OO$ such that any other point in the orbit takes the form $L_0^f(\varphi)$.
Moreover, let $(\rho_0,\varphi_0)$ be the coordinates of the reference point, i.e.\ so that the metric takes the form 
\begin{equation}
  ds^2(L_0) = d\rho_0^2 + \left(e^{\rho_0} - L_0(\varphi_0)
  e^{-\rho_0}\right)^2 d\varphi_0^2
\end{equation}
We then construct a local diffeomorphism $\tilde f$ as follows:
\begin{equation}
  \label{eq:ftdef}
  \tilde f = z_{L_0}^{-1} \circ z_{L_0^f} \colon (S, ds^2(L_0^f)) \to
  (\HH,ds^2_{\HH}) \to (S, ds^2(L_0)).
\end{equation}

The definition of $\tilde f$ assumes that the images of $S$ under $z_{L_0}$
and $z_{L_0^f}$ have a non-trivial intersection in $\HH$.
Without giving a technical argument, one can see directly from the
explicit calculations in Section \ref{sec:geometry} that for $\rho$
large enough such a non-trivial intersection always exists. 
Given this, 
the key property \eqref{eq:ftllf} of $\tilde{f}$ 
follows immediately from the definition \eqref{eq:ftdef} and the 
defining property \eqref{eq:zlhl} of the uniformisation map $z_L$, 
namely that it relates the metric $ds^2(L)$ to $ds^2_\HH$, 
\be
\label{eq:ftllf2}
ds^2(L_0^f) = z_{L_0^f}^* ds^2_{\HH} = (z_{L_0}\circ \tilde f)^*ds^2_\HH = 
\tilde{f}^* ds^2(L_0) 
\ee
We will also use $\tilde{f}$ to determine and read off the coordinate
transformation from the FG coordinates $(\rho,\varphi)$ to the standard
coordinates $(\rho_0(\rho,\varphi),\varphi_0(\rho,\varphi))$ of some reference metric,
by comparing the images 
\begin{equation}
\label{eq:zfz}
  z_{L_0^f}(\rho,\varphi) = z_{L_0}(\rho_0,\varphi_0)
\end{equation}
For that  it is useful to establish from the outset that $\tilde{f}$ has all
the local properties required to define a valid coordinate transformation, 
and this is what we will do in the remainder of this Section. 

First of all, 
since both $z_{L_0}$ and $z_{L_0^f}$ are only uniquely defined up to fractional
linear transformations \eqref{eq:z_change_of_basis}, we need to be more
specific. The choice we will make in the above definition of $\tilde{f}$ is that
if $z_{L_0}$ is defined in terms
of the Hill solution vector $\Psi_{L_0}$, then $z_{L_0^f}$ is to be 
defined by the $f$-transformed solution vector 
$\left( \Psi_{L_0} \right)^f$ \eqref{eq:Diff_action_on_Hill1}. 
We will now show that with this choice $\tilde{f}$ has all 
the desirable properties: 

\begin{enumerate}
  \item $\tilde f$ is locally invertible

    We have remarked before that for $\rho$ large enough the images of $S$ under $z_L$ and $z_{L^{f}}$ have non-trivial intersection in $\HH$ so that $\tilde f$ exists on this intersection. 
    It follows immediately from its definition \eqref{eq:ftdef} that then also $\tilde f^{-1}$ exists on this intersection.

  \item $\tilde{f}$ is independent of the choice of $\Psi$

    The above definition of $\tilde{f}$
    may still appear to depend on the choice of solution vector
    $\Psi_{L_0}$,  but it is actually independent of this choice. In order to see
    that, note that if 
    $\hat z_{L_0}$ denotes the uniformisation map constructed from
    $\hat\Psi_{L_0} = S\Psi_{L_0}$, for some $S \in \SL(2,\RR)$, then 
    from \eqref{eq:z_change_of_basis} one has 
    \be
    \hat z_{L_0} = S\cdot z_{L_0}
    \ee
    Moreover, since 
    Since $(\hat\Psi)^f = (S\Psi)^f = S(\Psi)^f$, we find from
    \eqref{eq:z_change_of_basis} that $z_{L_0^f}$ also transforms as
    \be
    \hat z_{L_0^f} = S\cdot z_{L_0^f} 
    \ee
    and therefore $S$ drops out of the definition of $\tilde{f}$, 
    \begin{equation}
      \hat{\tilde{f}} \equiv   \hat z_{L_0}^{-1} \circ \hat z_{L_0^f} = z_{L_0}^{-1} \circ
      \left(
      S^{-1} \cdot S  \cdot z_{L_0^f} \right) = z_{L_0}^{-1} \circ z_{L_0^f} =
      \tilde{f}
    \end{equation}

  \item $\tilde{f}$ is periodic

    By construction, the map $\tilde f = (\rho_0(\rho,\varphi),\varphi_0(\rho,\varphi))$ satisfies the periodicity condition
    \begin{equation}
      \rho_0(\rho,\varphi+2\pi) = \rho_0(\rho,\varphi) \quad , \quad \varphi_0(\rho,\varphi+2\pi) = \varphi_0(\rho,\varphi) + 2\pi
      \label{eq:tilde_f_periodic}
    \end{equation}
    In order to establish this, note that the monodromy is
    $\Diff^+(S^1)$-invariant \eqref{eq:M_Psi_L_f = M_Psi_L},
    $M_{\Psi_{L_0}} = M_{(\Psi_{L_0})^f}$.
    Therefore,
    \begin{equation}
      z_{L_0^f}(\rho,\varphi + 2\pi) = M_{(\Psi_{L_0})^f} \cdot
      z_{L_0^f}(\rho,\varphi) =  M_{\Psi_{L_0}} \cdot
      z_{L_0^f}(\rho,\varphi) 
    \end{equation}
    and hence
    \begin{multline}
      z_{L_0}(\rho_0(\rho,\varphi),\varphi_0(\rho,\varphi) + 2\pi) =
      M_{\Psi_{L_0}} \cdot
      z_{L_0}(\rho_0(\rho,\varphi),\varphi_0(\rho,\varphi)) \\
      = M_{\Psi_{L_0}} \cdot
      z_{L_0^f}(\rho,\varphi) 
      = z_{L_0^f}(\rho,\varphi + 2\pi)
    \end{multline}
    Equation \eqref{eq:tilde_f_periodic} follows immediately by taking
    $z_{L_0}^{-1}$ on both sides.

  \item $\tilde{f}$ is an extension of $f$ 

    The local diffeomorphism $\tilde f$ extends $f\in\Diff^+(S^1)$ from the ideal boundary $\del_{\infty}S$ to the bulk.
    To see this, consider the limit $\rho \to \infty$ in the defining equation
    of $z_{L_0}$
    \begin{equation}
      z_{L_0}(\rho,\varphi) =  \frac{e^{\rho} \psi_2(\varphi) + i \psi_2'(\varphi)}{e^{\rho}
    \psi_1(\varphi) + i \psi_1'(\varphi)}
  \end{equation}
  Recall that in this limit 
  \begin{equation}
    \lim_{\rho \to \infty} z_{L_0}(\rho,\varphi) =
    \frac{\psi_2(\varphi)}{\psi_1(\varphi)} = F_{L_0}(\varphi)
  \end{equation}
  and at the same time
  \begin{equation}
    F_{L_0^f}(\varphi) = \lim_{\rho \to \infty} z_{L_0^f}(\rho,\varphi) =
    \frac{\psi^f_2(\varphi)}{\psi^f_1(\varphi)} =
    \frac{\psi_2(f(\varphi))}{\psi_1(f(\varphi))} = F_{L_0}(f(\varphi))
  \end{equation}
  where we used \eqref{eq:Diff_action_on_Hill1} in the second to last equality.
  Therefore, locally we find
  \begin{equation}
    \tilde f \rvert_{\del_{\infty}S} = F_{L_0}^{-1} \circ F_{L_{0}^f} \colon
    \varphi  \mapsto f(\varphi)
    \label{eq:tilde_f_restricts_to_f}
  \end{equation}
  This shows in particular that $\tilde f(\rho,\varphi)$ can be seen as an extension of the
  boundary diffeomorphism $f(\varphi)$ into the bulk in the same way that the vector
  fields $\xi_v(L)$ \eqref{eq:xi} can be regarded as an extension of the bulk
  vector field $v(\varphi)\del_{\varphi}$. 
  In fact, one can check that if $f$ integrates a vector field $v$ on
  the boundary, then $\tilde f$ integrates the bulk vector field $\xi_v(L)$.
  We will give explicit examples in Appendix \ref{app:explicit_calc}.

\end{enumerate}

Up to this point, the construction of the bulk 
diffeomorphism $\tilde{f}$ extending the boundary diffeomorpbism $f\in
\Diff^+(S^1)$ has been general, i.e.\ valid for any orbit. It turns out to be 
particuarly useful, however, for those orbits which contain a constant 
representative $L_0(\varphi)=L_0$. In this case, 
the existence of the extension $\tilde f$ shows that the metric $ds^2(L_0^f)$ defined by a point $L_0^f(\varphi) \in \OO_{L_0}$, can be put into one of the following standard forms, discussed in Section \ref{sec:examples}:
\begin{itemize}
  \item $L_0 = L_{0,n} = n^2 /4$, $n \in \NN$: \[ds^2(L_{0,n}) = d\rho_0^2 +
    n^2 \sinh^2(\rho_0) d\varphi_0^2\]

  \item $L_0 = L_{\alpha,n_0} = (\alpha+n_0)^2 /4$, $\alpha \in (0,1)$,
    $n_0\in\NN_0$:
    \[ds^2(L_{\alpha,n_0}) = d\rho_0^2 +
    (\alpha + n_0)^2 \sinh^2(\rho_0) d\varphi_0^2\]

  \item $L_0 = L_{\ell,0} = - \ell^2/4$: \[ds^2(L_{\ell,0}) = d\rho_0^2 + \ell^2
    \cosh^2(\rho_0) d\varphi_0^2\]

  \item $L_0 = L_{+,0} = 0$: \[ds^2(L_{+,0}) = d\rho_0^2 + e^{2\rho_0}
    d\varphi_0^2\]

\end{itemize}
In Appendix \ref{app:explicit_calc}, we will construct the diffeomorphism $\tilde f$ explicitly for theses cases.
This generalises the results obtained previously in other ways 
for the
disc $L_0=1/4$ \cite{Choi_Larsen} and for the cusp $L_0=0$ \cite{GNW} 
to all Virasoro orbits that admit a constant representative $L_0$.

\subsection{Bulk \texorpdfstring{$\Diff^+(S^1)$}{Diff(S1)} 
Action II: Gauge Theory Perspective} 
\label{sec:gbulk}

We now look at the existence and construction of $\tilde{f}$ from a 
gauge theoretic point of view. Recall that in Section 
\ref{sec:Hill_and_gauge_theory} for any flat gauge field $A(L)$ 
\eqref{eq:A} encoding a hyperbolic geometry we had constructed a 
group valued field $g_L=g_L(\rho,\varphi)$ \eqref{eq:g} such that 
$A(L) = (g_{L})^{-1}dg_{L}$. Thus given any two flat connections 
$A(L_0)$ and $A(L_1)$, say (later we will specialise to $L_1 = L_0^f$), 
we have 
\be
A(L_0) =(g_{L_0})^{-1} dg_{L_0} \quad,\quad 
A(L_1) = (g_{L_1})^{-1} dg_{L_1} 
\ee
and thus $A(L_0)$ and $A(L_1)$ are related by what at first sight looks like 
a gauge transformation, 
\be
\label{eq:hdef}
h_{L_0L_1} = (g_{L_0})^{-1} g_{L_1} 
\quad\Rightarrow\quad A(L_0)^{h_{L_0L_1}} = A(L_1) 
\ee
However, this ignores the monodromies (holonomies) of $g_L$ and $A_L$: gauge fields
with inequivalent holonomies cannot possibly be gauge equivalent. In the case at hand
this is reflected in the fact that if $g_{L_0}$ and $g_{L_1}$ have different
monodromies
\be
g_{L_{0,1}}(\rho,\varphi+2\pi) = M_{L_{0,1}} g_{L_{0,1}}(\rho,\varphi)
\ee
the would-be gauge transformation $h_{L_0L_1}$ is not periodic. If, on the other
hand, the monodromies are equal then $h_{L_0L_1}$ is periodic and thus a 
legitimate gauge transformation, 
\be
M_{L_0}=M_{L_1}\quad\Rightarrow \quad
h_{L_0L_1}(\rho,\varphi+2\pi) = h_{L_0L_1} (\rho,\varphi)
      \label{eq:hll2}
\ee
One situation when one has $M_{L_0}=M_{L_1}$ (and the one of interest
in this Section) is when $L_0$ and $L_1$ lie in the same Virasoro orbit, 
$L_1 = L_0^f$, 
\be
L_1 = L_0^f \quad\Rightarrow\quad  M_{L_0} = M_{L_1}
\ee
(see Appendix \ref{app:Hill_and_monodromy} and \eqref{eq:mfm2}). 
In that case certainly
\be
      h_{L_0L_0^f} = (g_{L_0})^{-1}\; g_{L_0^f}
      \label{eq:hll1}
      \ee
is a periodic gauge transformation and $A(L_0)$ and $A(L_0^f)$ 
are gauge equivalent. As anticipated in 
Section \ref{sec:Hill_and_gauge_theory}, it follows that 
the conjugacy class 
\be
[\Hol(A(L_0)] = [\Hol(A(L_0^f))] 
\ee 
only depends on the orbit $\OO_{L_0}$ and not on a representative. It 
provides the gauge theoretic description of the $(\mathrm{P})\SL(2,\RR)$ conjugacy 
class $\sigma$ appearing in the classification of Virasoro orbits by pairs 
$(\sigma,n_0)$ with $n_0\in \NN_0$ a winding number. 

The other situation of interest is when $L_0,L_1$ lie in distinct
Virasoro orbits labelled by the same $\PSL(2,\RR)$ conjugacy class
$\sigma$ (so that one can arrange $M_{L_0} = M_{L_1}$ by a suitable
change of solution vector $\Psi\to S\Psi$) but different winding
numbers.  This describes pairs of geometries which are gauge
equivalent but not diffeomorphic. We will look at this interesting
phenomenon separately in Section \ref{sec:G} below.

Let us return to the case $L_1= L_0^f$; we have just seen 
that the gauge fields $A(L_0)$ and $A(L_0^f)$ that 
encode in particular the hypberbolic metrics $ds^2(L_0)$ and 
$ds^2(L_0^f)$ are gauge equivalent, 
\be
A(L_0^f)= A(L_0)^{h_{L_0L_0^f}} 
\ee
On the other hand, in the previous Section we had seen that there is a
bulk diffeomorphism $\tilde{f}$ which relates these metrics \eqref{eq:ftllf}, 
\be
   \tilde f^* ds^2(L_0) = ds^2(L_0^{f}) 
\ee
This raises the question if it is actually true that $A(L_0)$ and $A(L_0^f)$
are also directly related by $\tilde{f}^*$. A quick look back at Section 
\ref{sec:Fred} reveals that this is too much to expect: while the bulk 
diffeomorphism is designed to preserve the metric FG gauge condition
\eqref{eq:FG_cond}, it need not satisfy the additional condition 
$e^1=d\rho$ \eqref{eq:gfg} fixing the $\SO(2)$ frame rotations. And indeed in 
\eqref{eq:XLR} we had shown explicitly that the infinitesimal 
gauge transformations and diffeomorphisms preserving the FG gauge
differ precisely by such an infinitesimal frame rotation. A fortiori, this 
will be true for finite rather than infinitesimal transformations, 
so that we will have
\be
\tilde{f}^*A(L_0)  = A(L_0^f)^R = A(L_0)^{(h_{L_0L_0^f}R)} 
\label{eq:tfA}
\ee
for some $R=R(\rho,\varphi)$,  an $\SO(2)$ transformation implementing 
the required frame rotation. At the level of the $g_L$ rather than the 
gauge fields $A(L)$ this is the relation
\be
\tilde{f}^*g_{L_0}=g_{L_0} h_{L_0L_0^f} R = g_{L_0^f} R 
\label{eq:tfg}
\ee
  Importantly, this allows us to define $z_{L_0^f}$ (within an appropriate domain) either using $g_{L_0^f}$ or $\tilde f^* g_{L_0}$
  \begin{equation}
    z_{L_0^f} = g_{L_0^f} \cdot i = \tilde f^* g_{L_0} \cdot i =
\tilde{f}^*z_{L_0} 
    \label{eq:z_via_gf_and_tf_g}
  \end{equation}
(since $R\in\SO(2)$ and $\SO(2)$ is precisely the stabiliser of the point $i\in
\HH$ under fractional linear
transformations).

In principle, either one of the equations \eqref{eq:tfA} or \eqref{eq:tfg} could be used to determine 
first $R$ and then 
$\tilde{f}$, but this is slightly more cumbersome than reading off 
$\tilde{f}$ from the uniformisation maps as in \eqref{eq:zfz}
(which sidesteps the issue of having to determine $R$ first). 

A direct relation between gauge transformations and 
the change of uniformisation maps can be obtained by writing 
\be
z_{L_0^f} = g_{L_0^f} \cdot i = 
g_{L_0^f} (g_{L_0})^{-1} g_{L_0}\cdot i = 
g_{L_0^f} (g_{L_0})^{-1} \cdot z_{L_0}
\label{eq:gauge_trafo_as_transition_fct}
\ee
and noting that the $\SL(2,\RR)$ transition function relating the two 
$\HH$-valued uniformisation maps (which we will interpret as hyperbolic charts
in Section \ref{sub:hs}) can be written as
\be
g_{L_0^f} ( g_{L_0})^{-1} = g_{L_0} h_{L_0L_0^f} (g_{L_0})^{-1}
\ee
and is thus conjugate to the gauge transformation $h_{L_0L_0^f}$ in 
$\SL(2,\RR)$.

\subsection{Global Properties I: Monodromy and Singularities}\label{sec:M}

So far we have seen that a hyperbolic metric in FG gauge $ds^2(L)$
is parametrised by a point $L$ in a Virasoro coadjoint orbit.
Moreover, the local geometry defined by $ds^2(L)$ can be nicely
described via the uniformisation map $z_L$.  In the uniformising
coordinates $z_L(\rho,\varphi)$, locally the metric then always
takes the form of the standard upper half plane metric $ds^2_{\HH}$.
We have also understood in Section \ref{sec:bulk_extension}, how
geometries $ds^2(L)$ and $ds^2(L^f)$ 
corresponding to different points in a given Virasoro
orbit are related: namely, they are diffeomorphic, with a diffeomorphism
that is non-trivial on the (ideal) boundary. It thus crucially
remains to understand, what distinguishes the geometries in different
orbits from each other. 

To that end, recall from Appendix \ref{app:virsl2r} 
that there are four types of 
conjugacy classes inside $\PSL(2,\RR)$, degenerate, elliptic, hyperbolic and
parabolic, 
\begin{equation}
  [\pm \mathbb{I}] \quad , \quad [\pm M_{\alpha}] \quad , \quad [\pm
  M_{\ell}] \quad , \quad [\pm M_{\epsilon}] 
\label{eq:cc}
\end{equation}
where $\alpha \in (0,1)$, $\ell \in \RR_+$, and $\epsilon \in \{+,-\}$,
which we label by 
$\sigma\in \{0,\alpha,\ell,\pm\}$ respectively, 
and that Virasoro orbits are classified by two parameters $(\sigma,n_0)$,
with $n_0\in\NN_0$ a kind of winding number. 

We will therefore first describe, in turn and in general terms, the
impact of the conjugacy class $\sigma$ (in this Section) and then that of 
the winding number $n_0$ (in Sections
\ref{sec:F} and \ref{sec:G}) on the global 
aspects of the geometry, and then turn to a more detailed and 
explicit description of the geometries on a case-by-case basis 
in Section \ref{sec:geometry}.

Locally, any smooth two-dimensional hyperbolic metric looks like (is isometric to) 
the upper half plane metric $ds^2_\HH$. For the hyperbolic 
metrics $ds^2(L)$ on the cylinder $S$ in the FG gauge we made 
this manifest  by constructing explicitly a corresponding uniformisation map 
$z_L$ (Section \ref{sec:uniformization_map}). On the other hand, 
we had already seen in Section
\ref{sec:examples} that $ds^2(L)$ may have singularities or may in other ways
differ globally from the upper half plane. 
The resolution of this apparent paradox is that, as 
we saw in \eqref{eq:glue_z}, the uniformising coordinate
$z_L(\rho,\varphi) = g_L(\rho,\varphi)\cdot i$ \eqref{eq:z_via_g}
satisfies 
\begin{equation}
  z_L(\rho,\varphi+2\pi) = M_{\Psi_L} \cdot z_L(\rho,\varphi)
\end{equation}
where $M_{\Psi_L}$ is the monodromy of the Hill problem associated to $L$, 
acting via fractional linear transformations \eqref{eq:frac_lin_on_z}. 
Thus the $\varphi \sim\varphi + 2\pi$ periodicity of the geometry on the cylinder $S$
implies that the uniformising coordinate $z_L$ 
is subject to the identification 
\begin{equation}
  z_L(\rho,\varphi+2\pi) = M_{\Psi_L} \cdot z_{L}(\rho,\varphi)
  \stackrel{!}{=}z_L(\rho,\varphi)
  \label{eq:identifications_z}
\end{equation}
The geometry defined by $L$ is hence modelled on the quotient space $\HH
/\{ z \sim M_{\Psi_L}\cdot z\}$ rather than on $\HH$ itself. 
It is this identification which leads to global identifications (and hence 
possibly also to quotient singularities) of the geometry described 
by the metric $ds^2(L)$ on $S$. 

This may look somewhat unfamiliar since 
usually the global aspects of a geometry are encoded 
in the single-valued transition functions of local charts, not in some kind of
monodromy. We will show in section \ref{sub:hs} 
how to translate the $\PSL(2,\RR)$-monodromy of $z_L$ into 
$\PSL(2,\RR)$-valued transition functions of $\HH$-valued charts, thus 
defining what is known as a \textit{hyperbolic structure} on $S$ in terms
of the usual local data. 

The detailed implications of the identification \eqref{eq:identifications_z}
depend crucially on the type of conjugacy class $\sigma$ of 
the monodromy. E.g.\ an elliptic monodromy ($\sigma = 
\alpha$ according to the above labelling) acts as a rotation about the 
point $i\in\HH$, and thus has a single fixed point $i$ in the interior of $\HH$, 
while a hyperbolic monodromy ($\sigma = \ell$) acts as a scaling on the upper half plane
coordinate, 
\begin{equation} M_\ell \cdot z = e^{2\pi \ell}z
\ee
which has two asymptotic fixed points at $z\to 0$ and $z \to i\infty$. 
We will therefore study the implications of the 
identification \eqref{eq:identifications_z} in greater detail
in sections \ref{sec:degenerate_monodromy_geometry} -- \ref{sub:par} below, 
when we determine the image of the FG-cylinder $S$ under the uniformisation map
for the various conjugacy classes. 

We close this section with two comments on the interplay between the monodromy
and the isometries of the Poincar\'e metric:

\begin{itemize}
\item Monodromy and orientation-preserving $\PSL(2,\RR)$ isometries

The group $\PSL(2,\RR)$ acts by 
orientation-preserving isometries of the Poincar\'e metric on $\HH$. 
Under this action, the monodromy matrix $M_{\Psi_L}$ is conjugated inside $\PSL(2,\RR)$, 
while its $\PSL(2,\RR)$ conjugacy class $[\pm M_{\Psi_L}]$ is 
invariant. 

\item Monodromy and the orientation-reversing isometry $P:z\mapsto -\bar{z}$

The orientation-reversing isometry $P$ \eqref{eq:P}, on the other hand, acts 
non-trivially on the conjugacy classes. Using the 
explicit representatives of the non-degenerate $\PSL(2,\RR)$ conjugacy classes given in
Appendix \ref{app:virsl2r}, it is easy to see that the monodromy 
changes according to 
\be
\label{eq:Pmon}
z\mapsto -\bar{z}\quad\Rightarrow\quad
\begin{cases}
M_{\alpha} &\mapsto M_{-\alpha} \\
M_{\ell} &\mapsto M_{\ell} \\
M_{\pm} &\mapsto M_{\mp} 
\end{cases}
\ee
Thus in the elliptic case, since for $\alpha\in (0,1)$ one has $1-\alpha
\in  (0,1)$ and $M_{1-\alpha} = -M_{-\alpha}$, at the level of conjugacy
classes one has $P: [\pm M_\alpha] \mapsto [\pm M_{1-\alpha}] $ (which
is a non-trivial action for $\alpha \neq 1/2$), and in the parabolic
case $P$ simply exchanges the two conjugacy classes.

Geometrically, the former is easy to understand: an orientation reversal of a
conical singularity with opening angle $2\pi\alpha$ is simply a conical
singularity with with opening angle $2\pi(-\alpha)$. The parabolic case 
is interesting, as it sheds light on the somewhat elusive exotic cusps
described by the parabolic orbits $(\pm,n)$ with non-trivial winding, 
and shows that the $\pm$-geometries are chiral partners of each other 
under orientation-reversal in $\HH$. This will be discussed further in 
Section \ref{subsub:exc}. 

\end{itemize}

\subsection{Global Properties II: Windings and Coverings}\label{sec:F}

Having understood (or at least anticipated) the significance of the monodromy 
on the geometry, and thus that of the parameter $\sigma$ in the classification 
of Virasoro orbits by pairs $(\sigma,n_0)$, we now turn to the implications and
manifestations of the winding number $n_0\in\NN_0$. 

Recall from Appendix \ref{app:F_winding} that one interpretation of this 
winding number is that of the winding number of the prepotential, 
regarded as a map
\begin{equation}
  F_{\Psi_L}= \frac{\psi_2}{\psi_1}: \RR \to \RP^1 \cong S^1 
\end{equation} 
Equivalently, it can be regarded as the winding number 
of the corresponding Wronskian matrix $W_{\Psi_L}$ around the $\SO(2)\subset
\PSL(2,\RR)$ subgroup. 

It turns out that one implication of the winding of some prepotential 
$F_{\sigma,n}$ with monodromy class $\sigma$ and non-zero winding number $n\in
\NN$ is 
that the geometries with $n>1$ are $n$-fold coverings (in a sense we make 
precise) of the geometries with $n=1$. 
The crucial observation is that, 
given any $F_{\sigma,1}$ with winding number $n=1$, an $F_{\sigma,n}$ (with the
same monodromy but winding number $n$) can be constructed as 
\begin{equation}
  F_{\sigma,n}(\varphi) = F_{\sigma/n,1}(n\varphi)
  \label{eq:F_cover}
\end{equation}
(for $\sigma= 0$ or $\sigma = \alpha$ or $\sigma = \ell$, the meaning of $\sigma/n$ is clear;
for the parabolic case, $M_{\pm/n}$ refers to the matrix with lower off-diagonal
entry $\pm 1/n$, 
which is of course conjugate to $M_\pm$). 
Indeed, this $F_{\sigma,n}$ clearly has monodromy 
\begin{equation}
\label{eq:MMn}
  (M_{\sigma/n})^n=M_\sigma
\end{equation}
and winding number $n$, as required. 

There are two simple ways 
to see that this leads to an $n$-fold cover of the associated geometry:
\begin{enumerate}
\item from the metric $ds^2(L)$ in the FG gauge

  From the composition law \eqref{Scomp} of the Schwarzian
  or by direct calculation it follows that 
  \begin{equation}
    \Sch(F_{\sigma,n}(\varphi)) = \Sch(F_{\sigma/n,1}(n\varphi)) = n^2
    \Sch(F_{\sigma/n,1})(n\varphi)
  \end{equation}
  Substituting this into the expression of the metric in FG gauge,
  $ds^2(L_{\sigma,n})$, for $L_{\sigma,n} = \frac12 \Sch(F_{\sigma,n})$, we find
  \begin{equation}
    \begin{split} 
      ds^2(L_{\sigma,n}) &= d\rho^2 + \left( e^{\rho} - L_{\sigma,n}(\varphi)
      e^{-\rho} \right)^2 d\varphi^2 \\
      &= d\rho^2 + \left( e^{\rho} - \frac12 \Sch(F_{\sigma,n})(\varphi)
      e^{-\rho} \right)^2 d\varphi^2 \\
      &= d\tilde\rho^2 + \left( e^{\tilde\rho} - \frac12
      \Sch(F_{\sigma/n,1})(n\varphi)
      e^{-\tilde\rho} \right)^2 d(n\varphi)^2
    \end{split}
  \end{equation}
  where $\tilde\rho =\rho + \log(n)$.
  Hence the metric $ds^2(L_{\sigma,n})$ is the same as the metric
  $ds^2(L_{\sigma/n,1})$ with $\varphi$ replaced by $n\varphi$.
  Thus, as claimed, the geometry defined by $ds^2(L_{\sigma,n})$ describes an
  $n$-fold cover of the geometry described by $ds^2(L_{\sigma/n,1})$.

\item from the uniformisation map:

  The geometrical interpretation of $n > 1$ geometries as coverings of the
  geometries for $n=1$ could also be made directly from the uniformisation map.
  Notice that since $F_{\sigma,n}$ determines (up to $\PSL(2,\RR)$
  transformations) $L_{\sigma,n}$ and an associated Wronskian-normalised basis
  $\psi_{1,2}$ according to \eqref{eq:psi_from_F}, equation \eqref{eq:F_cover} implies the
  relation
  \begin{equation}
    z_{L_{\sigma,n}}(\rho,\varphi) = z_{L_{\sigma/n,1}}(\rho,n\varphi) 
    \label{eq:n_branched_coverings}
  \end{equation}
  This shows directly that the image of the FG cylinder $S$ under $z_{L_{\sigma,n}}$
  covers its image under $z_{L_{\sigma/n,1}}$ $n$-times and hence describes its
  $n$-fold cover.

  \end{enumerate}

  \subsection{Global Properties III: Windings and Large Gauge Transformations}\label{sec:G}

  While the geometries associated to the $F_{\sigma,n}$ for different values of
  $n$ are thus evidently not diffeomorphic, they are gauge equivalent in
  $\PSL(2,\RR)$. This follows from the fact that the construction of the gauge
  transformation $h_{LL^f}$ \eqref{eq:hll1}
  in Section \ref{sec:gbulk} remains valid when the pair
  $(L,L^f)$ in the same Virasoro orbit is replaced by the pair
  $(L_{\sigma,n_0},L_{\sigma,p_0})$ for $n_0,p_0\in\NN_0$ 
  with $p_0\neq n_0$, and thus in distinct Virasoro
  orbits. 
  The key point is that, because both have the same $\PSL(2,\RR)$ monodromy, the
  argument leading to the single-valuedness of $h$ \eqref{eq:hll2} still applies, 
  and thus $A(L_{\sigma,n_0})$ and $A(L_{\sigma,p_0})$ are gauge equivalent.  

  The reason for talking about $\PSL(2,\RR)$ rather than $\SL(2,\RR)$ gauge
  transformations in the above argument is the following: by construction all the 
  $F_{\sigma,n_0}$ for fixed $\sigma$ have the same $\PSL(2,\RR)$-holonomy
  $M_\sigma$. 
  Their lift to $\SL(2,\RR)$, however, provided by the monodromy $M_{\sigma,n_0}$ 
  of the solution 
  vector $\Psi_{\sigma,n_0}$ constructed e.g.\ according to the prescription 
  \eqref{eq:psi_from_F}
  may a priori be equal to $M_{\sigma,n_0} = \pm M_\sigma$, and indeed we show 
  in Section \ref{sec:geometry} that non-trivial signs occur in the degenerate and elliptic case. 
  Now the monodromy of the corresponding Wronskian matrix $W_{\sigma,n_0}$ is equal to that 
  of $\Psi_{\sigma,n_0}$. From \eqref{eq:def_g} and \eqref{eq:hll1} one finds that 
  the required gauge transformation 
  $h_{n_0,p_0}\equiv h_{L_{\sigma,n_0}L_{\sigma,p_0}}$ is explicitly given by 
  \begin{equation}
    h_{n_0,p_0} = (g_{\sigma,n_0})^{-1} g_{\sigma,p_0} = s(\rho)^{-1}
    (W_{\sigma,n_0})^{-1}W_{\sigma,p_0} s(\rho) 
  \end{equation}
  When $M_{\sigma,n_0}=-M_{\sigma,p_0}$, the monodromy cancels only up to a sign, and
  thus $h_{n_0,p_0}$ is best viewed as a genuine $\PSL(2,\RR)$-valued gauge
  transformation. 

  Given the infinitesimal equivalence of diffeomorphisms and gauge transformations
  explained in Section \ref{sec:Fred}, this gauge transformation
  must then be a \textit{large} gauge transformation,
  i.e.\ with non-trivial winding around the compact $\SO(2)\subset(\mathrm{P})\SL(2,\RR)$ 
  subgroup. This is indeed the case: 
  since the Wronskian matrix $W_{\sigma,n_0}$ has $\PSL(2,\RR)$ 
  winding number $n_0$ (Appendix \ref{app:F_winding}), $h_{n_0,p_0}$ has $\PSL(2,\RR)$-winding number 
  $p_0-n_0 \neq 0$, as anticipated. 

  One can also confirm this by explicit calculations, using the 
  Wronskians associated to the 
  solution vectors $\Psi_{\sigma,n_0}$ 
  constructed from the $F_{\sigma,n_0}$ according to the 
  prescription \eqref{eq:psi_from_F}. In the elliptic case, for example, 
  one then finds that the winding part of 
  $h_{n_0,p_0}$ has the form 
  \be
  h_{n_0,p_0} \sim 
  \begin{pmatrix}
    \cos (p_0-n_0)\varphi/2&
    -\sin (p_0-n_0)\varphi/2\\
    \sin (p_0-n_0)\varphi/2&
    \cos (p_0-n_0)\varphi/2
  \end{pmatrix}
  \ee
  which indeed has winding number $p_0-n_0$ in $\PSL(2,\RR)$. Note also that, in
  agreement with the above comments,  for $p_0-n_0$ odd this is odd under $\varphi\to
  \varphi + 2\pi$, and should thus be viewed as living inside $\PSL(2,\RR)$. 

  This provides a nice illustration of the inequivalence of large gauge
  transformations and diffeomorphisms in gauge theories of gravity, pointed 
  out in a related (albeit Lorentzian) 1+1 dimensional context a long time ago in 
  \cite{SchallerStrobl}. Note also that a relation between large (0+1)-dimensional 
  gauge transformations and exotic Virasoro orbits was pointed out in \cite{GRS} in a
  slightly different context.

  \section{A Tour through the Virasoro Zoo of Hyperbolic Geometries}\label{sec:geometry}

    We study the geometries of the hyperbolic metrics 
  \eqref{eq:FG_gauge}
  \begin{equation}
    ds^2(L) = d\rho^2 + \left( e^{\rho} - L(\varphi) e^{-\rho} \right)^2d\varphi^2
  \end{equation}
  on the cylinder according to the classification of Virasoro orbits, and 
  using the local and global properties of the uniformisation map $z_L$ 
  obtained in the previous section.

  \subsection{Strategy}

  Recall that the uniformisation map 
  \eqref{eq:z}
  \begin{equation}
    z_L(\rho,\varphi) = 
    \frac{e^{\rho} \psi_2(\varphi) + i
  \psi_2'(\varphi)}{e^{\rho}
\psi_1(\varphi) + i \psi_1'(\varphi)}
  \end{equation}
  depends on a choice of basis of solutions of the associated Hill equation. In 
  order to obtain these in a uniform and simple manner for any Virasoro orbit, 
  we make use of the fact (explained in Appendix \ref{app:F}) that 
  for any Virasoro orbit $(\sigma,n_0)$ we can obtain a
  representative $L_{\sigma,n_0}$ and solutions $\psi_{1,2} = \psi_{1,2(\sigma,n_0)}$
  from a simple prepotential $F_{\sigma,n_0}(\varphi)$
  having $\PSL(2,\RR)$-monodromy of type $\sigma$ and (projective) winding 
  number $n_0\in \NN_0$. 

  Indeed, given such a prepotential, one obtains a point $L_{\sigma,n_0}$ in the orbit 
  $\OO_{\sigma,n_0}$ via \eqref{eq:LSchF}
  \begin{equation}
    \frac12 \Sch(F_{\sigma,n_0})(\varphi) = L_{\sigma,n_0}(\varphi)
    \label{eq:Lsn}
  \end{equation}
  and the corresponding solution vector $\Psi_{\sigma,n_0}$ of Hill's equation has
  components (suppressing the
  $(\sigma,n_0)$-labels on the $\psi_{1,2}$) \eqref{eq:psi_from_F}
  \begin{equation}
    \psi_1(\varphi) =
    \frac{1}{\sqrt{F'_{\sigma,n_0}(\varphi)}} \quad , \quad
    \psi_2(\varphi) = \frac{F_{\sigma,n_0}(\varphi)}{\sqrt{F'_{\sigma,n_0}(\varphi)}}
    \label{eq:psins}
  \end{equation}
  We give typical examples of $F_{\sigma,n_0}$ in Appendix
  \ref{app:classification_vir_orbits}. By construction, $F_{\sigma,n_0}$ has 
  the same $\PSL(2,\RR)$ monodromy $M_\sigma$ for all $n_0$. The monodromy
  $M_{\sigma,n_0}$ of
  $\Psi_{\sigma,n_0}$, however, provides a lift of this monodromy to $\SL(2,\RR)$ 
  and may therefore a priori be equal to $\pm M_\sigma$. We determine the sign
  below. 

  Given the $\psi_{1,2}$, we can then also determine the uniformisation map,  its image
  in the upper half plane $\HH$, and thus also the effect of the monodromy
  identification discussed in Section \ref{sec:M}. We also identify the isometry 
  group of the geometries, based on the discussion in Section \ref{sec:Isom}.
    Recall that via \eqref{eq:xi}, the Killing vectors of $ds^2(L)$ can be obtained from the
    stabilising vector fields $v$ of $L$. 
    As we recall in Appendix \ref{app:stab}, such $v$ can be expressed as a
    bilinear expression in $\psi_{1,2}$, where the periodicity of $v$ then
    strongly depends on the monodromy $M$ of $\Psi_L$.
    In particular, if $M = M_{\alpha}$, $M_{\ell}$ or $M_{\pm}$, then
      \begin{equation}
	\begin{split}
	  v^{ell}(\varphi) = \frac12 \left( \psi^2_1(\varphi) +
\psi_2^2(\varphi) \right) \quad , \quad 
	  v^{hyp}(\varphi) = \psi_1(\varphi)\psi_2(\varphi) \quad , \quad 
	  v^{par}(\varphi) = \psi_1^2(\varphi)	
	\end{split}
	\label{eq:v_bilinears}
      \end{equation}
      are periodic and generate infinitesimal elliptic / hyperbolic / parabolic fractional linear transformations on the prepotential $F_L$, and hence on the uniformisation map $z_L$. 
      The Killing vector fields they give rise to thus integrate to elliptic, hyperbolic and parabolic $\SL(2,\RR)$
transformations respectively.

  The remainder of this section is 
  organised according to the following pattern, based on the monodromy, 
  the terminology and notation
  following that of the
  classification of Virasoro orbits in Appendix
  \ref{app:classification_vir_orbits}
  \begin{equation}
    \begin{tabular}{c|c|c|c|c|c|c}
      \multicolumn{1}{c|}{degenerate} & \multicolumn{2}{c|}{elliptic} &
      \multicolumn{2}{c|}{hyperbolic} &  \multicolumn{2}{c}{parabolic}\\ \hline
      $F_{0,n}$ & $F_{\alpha,0}$ & $F_{\alpha,n}$  & $F_{\ell,0}$ & $F_{\ell,n}$ &
      $F_{+,0}$ & $F_{\pm,n}$
    \end{tabular}
  \end{equation}

  \subsection{Degenerate Monodromy: Covering Geometries of the
  Disc}\label{sec:degenerate_monodromy_geometry}

  Consider $L_{0,n} = \frac{n^2}{4}$, for $n \in \NN$. 
  Recall from the discussion in Section \ref{sec:examples} that 
  by defining new coordinates $\rho_0 = \rho -
  \log(n/2)$, $\varphi_0=\varphi$ we can put the hyperbolic metric
  $ds^2(L_{0,n})$ into the standard form
  \begin{equation}
    ds^2(L_{0,n}) = d\rho_0^2 + n^2 \sinh^2(\rho_0) d\varphi_0^2
    \label{eq:dsln}
  \end{equation}
  For $n=1$, this is the standard hyperbolic metric on the infinite disc
  while the geometry for $n>1$ is that of an $n$-fold covering of the
  standard $n=1$ metric. 

  The metric \eqref{eq:dsln} has a three-dimensional isometry group $\SL^{(n)}(2,\RR)$, the 
  $n$-fold covering of the isometry group of the standard $n=1$ metric, 
  generated by the Killing vector fields $\xi_{v_0}$ \eqref{eq:xi0}
  and $\xi_{v_1},\xi_{v_2}$ \eqref{eq:xi12}, 
  associated to the Virosoro stabilisers $v_0=1, v_1 = \cos n\varphi, v_2 = \sin
  n\varphi$. 
  For the purposes of this section, it will be convenient to use a
  the basis constructed from the bilinears \eqref{eq:v_bilinears} of solutions
to the associated Hill equation. 

To that end, note that for $L_{0,n} = n^2/4$ 
a suitable choice of prepotential $F_{0,n}$ and corresponding Wronskian
normalised solutions is given by 
\begin{equation}
  \psi_1(\varphi_0) = \frac{\cos(n \varphi_0 / 2)}{\sqrt{n/2}}
  \quad , \quad
  \psi_2(\varphi_0) = \frac{\sin(n \varphi_0 / 2)}{\sqrt{n/2}} \quad , \quad
  F_{{0,n}}(\varphi_0) = \tan\left(
  \frac{n\varphi_0}{2} \right)
  \label{eq:psi0_n}
\end{equation} 
so that the corresponding solution vector $\Psi_{0,n}$ has monodromy 
\begin{equation}
  M_{0,n} = (-1)^n\mat{1}{\phantom{+}0}{0}{\phantom{+}1} \equiv
  (-1)^n\mathbb{I}
  \label{eq:M_n}
\end{equation} 
Then e.g.\ for $n=1$ one finds from \eqref{eq:v_bilinears} the following linear combinations of
$v_0,v_1,v_2$, 
      \begin{equation}
	\begin{split}
	  &v^{ell}_{0,1}(\varphi) = \frac12 \left( \psi^2_1(\varphi) +
\psi_2^2(\varphi) \right) = 1 = v_0\\
	  &v^{hyp}_{0,1}(\varphi) = \psi_1(\varphi)\psi_2(\varphi) =
\sin(\varphi) = v_2\\
	  &v^{par}_{0,1}(\varphi) = \psi_1^2(\varphi) = 1 + \cos(\varphi) = v_0
+ v_1
	\end{split}
	\label{eq:stab_L_disc}
      \end{equation}

Turning now to the uniformisation map, which provides additional insight in
particular into the nature of the covering and the singular (branch) point, 
taking into account the shift in $\rho$, the 
  uniformisation map \eqref{eq:z} takes the form
  \begin{equation}
    z_{L_{0,n}}(\rho_0,\varphi_0) = \frac{\frac{n}{2}
    e^{\rho_0} \psi_2(\varphi_0) + i \psi_2'(\varphi_0)
  }{{\frac{n}{2} e^{\rho_0} \psi_1(\varphi_0) + i \psi_1'(\varphi_0) }}
  \label{eq:z_n}
\end{equation}
With the solutions $\psi_i$ given above, 
the uniformisation map \eqref{eq:z_n} is then explicitly 
\begin{equation}
  z_{L_{0,n}}(\rho_0,\varphi_0) = \frac{e^{\rho_0} \sin\left(
  \frac{n\varphi_0}{2} \right) + i \cos\left(
  \frac{n\varphi_0}{2}  \right)}{e^{\rho_0}\cos\left( \frac{n\varphi_0}{2}
  \right) - i \sin\left( \frac{n\varphi_0}{2}  \right)}
\end{equation}
In the case at hand (and for the conical singularities to be discussed below), 
it turns out to be slightly more convenient to study the problem not in the
upper half plane $\HH$ but in the Poincar\'e disc 
\begin{equation}
  \DD = \{ w \in \CC \mid \lvert w \rvert < 1 \} 
\end{equation}
endowed with the standard hyperbolic metric
\begin{equation}
  ds^2_{\DD} = \frac{4 dw d\bar w}{(1 - \lvert w \rvert^2)^2}
\end{equation}
In order to pass from $\HH$ to $\DD$, we apply a Cayley transformation 
\begin{equation}
  C \colon \HH \to \DD \quad , \quad w \mapsto C \cdot w \quad , \quad C = \mat{i}{\phantom{-}1}{i}{-1}
  \label{eq:Cayley}
\end{equation}
defined by the fractional linear transformation \eqref{eq:frac_lin_on_z}.

We denote the image of $z_{L_{0,n}}$ under the Cayley transformation by
$w_{L_{0,n}}$.
Concretely, we obtain
\begin{equation}
  w_{L_{0,n}}(\rho_0,\varphi_0) = \frac{i - z_{L_{0,n}}(\rho_0,\varphi_0)}{i
    + z_{L_{0,n}}(\rho_0,\varphi_0)}
  \end{equation}
  Direct calculation shows that 
  \begin{equation}
    w_{L_{0,n}}(\rho_0,\varphi_0) = \tanh\left( \frac{\rho_0}{2}
    \right) e^{in\varphi_0}
    \label{eq:w}
  \end{equation}
  which is precisely the coordinate transformation \eqref{eq:trafo_exc}.
  The range of $w_{L_{0,n}}$ is readily computed from \eqref{eq:w} as $\rho_0\in(0,\infty)$ and $\varphi_0 \in [0,2\pi)$,
    \begin{equation}
      \lvert w_{L_{0,n}} \rvert \in (0,1) \quad , \quad \arg w_{L_{0,n}} \in [0, 2\pi n)
      \end{equation}

      Note that as $\varphi_0 \to \varphi_0 + 2\pi $, $\arg w_{L_{0,n}}$ transforms by a shift of $2\pi n$ which shows that $w_{L_n}$ defines a coordinate on the disc for $n=1$, resp.\ on the $n$-fold cover of the disc for $n>1$.
      Since the coordinate $w_{L_n}$ has a winding number $n$, one is tempted to
      define a coordinate $\zeta_{n}$ which winds only once by 
      $(w_{L_{0,n}})^{1/n} = \zeta_{n}$.
      The coordinates $\zeta_n$ define standard coordinates on the unit disc $\DD$.
      However, the metric now takes the form
      \begin{equation}
	ds^2_{\DD^{(n)}} = \frac{4 n^2 \lvert \zeta_{n} \rvert^{2n - 2}d\zeta_{n} d\bar
	\zeta_{n}}{(1 - \lvert \zeta_{n} \rvert^{2n})^2}
	\label{eq:dsdn}
      \end{equation}
      The map 
      \be
      \zeta_n \to (\zeta_n)^n= w_{L_{0,n}} 
      \ee
      shows explicitly that this is a hyperbolic metric on the 
      $n$-fold branched cover of the unit disc $\DD$ (with branch point the origin). 

      \subsection{Elliptic Monodromy: Conical Geometries}
      \subsubsection{Deficit Conical Geometries}
      The discussion of the geometry of the uniformisation for orbits with elliptic
      monodromy follows closely the discussion of orbits with degenerate monodromy.

      Consider $L_{\alpha,0} = \frac{\alpha^2}{4}$ for $\alpha \in (0,1)$.
      Recall from Section \ref{sec:examples} that in the coordinates $\rho_0 = \rho +
      \log(\alpha/2)$, $\varphi_0 = \varphi$, the metric looks like 
      \begin{equation}
	ds^2(L_{\alpha,0}) = d\rho_0^2 + \alpha^2 \sinh^2(\rho_0)d\varphi_0^2	
      \end{equation}
      In contrast to the disc geometry, this metric only has a one-dimensional
      compact isometry group $\SO(2) \subset \SL(2,\RR)$, namely 
      $\varphi_0$-translations generated by the vector field $\xi_{v_0}=\del_\varphi$. 

      We had already argued in Section \ref{sec:examples} that this metric defines 
      a conical hyperbolic geometry with opening angle $2\pi \alpha$. Here we 
      establish this from the perspective of the uniformisation map and its 
      monodromy. 
      The uniformisation map \eqref{eq:z} now reads
      \begin{equation}
	z_{L_{\alpha,0}}(\rho_0,\varphi_0) = \frac{\frac{\alpha}{2}
	e^{\rho_0} \psi_2(\varphi_0) + i \psi_2'(\varphi_0)
      }{{\frac{\alpha}{2} e^{\rho_0} \psi_1(\varphi_0) + i \psi_1'(\varphi_0) }}
      \label{eq:z_alpha}
    \end{equation}
    Moreover, for $L_{\alpha,0} = \frac{\alpha^2}{4}$, a Wronskian normalised basis of the associated Hill problem is given by 
    \begin{equation}
      \psi_1(\varphi_0) = \frac{\cos(\alpha \varphi_0 / 2)}{\sqrt{\alpha/2}}
      \quad , \quad
      \psi_2(\varphi_0) = \frac{\sin(\alpha \varphi_0 / 2)}{\sqrt{\alpha/2}}  \quad , \quad
      F_{{\alpha,0}}(\varphi_0) = \tan\left(
      \frac{\alpha\varphi_0}{2} \right)
      \label{eq:psi0_alpha}
    \end{equation}
    with monodromy
    \begin{equation}
      M_{\alpha}=\mat{\cos(\pi\alpha)}{-\sin(\pi\alpha)}{\sin(\pi\alpha)}{\phantom{-}\cos(\pi\alpha)}
      \label{eq:M_alpha}
    \end{equation}   
    which is an elliptic element in $\PSL(2,\RR)$.
    It is again convenient to study the problem in the Poincar\'e disc.
    As before, we obtain the disc coordinate $w_{L_{\alpha,0}}$ from
    $z_{L_{\alpha,0}}$ by a Cayley transformation.
    The result is
    \begin{equation}
      w_{L_{\alpha,0}}(\rho_0,\varphi_0) = \tanh\left( \frac{\rho_0}{2}
      \right)e^{i\alpha\varphi_0}
    \end{equation}
    It follows directly from \eqref{eq:Cayley} that as $z_{L_{\alpha,0}} \to
    M_{\alpha}\cdot z_{L_{\alpha,0}}$ one
    has $w_{L_{\alpha,0}} \mapsto (CM_{\alpha}C^{-1})\cdot w_{L_{\alpha,0}}$.
    Since 
    \begin{equation}
      CM_{\alpha}C^{-1} = \mat{e^{-i\pi\alpha}}{0}{0}{e^{i\pi\alpha}}
    \end{equation}
    the coordinate $w_{L_{\alpha,0}}$ is subject to the identification
    \begin{equation}
      w_{L_{\alpha,0}}(\rho_0, \varphi_0 + 2\pi) \sim e^{2\pi i \alpha}
      w(\rho_0,\varphi_0)
      \label{eq:glue_w}
    \end{equation}
    When glueing according to \eqref{eq:glue_w}, we obtain a cone. 
    In Figure \ref{fig:cone}, we show the image of $w_{L_{\alpha,0}}$ inside the unit disc.
    Remarkably, it traces out precisely once 
    a typical fundamental domain for the action of $CM_{\alpha}C^{-1}$ on $\DD$ in this case, namely a wedge whose boundary edges form the angle $2\pi\alpha$.
    The cone is obtained by identifying the boundaries of the fundamental domain.

    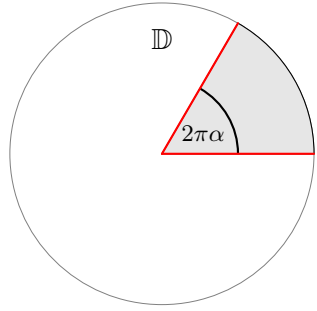
\begin{figure}[htb]
      \centering
      \begin{tikzpicture}
	\def\in{0}
	\def\out{60}

	\tkzDefPoint(0,1.8){label}

	\tkzDefPoint(0,0){C}
	\tkzDefPoint(2,0){P}
	\tkzDefPoint(\in:2){Q1}
	\tkzDefPoint(\out:2){Q2}
	\tkzDefPoint(\in:1){alpha1}
	\tkzDefPoint(\out:1){alpha2}
	\tkzDefPoint(1.1*\out:0.3){alpha}

	\tkzDrawCircle(C,P)
	\begin{scope}
	  \tkzDrawSector[fill=gray!20](C,Q1)(Q2)
	\end{scope}

	\tkzLabelPoint[anchor=west](alpha){\footnotesize $2\pi\alpha$}
	\tkzDrawArc[thick](C,alpha1)(alpha2)

	\tkzDrawSegment[thick,red](C,Q1)
	\tkzDrawSegment[thick,red](C,Q2)

	\tkzLabelPoint[anchor=north](label){$\mathbb D$}

      \end{tikzpicture}
      \caption{Fundamental domain of $CM_{\alpha}C^{-1}\colon w \mapsto e^{2\pi
      i\alpha} w$, for $\alpha < 1$, in the Poincar\'e disc $\DD$. The cone angle is
      defined to be the angle between the two boundaries (red) of the fundamental
    domain. It is given by $2\pi\alpha$. The conical singularity is found at $\rho_0 = 0$ (origin of the disc).}
    \label{fig:cone}
  \end{figure}

  When gluing according to \eqref{eq:glue_w}, one needs to be careful with the topology one assigns to the resulting quotient space. 
  If $\alpha$ is irrational, the quotient topology would define a non-Hausdorff space.
  However, the resulting space is simply a cone, thus a well-defined
  topological space---one merely has to equip it with a topology different from the quotient topology. 
  On the other hand, for $\alpha \in \QQ$, the subgroup generated by the
  monodromy \eqref{eq:M_alpha} is discrete and hence the quotient space $\HH /
  \{ z \sim e^{2\pi i\alpha} z \}$ endowed with the quotient topology does define a
  (singular) manifold known as an orbifold. 

  We can formally obtain coordinates which are invariant under the shift
  $\varphi_0 \to \varphi_0 + 2\pi$ by defining a coordinate $\zeta_{\alpha}$
  taking values in the punctured disc $\DD^*$ such that
  $\zeta_{\alpha} = (w_{L_{\alpha,0}})^{1/\alpha}$.
  In these coordinates, the metric takes the form  
  \begin{equation}
    ds^2(L_{\alpha,0}) = \frac{4 \alpha^2 \lvert \zeta_{\alpha} \rvert^{2\alpha -
    2}d\zeta_{\alpha} d\bar \zeta_{\alpha}}{(1 - \lvert \zeta_{\alpha}
    \rvert^{2\alpha})^2}
    \label{eq:ds_alpha_punctured_disc2}
  \end{equation}
  The metric \eqref{eq:ds_alpha_punctured_disc2} describes a conical singularity
  at the origin and describes the cone $\DD^*_{\alpha}$ of opening angle $2\pi \alpha$.

  \subsubsection{Excess Conical Geometries}
  \label{sub:excess} 

  The discussion of $L_{\alpha,n} = \frac{(\alpha + n)^2}{4}$ with $\alpha \in
  (0,1)$ and $n \in \NN$ is quite analogous to that for $L_{\alpha,0}$.
  One just replaces $\alpha$ by $\alpha + n$.
  In particular, the metric is
  \begin{equation}
    ds^2(L_{\alpha,n}) = d\rho_0^2 + (\alpha+n)^2 \sinh^2(\rho_0)d\varphi_0^2	
  \end{equation}
  also with the one-dimensional compact 
  isometry group $\SO(2) \subset \SL(2,\RR)$ of 
  $\varphi_0$-translations generated by the vector field $\xi_{v_0}=\del_\varphi$, 
  and one has the prepotential
  \begin{equation}
    F_{\alpha,n} = \tan\left( \frac{(\alpha + n)\varphi}{2} \right)
  \end{equation}

  The corresponding solution vector $\Psi_{\alpha,n}$ constructed according to 
  \eqref{eq:psins} is just like that of \eqref{eq:psi0_alpha} with
  $\alpha\to\alpha+n$, and thus has monodromy
  \be
  M_{\alpha,n} = M_{\alpha+n} = (-1)^n M_\alpha
  \label{eq:Man}
  \ee 
  From this one ultimately derives
  \begin{equation}
    w_{L_{\alpha,n}}(\rho_0,\varphi_0) = \tanh\left( \frac{\rho_0}{2} \right)
    e^{i(\alpha+n)\varphi_0}
    \label{eq:w_excess}
  \end{equation}
  However, now the coordinate $w_{L_{\alpha,n}}$ has non-trivial winding $n \geq
  1$ and the conical singularity at $\rho_0=0$ is described by an excess angle.
  In particular, the image of the FG cylinder under the uniformisation map 
  covers the punctured disc $n$ times before one makes the conical
  identification.  
  We illustrate how to glue the resulting cone from the disc in Figure \ref{fig:excess_cone}.

  Moreover, since
  \begin{equation}
    F_{\alpha,n}(\varphi) = \tan\left( \frac{(\alpha + n)\varphi}{2} \right) = \tan\left(
    \frac{\left(\frac{\alpha}{n} + 1\right) n \varphi}{2} \right) =
    F_{\alpha/n,1}(n\varphi)
  \end{equation}
  it follows from the general discussion in Section \ref{sec:F} that the excess geometry
  described by $L_{\alpha,n}$ with $n+ \alpha \in (n,n+1)$ 
  is a $n$-fold covering of the basic excess geometry described by
  $L_{\alpha/n,1}$ with $1 + \alpha/n \in (1,2)$, 
  \begin{equation}
    \DD_{n+\alpha} = \left(\DD_{1+\alpha/n}\right)^{(n)}
  \end{equation}
  In words: the excess geometry 
  with opening angle $2\pi (n+\alpha) \in (2\pi n, 2\pi(n+1))$ is an $n$-fold
  covering of the excess geometry with opening angle $2\pi (1+ \alpha/n) \in (2\pi,4\pi)$.
  This can also be seen explicitly from the uniformisation coordinate
  \begin{equation}
    w_{L_{\alpha,n}}(\rho_0,\varphi_0) = \tanh\left( \frac{\rho_0}{2} \right)
    e^{i(\alpha+n)\varphi_0} =  \tanh\left( \frac{\rho_0}{2} \right)
    e^{i(\alpha/n+1)n\varphi_0} = w_{L_{\alpha/n,1}}(\rho_0,n\varphi_0)
  \end{equation}

  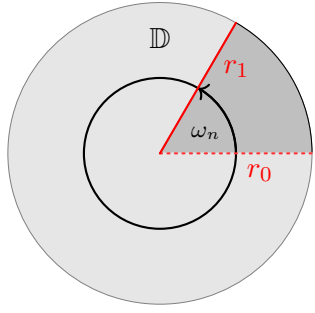
\begin{figure}[htb]
    \centering
    \begin{tikzpicture}
      \def\in{0}
      \def\out{60}

      \tkzDefPoint(0,1.8){label}

      \tkzDefPoint(0,0){C}
      \tkzDefPoint(2,0){P}
      \tkzDefPoint(\in:2){Q1}
      \tkzDefPoint(\out:2){Q2}
      \tkzDefPoint(\in:1){alpha1}
      \tkzDefPoint(\out:1){alpha2}
      \tkzDefPoint(1.1*\out:0.3){alpha}

      \tkzDrawCircle[fill=gray!20](C,P)
      \begin{scope}
	\tkzDrawSector[fill=gray!50](C,Q1)(Q2)
      \end{scope}
      \tkzDrawSegment[thick,gray!20](C,Q1)

      \tkzLabelPoint[anchor=west,xshift=.4em](alpha){\footnotesize $\omega_n$}
      \tkzDrawArc[thick,black,->](C,alpha1)(alpha2)
      \tkzDrawCircle[thick,black](C,alpha1)

      \tkzDrawSegment[thick,dotted,red!80](C,Q1)
      \tkzDrawSegment[thick,red](C,Q2)

      \tkzLabelSegment[red,below right](C,Q1){$r_0$}
      \tkzLabelSegment[red,above right,xshift=0.5em](C,Q2){$r_1$}

      \tkzLabelPoint[anchor=north](label){$\mathbb D$}

    \end{tikzpicture}
    \caption{Representation of an excess cone angle $\omega_n = 2\pi(\alpha + n)$: we start with the ray $r_0$ and sweep out the disc as we move $r_0$ in counterclockwise direction. 
      We continue to do so until the disc is 
      covered $n$-times before identifying the ray $r_0$ with 
    the ray $r_1$ at angle $2\pi\alpha$.}
    \label{fig:excess_cone}
  \end{figure}

    Notice in particular that the uniformisation map $w_{L_{\alpha,n}}$ covers the \emph{entire} Poincar\'e disc, as is clear from \eqref{eq:w_excess}, while the uniformisation map $w_{L_{\alpha,0}}$ covers merely a fundamental domain of the action of $M_{\alpha}$ on $\DD$.

  \subsection{Hyperbolic Monodromy: Annular Geometries}

  \subsubsection{Hyperbolic Cylinders and Funnels}

  Consider $L_{\ell,0} = -\frac{\ell^2}{4}$, $\ell \in \RR_+$. 
  Setting $\rho_0 = \rho - \log(\ell/2)$, $\varphi_0 = \varphi$, 
  the metric \eqref{eq:FG_gauge} takes the standard form 
  \begin{equation}
    ds^2(L_{\ell,0}) = d\rho_0^2 + \ell^2 \cosh^2(\rho_0) d\varphi_0^2.
  \end{equation}
  Like the conical gemetries described above, this metric has a one-dimensional 
  compact $\SO(2)$ isometry group of $\varphi_0$-translations.
  As we saw in Section \ref{sec:examples}, it
  describes a hyperbolic cylinder, a wormhole-like geometry interpolating between
  two hyperbolic discs for $\rho_0\in(-\infty,+\infty)$ and a funnel geometry 
  for $\rho_0\in [0,\infty)$. 

    This can also nicely be seen from the gluing properties of the uniformisation map.
    The equation for the uniformisation map \eqref{eq:z} becomes
    \begin{equation}
      z_{L_{\ell,0}}(\rho_0,\varphi_0) = \frac{\frac{\ell}{2}e^{\rho_0}
    \psi_2(\varphi_0) + i \psi_2'(\varphi_0)}{\frac{\ell}{2}e^{\rho_0}
  \psi_1(\varphi_0) + i \psi_1'(\varphi_0)}.
  \label{eq:zhyp}
\end{equation}
A Wronskian normalised basis of the Hill problem associated to $L_{\ell,0}$ is given by 
\begin{equation}
  \psi_1(\varphi_0) = \frac{e^{-\ell \varphi_0 / 2}}{\sqrt{\ell}}\quad , \quad
  \psi_2(\varphi_0) = \frac{e^{\ell \varphi_0 / 2}}{\sqrt{\ell}} \quad , \quad
  F_{{\ell,0}}(\varphi_0) = e^{\ell\varphi_0}
  \label{eq:psi_hyp}
\end{equation}
The solution vector $\Psi_{\ell,0}$ has monodromy
\begin{equation}
  M_{\ell} = \mat{e^{-\ell\pi}}{0}{0}{e^{\ell\pi}}
\end{equation} 
which is a hyperbolic element in $\PSL(2,\RR)$.
By substituting \eqref{eq:psi_hyp} into \eqref{eq:zhyp}, we obtain
\begin{equation}
  z_{L_{\ell,0}}(\rho_0,\varphi_0) = e^{\ell\varphi_0}
  \frac{e^{\rho_0} + i}{e^{\rho_0} - i}
  \label{eq:z_tilde_hyp}
\end{equation}
which is precisely the coordinate transformation given in \eqref{eq:trafo_hyp}.
When $\varphi_0$ is shifted by $2\pi$, the coordinate
$z_{L_{\ell,0}}(\rho_0,\varphi_0)$ is scaled: 
\begin{equation}
  z_{L_{\ell,0}}(\rho_0,\varphi_0+2\pi) = e^{2\pi\ell}
  z_{L_{\ell,0}}(\rho_0,\varphi_0) = M_{\ell}\cdot
  z_{L_{\ell,0}}(\rho_0,\varphi_0)
\end{equation}
The identification $\varphi_0 \sim \varphi_0 + 2\pi$ thus enforces us to glue the
coordinate $z_{L_{\ell,0}}(\rho_0,\varphi_0)$ according to
$z_{L_{\ell,0}}(\rho_0,\varphi_0) \sim M_{\ell} \cdot
z_{L_{\ell,0}}(\rho_0,\varphi_0)$.

Equation \eqref{eq:z_tilde_hyp} allows us to compute the range of
$z_{L_{\ell,0}}$: for $\rho_0 \in (-\infty,\infty)$ and $\varphi_0 \in [0,2\pi)$ we find
  \begin{equation}
    1 \leq \lvert z_{L_{\ell,0}} \rvert < e^{2\pi \ell}, \qquad 0 \leq \arg
    z_{L_{\ell,0}} < \pi
  \end{equation}
  whereas for the funnel with $\rho_0 \in [0,+\infty)$ we obtain 
    \begin{equation}
      1 \leq \lvert z_{L_{\ell,0}} \rvert < e^{2\pi \ell}, \qquad 0 \leq \arg
      z_{L_{\ell,0}} \leq \pi/2
    \end{equation}
    This is illustrated in 
    in Figure \ref{fig:funnel}, where 
    we show the image of $z_{L_{\ell,0}}$ inside the upper half plane.
    Interestingly, also in this case, $z_{L_{\ell,0}}$ traces out precisely 
    a fundamental domain of the action of $M_{\ell}$ on $\HH$.
    By identifying its boundaries according to $z \sim e^{2\pi\ell} z$, 
    we obtain the hyperbolic cylinder (funnel) 
    with its unique periodic geodesic sitting at the throat (boundary).

    \begin{figure}[htb]
      \centering
      \begin{tikzpicture}
	\tkzDefPoint(0,2){i}
	\tkzDefPoint(-1,3){label}
	\tkzDefPoint(0,1){ii}

	\tkzDefPoint(0,0){O}
	\tkzDefPoint(0,3){I}
	\tkzDefPoint(-3,0){A}
	\tkzDefPoint(3,0){B}
	\tkzDefPoint(1,0){Q} 
	\tkzDefPoint(2,0){P}

	\tkzDefPoint(1,1.3){rho_pos}
	\tkzDefPoint(-1,1.3){rho_neg}

	\begin{scope}
	  \tkzClipCircle(O,P)
	  \tkzClipCircle[out](O,Q)
	  \path[fill=gray!20] (-3,0) -- (-3,3) -- (3,3) -- (3,0) -- cycle;
	\end{scope}

	\tkzLabelPoint(rho_pos){\footnotesize $\rho_0 > 0$}
	\tkzLabelPoint(rho_neg){\footnotesize $\rho_0 < 0$}

	\tkzDrawSegment(O,I)
	\tkzDrawSegment(A,B)

	\tkzDrawSemiCircle[thick,red](O,P)
	\tkzDrawSemiCircle[thick,red](O,Q)

	\tkzInterLC(O,I)(O,Q)\tkzGetPoints{z1}{z2}
	\tkzInterLC(O,I)(O,P)\tkzGetPoints{w1}{w2}
	\tkzDrawSegment[thick,blue](z2,w2)

	\tkzLabelPoint[anchor=north west](ii){\footnotesize $i$}
	\tkzLabelPoint[anchor=south west](i){\footnotesize $i e^{2\pi\ell}$}
	\tkzLabelPoint(label){$\HH$}

	\tkzDrawPoints(i,ii)

      \end{tikzpicture}
      \caption{Fundamental domain of $M_{\ell} \colon z \mapsto e^{2\pi\ell} z$ in
      the upper half plane $\HH$. The periodic geodesic is defined by $\rho_0 = 0$ and
      is given by the straight (vertical) line between $i$ and $ie^{2\pi\ell}$. The
    regions $\rho_0 > 0$ and $\rho_0 < 0$ each describe a funnel.}
    \label{fig:funnel}
  \end{figure}
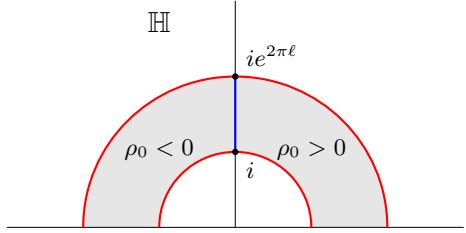
  As in the examples before, the uniformisation coordinate $z_{L_{\ell,0}}$ is not invariant under shifts of $\varphi_0 \to \varphi_0 + 2\pi$. 
  However, it is possible to define an invariant coordinate via 
  \begin{equation}
    \zeta_{\ell}  = e^{i \log z_{L_{\ell,0}} / \ell}
    \label{eq:zeta_hyp}
  \end{equation}
  This coordinate transformation maps the upper half plane to the annulus 
  \begin{equation}
    \mathcal A_{\ell} = \{ \zeta \in \CC \mid e^{-\pi / \ell} < \lvert \zeta \rvert < 1 \}
    \label{eq:annulus2}
  \end{equation}
  in the complex plane.
  Notice that from \eqref{eq:z_tilde_hyp} and \eqref{eq:zeta_hyp}, we find that 
  \begin{equation}
    \lim_{\rho_0 \to \infty} z_{L_{\ell,0}}(\rho_0,\varphi_0) = e^{\ell \varphi_0} \implies \lim_{\rho_0 \to \infty} \lvert\zeta_{\ell}(\rho_0,\varphi_0)\rvert = 1
  \end{equation}
  and 
  \begin{equation}
    \lim_{\rho_0 \to -\infty} z_{L_{\ell,0}}(\rho_0,\varphi_0) = - e^{\ell \varphi_0} = e^{\ell \varphi_0 + i \pi} \implies \lim_{\rho_0 \to \infty} \lvert \zeta_{\ell}(\rho_0,\varphi_0) \rvert = e^{-\pi / \ell}
  \end{equation}
  Both asymptotic boundaries of $S$ are hence mapped to the two boundaries of $\mathcal A_{\ell}$, with the ideal boundary $\del_{\infty}S$ sitting at $\lvert \zeta_{\ell} \rvert = 1$. 
  Moreover, in the coordinates \eqref{eq:z_tilde_hyp}, the unique periodic geodesic $\gamma$ sits at $\rho_0 = 0$: $\gamma(\varphi_0) = e^{\ell \varphi_0 + i \pi / 2}$.
  Consequently, with \eqref{eq:zeta_hyp}, the periodic geodesic sits at the
  center of the annulus, the curve defined by $\lvert \zeta_{\ell} \rvert =
  e^{-\pi/2\ell}$, and the image $\mathcal{F}_\ell$ 
  of the funnel in the complex plane is simply
  given by one half of the annulus $\mathcal{A}_\ell$, 
  \be
  \mathcal{F}_\ell = 
  \{ \zeta \in \CC \mid e^{-\pi /2 \ell} \leq  \lvert \zeta \rvert < 1 \}
  \label{eq:funnel}
\end{equation}

Finally, in terms of the invariant coordinate $\zeta_{\ell} $,
the metric becomes
\begin{equation}
  ds^2(L_{\ell,0}) = ds^2_{\Al} = \frac{\ell^2 d\zeta_{\ell}  d\bar \zeta_{\ell} }{\lvert \zeta_{\ell}  \rvert^2
  \sin^2(\ell \log \lvert \zeta_{\ell}  \rvert)}
  \label{eq:ds_annulus2}
\end{equation}
which is smooth on the annulus $\mathcal A_{\ell}$, since by
\eqref{eq:annulus2} 
\begin{equation}
  -\pi < \ell \log\lvert \zeta_{\ell} \rvert < 0
\end{equation}

\subsubsection{Exotic Funnels} \label{subsub:exf}

Exotic funnels are geometries that arise from 
\begin{equation}
  F_{\ell,n}(\varphi) = e^{\ell\varphi} \tan\left( \frac{n\varphi}{2} \right)
  =F_{\ell/n,1}(n\varphi)
  \label{eq:Fln}
\end{equation}
which has winding number $n$ and monodromy 
\begin{equation}
  F_{\ell,n}(\varphi + 2\pi) = e^{2\pi \ell} F_{\ell,n}(\varphi) =
  M_{\ell}\cdot F_{\ell,n}(\varphi) \quad , \quad M_{\ell} = \mat{e^{-\pi\ell}}{0}{0}{e^{\pi\ell}}
\end{equation}

It follows that 
\begin{equation}
  L_{\ell,n}(\varphi) = \frac12 \Sch(F_{\ell,n}) = -\frac{\ell^2}{4} -\frac12 \frac{n(n^2 + \ell^2)}{n + \ell\sin(n\varphi)} + \frac{3}{4} \frac{n^2(n^2 - \ell^2)}{(n + \ell \sin(n\varphi))^2}
\end{equation}
defines a point in an exotic Virasoro coadjoint orbit of hyperbolic type. 
One
can check that the corresponding solution vector $\Psi_{\ell,n}$ has the same
monodromy as $\Psi_{\ell,0}$, i.e.
\be
M_{\ell,n} = M_{\ell}
\label{eq:Mln}
\ee
(no factor of $(-1)^n$, unlike in the degenerate and elliptic case). 

The monodromy of $F_{\ell,n}$ translates to the same
monodromy of the uniformisation map so that
\begin{equation}
  z_{L_{\ell,n}}(\rho,\varphi + 2\pi) =  M_{\ell} \cdot
  z_{L_{\ell,n}}(\rho,\varphi)
  \label{eq:glue_z_ell_n}
\end{equation}
which we identify accordingly.
Hence the image of the FG-cylinder $S$ under the uniformisation map $z_{L_{\ell,n}}$ (where defined) is given by $\HH / \Gamma(M_{\ell})$ where $\Gamma(M_{\ell}) \subset \PSL(2,\RR)$ denotes the group freely generated by $M_{\ell}$.
Due to this identification \eqref{eq:glue_z_ell_n}, locally, we obtain the same space as for $z_{L_{\ell,0}}$, namely an annulus in the upper half plane, cf.\ Figure \ref{fig:funnel}.

Despite these similarities, there are several crucial differences between the
standard and exotic funnels (and among the exotic funnels with different winding
numbers):
\begin{enumerate}

\item Asymptotics of the Uniformisation map $z_{L_{\ell,n}}$

  As in \eqref{eq:z_asymptotics}, one finds from \eqref{eq:zhyp} that, away from the zeros of $\psi_{1,2}$, the asymptotics of the uniformisation map $z_{L_{\ell,n}}$ as $\rho \to \infty$ is given by:

  \begin{equation}
    \begin{split} 
      z_{L_{\ell,n}}(\rho,\varphi) &\mathop{\sim}_{\rho\to\infty} F_{\ell,n}(\varphi) + \frac{2i}{\ell} e^{-\rho} F'_{\ell,n}(\varphi) + \OO\left( e^{-2\rho} \right)
      \\
      &= e^{\ell\varphi} \tan\left( \frac{n\varphi}{2} \right) + i e^{-\rho} \left( 2 e^{\ell\varphi} \tan\left( \frac{n \varphi}{2} \right) + \frac{n}{\ell}\frac{e^{\ell\varphi}}{\cos^2(n \varphi / 2)} \right) + \OO\left( e^{-2\rho} \right)
    \end{split}
    \label{eq:z_hyp_exotic_asymp}
  \end{equation}

  To leading order in $\rho$, the constant-$\varphi$ curves, $c_{\varphi}(\rho) = a_{\varphi} + i b_{\varphi} e^{-\rho}$, are vertical lines. 
    In fact, the proportionality factor of the imaginary part is everywhere strictly positive, $b_{\varphi} \geq b_0 > 0$, and $\Im(c_{\varphi}(\rho)) \to 0$ as $\rho\to\infty$. 
    Moreover, as one varies $\varphi$ in $[-\pi,\pi]$, the base point $a_{\varphi}$ covers all of $\RR$.
    This shows that the asymptotics of $z_{L_{\ell,n}}$ cover a neighbourhood of the \emph{entire} ideal boundary, while the asymptotics of $L_{\ell,0}$ only cover a neighbourhood of a small part of it.

  \item Boundary value of the Uniformisation Map $z_{L_{\ell,n}}$

    Another important difference is the boundary value of $z_{L_{\ell,n}}$ in
    comparison to $z_{L_{\ell,0}}$.
    In Section \ref{sec:properties_z}, we have shown that for general $z_{L}$, 
    \begin{equation}
      \lim_{\rho \to \infty} z_{L}(\rho,\varphi) = F_L(\varphi)
    \end{equation}
    Hence, the geometries defined by $z_{L_{\ell,0}}$ and $z_{L_{\ell,n}}$ differ by
    their asymptotic behaviours. In the standard case we have 
    \begin{equation}
      \lim_{\rho \to \infty} z_{L_{\ell,0}}(\rho,\varphi) = e^{\ell\varphi} 
    \end{equation}
    while the exotic funnels with winding number $n$ are characterised by the fact
    that 
    \begin{equation}
      \lim_{\rho \to \infty} z_{L_{\ell,n}}(\rho,\varphi) = e^{\ell\varphi}
      \tan\left( \frac{n \varphi}{2} \right)
      \label{eq:asymptotic_z_hyp_exotic}
    \end{equation}
    In Section \ref{sec:moduli_spaces}, we will reinterpret this in terms of 
    inequivalent  projective structures (with the same $\PSL(2,\RR)$ monodromy) 
    on the asymptotic boundary. 

  \item Range of the Uniformisation Map $z_{L_{\ell,1}}$ 

    Recall that in the standard funnel geometry obtained from $L_{\ell,0}$, the image of the FG cylinder under the uniformisation map $z_{L_{\ell,0}}$ is given by precisely one fundamental domain of the action $M_{\ell}$ on $\HH$, cf.\ Figure \ref{fig:funnel}.
    This changes when considering exotic funnels.  
    Since the geometries obtained for $n>1$ are $n$-fold branched coverings of a
    geometry obtained from $n=1$, cf.\ \eqref{eq:n_branched_coverings}, it
    suffices to consider the case $n=1$.

    As we have seen above, the asymptotics of $z_{L_{\ell,1}}$ cover a neighbourhood of the entire ideal boundary.
    Consequently, $z_{L_{\ell,1}}$ covers  necessarily a fundamental domain of the action of $M_{\ell,1} = M_{\ell}$ on $\HH$.
    Indeed, we may consider a geodesic $\gamma_{\varepsilon}$ in $\HH$ which connects asymptotically the points $(\pm\varepsilon,0)$ for $\varepsilon$ small enough.
    Then a fundamental domain is given by $\{ z \in \HH \mid \lvert \gamma_{\varepsilon} \rvert \leq \lvert z \rvert \leq \lvert M_{\ell}\cdot \gamma_{\varepsilon} \rvert\}$.
    Since $M_{\ell}$ acts isometrically on $\HH$, it is enough to understand the geometry in a chosen fundamental domain. 
    Indeed, we can extend the geometry obtained from $L_{\ell,1}$ beyond the validity of the FG coordinates by moving the fundamental domain via the action of $M_{\ell}$ into regions where the FG coordinates break down. 
    The underlying topological space of the geometry described by $L_{\ell,1}(\varphi)$ looks therefore the same as for the standard funnel, namely $\HH / \{ z \sim e^{2\pi\ell} z \}$.

  \item Isometry Group

    In contrast to the case of standard funnels, in this case the isometry 
    group is neither compact nor connected. Indeed,   
    \be
    \Isom(ds^2(L_{\ell,n})) = \Stab (L_{\ell,n}) = \RR \times \ZZ_n, 
    \ee
    the non-compact continuous factor $\RR$ being generated by the vector field 
    $\xi_v$ \eqref{eq:xi} corresponding to the Virasoro stabiliser $v(\varphi)$ given in
    \eqref{eq:vln}. The 
    additional invariance of $L_{\ell,n}$ (and hence of the metric) under 
    $\varphi \to \varphi + \frac{2\pi}{n}$ accounts for the second factor $\ZZ_n$.

      It is clear that for $\ell \to 0$ one finds $L_{\ell,n}\to L_{0,n}$ 
      \cite{Balog_Feher_Palla}, and this extends to the corresponding metrics,
i.e.\  the family of metrics $ds^2(L_{\ell,n})$ converges to
      $ds^2(L_{0,n})$. 
      The exotic funnels can thus be seen as deformations of the degenerate geometries obtained from $L_{0,n}$.
      This deformation breaks the isometry group $\SL^{(n)}(2,\RR)$ to a
hyperbolic subgroup.
      This can be seen as follows (for simplicity, we restrict ourselves to the
      case $n=1$):

      As $\ell \to 0$ one has $F_{\ell,1} \to F_{0,1}$ and hence $F_{\ell,1}$ can be seen as a deformation of $F_{0,1}$.
      Consequently, expanding the corresponding expressions \eqref{eq:psins} for
      $\psi_{1,2}$ in $\ell$, one can expand the Virasoro stabilisers given by the 
bilinears \eqref{eq:v_bilinears} as
      \begin{equation}
	\begin{split}
	  &v^{ell}_{\ell,1}(\varphi) = 1 + \left( \sin(\varphi) - \varphi \cos(\varphi) \right) + \OO(\ell^2) \\
	  &v^{hyp}_{\ell,1}(\varphi) = \sin(\varphi)(1 - \ell \sin(\varphi)) + \OO(\ell^2)  \\
	  &v^{par}_{\ell,1}(\varphi) = 1 + \cos(\varphi) - 2\ell\left( \cos^2\left( (\varphi + \sin(\varphi))\frac{\varphi}{2} \right) \right) + \OO(\ell^2)
	\end{split}
      \end{equation}
To leading order, one of course recovers the vector fields \eqref{eq:stab_L_disc} 
generating the isometry algebra of the disc geoemtry $ds^2(L_{0,1})$. The only 
deformation that is periodic in $\varphi$ is that corresponding to 
$v^{hyp}_{\ell,1}$, and thus only this one gives rise to a Killing vector
field of the deformed geometry. This is indeed also the vector field obtained by
the expansion of the stabiliser \eqref{eq:vln} of $L_{\ell,1}$, 
      \begin{equation}
	v_{\ell,1} = \frac{\sin(\varphi)}{1 + \ell \sin(\varphi)} = \sin(\varphi)(1 - \ell\sin(\varphi)) + \OO(\ell^2) 
      \end{equation}
      The deformation of $F_{0,1}$ to $F_{\ell,1}$ therefore breaks the elliptic and parabolic isometries of
      $ds^2(L_{0,1})$ while the hyperbolic isometry remains
      an isometry even after the deformation.

    \end{enumerate}

    \subsection{Parabolic Monodromy: Cuspidal Geometries}\label{sub:par}

    \subsubsection{Cusps}

    Consider $L_{+,0} = 0$, and coordinates $(\rho_0,\varphi_0)$ for which the metric takes the form 
    \begin{equation}
      ds^2({L_{+,0}}) = d\rho_0^2 + e^{2\rho_0} d\varphi_0^2
    \end{equation}
    with the compact $\SO(2)$ isometry group of $\varphi_0$-translations.
    A pair of linear independent Wronskian normalised solution of the associated Hill problem 
    \begin{equation}
      \psi''(\varphi_0) + L_{+,0} \psi(\varphi_0) = \psi''(\varphi_0) = 0
    \end{equation}
    is given by
    \begin{equation}
      \psi_1(\varphi_0) = \sqrt{2\pi} \quad ,\quad 
      \psi_2(\varphi_0) = \frac{\varphi_0}{\sqrt{2\pi}} \quad , \quad
      F_{+,0}(\varphi_0) = \frac{\varphi_0}{2\pi}
    \end{equation}
    The normalisation of $\psi_1$ and $\psi_2$ is chosen for later convenience.
    The monodromy of the solution vector $\Psi = (\psi_1~\psi_2)^t$ is given by 
    \begin{equation}
      M_{+} = \mat{1}{\phantom{+}0}{1}{\phantom{+}1}
      \label{eq:M_plus}
    \end{equation} 
    which is a parabolic element in $\SL(2,\RR)$.

    By the discussion in Section \ref{sec:examples}, 
    we expect the geometry to have a cuspidal singularity at $\rho_0 = - \infty$.
    This can also be seen by studying the image of $S$ under the uniformisation map.
    In the present case, from \eqref{eq:z} we obtain the upper half plane coordinate
    \begin{equation}
      z_{L_{+,0}}(\rho,\varphi) \equiv z_{L_{+,0}}(\rho_0,\varphi_0) =  \frac{1}{2\pi}\left( \varphi_0 + i
      e^{-\rho_0} \right)
      \label{eq:z_tilde_par}
    \end{equation}
    which is precisely the coordinate transformation \eqref{eq:trafo_par}.
    Note that 
    \begin{equation}
      \lim_{\rho_0 \to -\infty} z_{L_{+,0}}(\rho_0,\varphi_0) = i\infty,
    \end{equation} 
    so that we expect the cuspidal singularity to appear at $i\infty \in \HH$.
    As anticipated in the discussion around \eqref{eq:glue_z}, 
    \begin{equation}
      z_{L_{+,0}}(\rho_0,\varphi_0+2\pi) \sim z_{L_{+,0}}(\rho_0,\varphi_0) + 1 =
      M_{+}\cdot z_{L_{+,0}}(\rho_0,\varphi_0).
      \label{eq:parabbolic_cst_gluing}
    \end{equation}

  We read off the range of $z_{L_{+,0}}$ from \eqref{eq:z_tilde_par}. 
  As $\rho_0 \in (-\infty,\infty)$ and $\varphi_0 \in [0,2\pi)$ we have 
  \begin{equation}
    0 \leq {\rm Re}(z_{L_{+,0}}) < 1 \quad , \quad 0 < {\rm Im}(z_{L_{+,0}}) < \infty 
  \end{equation}
  in $\HH$.
      In Figure \ref{fig:cusp}, we show the image of $z_{L_{+,0}}$ in the upper half plane.
    As in the examples studied before, $z_{L_{+,0}}$ traces out a fundamental region of the action of $M_{+}$.
      The image of $S$ under the uniformisation map $z_{L_{+,0}}$ can therefore be identified with the quotient
      space $\HH / \{ z \sim z + 1 \}$, which has indeed a cuspidal singularity at $i\infty$.

    Note that the reflection $P$ \eqref{eq:P} maps
    \begin{equation}
      P \colon z_{L_{+,0}}(\rho,\varphi) \mapsto   - \bar z_{L_{+,0}}(\rho,\varphi) = z_{L{+,0}}(\rho,-\varphi)
    \end{equation}
    so that 
    \begin{equation}
      (Pz_{L_{+,0}})(\rho,\varphi+2\pi) = M_{-}\cdot (Pz_{L_{+,0}})(\rho,\varphi) = (Pz_{L_{+,0}})(\rho,\varphi)  - 1
    \end{equation}
where $M_-$ is the parabolic monodromy matrix
\be
M_{-} = \mat{\phantom{-}1}{\phantom{+}0}{-1}{\phantom{+}1}
\ee
Hence the standard cusp can equally well be realised via 
$Pz_{L_{+,0}}(\rho,\varphi)$ as the quotient space $\HH / \{ z \sim z - 1 \}$.
This observation will turn out to be useful in Section \ref{sub:hs} in order
to define hyperbolic structures for exotic cusps of the type $(-,n)$. 

      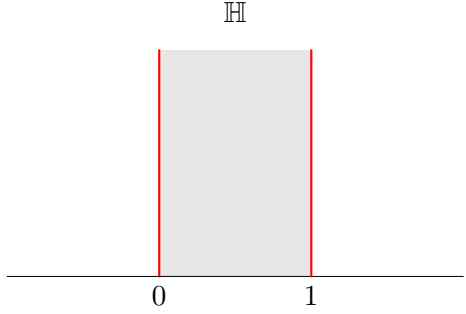
\begin{figure}[htb]
	\centering 
	\begin{tikzpicture}[baseline={([yshift=-.8ex]current bounding box.center)}]
	  \tkzDefPoint(0,3.5){label}
	  \tkzDefPoint(-1,0){O}
	  \tkzDefPoint(-1,3){I}
	  \tkzDefPoint(-3,0){A}
	  \tkzDefPoint(3,0){B}
	  \tkzDefPoint(1,0){P} 
	  \tkzDefPoint(1,3){Q} 

	  \path[fill=gray!20] (O) -- (I) -- (Q) -- (P) -- cycle;

	  \tkzDrawSegment[red,thick](O,I)
	  \tkzDrawSegment(A,B)
	  \tkzDrawSegment[red,thick](P,Q)

	  \tkzLabelPoint[anchor=center](label){$\mathbb H$}
	  \tkzLabelPoint[anchor=north](O){$0$}
	  \tkzLabelPoint[anchor=north](P){$1$}
	\end{tikzpicture}
	\caption{Fundamental region of $M_{+} \colon z \mapsto z+1$ in the upper half plane $\HH$. The cuspidal singularity lies at $i\infty$.}

	\label{fig:cusp}
      \end{figure} 

      We notice again that the coordinate \eqref{eq:trafo_par} is not invariant
      under $\varphi_0 \to \varphi_0 + 2\pi$.
      Again it is possible to define an invariant coordinate, namely
      \begin{equation}
	\zeta_0 = e^{2\pi i z_{L_{+,0}}} 
      \end{equation}
      which identifies the upper half plane with the punctured unit disc $\DD^*_0$.
      In terms of $\zeta_0$, the metric takes the form
      \begin{equation}
	ds^2({L_{+,0}}) =ds^2_{\DD^*_0} = 
	\frac{d\zeta_0 d\bar \zeta_0}{\lvert \zeta_0 \rvert^2 (\log \lvert
	\zeta_0 \rvert)^2}
	\label{eq:ds_par_punctured_disc2}
      \end{equation}
      describing a cuspidal singularity at the origin \cite{Wolpert}.

      In Section \ref{sec:examples} we saw in terms of FG coordinates 
      that the cuspidal singularity can be seen as the limit of 
      a conical singularity (as the opening angle
      $\alpha\rightarrow 0$) or a funnel geometry (as the length parameter
      $\ell\rightarrow 0$) \eqref{eq:limitcusp}. 
      A short computation shows that this
      is also true for the standard metrics \eqref{eq:ds_alpha_punctured_disc2}
      on the punctured disc $\DD^*_\alpha$ and 
      \eqref{eq:ds_annulus2} on the annulus $\Al$: in the limit they both
      develop a cuspidal singularity with metric 
      \eqref{eq:ds_par_punctured_disc2}, i.e.\ 
      \begin{equation}
	\lim_{\alpha \to 0} (\DD^*_{\alpha},ds^2_{\DD^*_{\alpha}}) = \lim_{\ell
	\to 0} (\mathcal A_{\ell},ds^2_{\mathcal A_{\ell}}) =
	(\DD^*_0,ds^2_{\DD^*_0})
      \end{equation}

      \subsubsection{Exotic Cusps} \label{subsub:exc}

      Exotic cusps arise as the geometry described by
      \begin{equation}
	F_{\pm,n}(\varphi) = \pm \frac{\varphi}{2\pi} + \tan\left( \frac{n\varphi}{2} \right)
	\label{eq:Fpmn}
      \end{equation}
      so that
      \begin{equation}
	F_{\pm,n}( \varphi + 2\pi ) = F_{\pm,n}(\varphi) \pm 1 = M_{\pm,n}
	\cdot F_{\pm,n}(\varphi) \quad , \quad M_{\pm} =
	\mat{\phantom{\pm}1}{\phantom{+}0}{\pm1}{\phantom{+}1}
      \end{equation}
      The corresponding representative of the exotic Virasoro coadjoint orbit of parabolic type is then given by 
      \begin{equation}
	L_{\pm,n}(\varphi) = \frac{1}{2}\Sch(F_{\pm,n}) = \frac{n^3}{8}\left( \frac{3\left(
	  \frac{n}{2}\pm\frac{1}{2\pi} \right)}{\left( \frac{n}{2} \pm
	  \frac{1}{2\pi}\cos^2\left( \frac{n \varphi}{2} \right) \right)^2} -
	  \frac{2}{\frac{n}{2} \pm \frac{1}{2\pi}\cos^2\left( \frac{n \varphi}{2} \right)} \right)
	\end{equation}
	For the corresponding solution vector $\Psi_{\pm,n}$ constructed from
	\eqref{eq:psins} one finds that 
	\be
	\psi_1(\varphi+2\pi)=\psi_1(\varphi) \quad,\quad
	\psi_2(\varphi+2\pi)=\psi_2(\varphi)  \pm \psi_1(\varphi) 
	\ee
	for all $n$, so that the monodromy of $\Psi_{\pm,n}$ is 
	\be
	M_{\pm,n} = M_{\pm}
	\label{eq:Mpmn}
	\ee 

	The monodromy $M_\pm$ of $F_{\pm,n}$ implies 
	\begin{equation}
	  z_{L_{\pm,n}}(\rho,\varphi+2\pi) = M_{\pm}\cdot z_{L_{\pm,n}}(\rho,\varphi)
	  = z_{L_{\pm,n}}(\rho,\varphi)  \pm 1
	  \label{eq:gluing_z_par_exotic}
	\end{equation}

	As in the case of standard versus exotic funnels 
	discussed above, there are again several crucial differences between the
	standard and exotic cusps (and among the exotic cusps with different
	winding numbers):
	\begin{enumerate}

      \item Asymptotics of the Uniformisation Map $z_{L_{\pm,n}}$

	Away from the zeros of $\psi_{1,2}$, the asymptotics of $z_{L_{\pm,n}}$ is easily determined from \eqref{eq:z_asymptotics} as $\rho\to\infty$:
	\begin{equation}
	  \begin{split} 
	    z_{L_{\pm,n}} &\mathop{\sim}_{\rho\to\infty} F_{\pm,n}(\varphi) + i e^{-\rho} F_{L_{\pm,n}}'(\varphi) + \OO\left( e^{-2\rho} \right)\\
	    &= \pm\frac{\varphi}{2\pi} + \tan\left( \frac{n \varphi}{2} \right) + i e^{-\rho} \left( \pm \frac{1}{2\pi} + \frac{n}{2} \frac{1}{\cos^2(n \varphi / 2)}\right) + \OO\left( e^{-2\rho} \right)
	  \end{split}
	\end{equation}

	Now, to leading order in $\rho$, the constant-$\varphi$ curves,  $c_{\varphi}(\rho) = a_{\varphi} + i b_{\varphi} e^{-\rho}$, are vertical lines. 
	In fact, the proportionality factor of the imaginary part is everywhere strictly positive, $b_{\varphi} \geq b_0 > 0$, and $\Im(c_{\varphi}(\rho)) \to 0$ as $\rho\to\infty$.

	  Moreover as one varies $\varphi$ in $[-\pi,\pi]$, the base point $a_{\varphi}$ covers all of $\RR$.
	  This shows that the asymptotics of $z_{L_{\pm,n}}$ cover a neighbourhood of the \emph{entire} ideal boundary, while the asymptotics of $L_{+,0}$ only cover a neighbourhood of a small part of it.

      \item Boundary value of the Uniformisation Map $z_{L_{\pm,n}}$

	For the standard cuspidal geometries one has 
	\begin{equation}
	  \lim_{\rho\to\infty}z_{L_{+,0}}(\rho,\varphi) = \frac{\varphi}{2\pi}
	\end{equation}
	while for the exotic cuspidal geometries one has 
	\begin{equation}
	  \lim_{\rho\to\infty}z_{L_{\pm,n}}(\rho,\varphi) = \pm\frac{\varphi}{2\pi} +
	  \tan\left( \frac{n\varphi}{2} \right) 
	  \label{eq:asymptotic_z_par_exotic}
	\end{equation}
	In Section \ref{sec:moduli_spaces}, we will reinterpret this in terms of 
	inequivalent  projective structures (with the same $\PSL(2,\RR)$ monodromy) 
	on the asymptotic boundary. 

	\item Range of the Uniformisation Map $z_{L_{\pm,1}}$

	  Recall that in the standard cusp geometry obtained from $L_{+,0}$, the image of the FG cylinder under the uniformisation map $z_{L_{+,0}}$ is given by precisely one fundamental domain of the action $M_{+}$ on $\HH$, cf.\ Figure \ref{fig:cusp}.
	Similarly to the exotic funnels, this changes when considering exotic cusps.  
	Indeed, one can show that apart from a vertical strip  of finite length,
	$z_{L_{\pm,1}}$ covers $\HH$ fully.
	The explicit calculations are more involved and are given explicitly in
	for $L_{+,1}$ in Appendix \ref{app:exotic_cusp}.
	The other cases are similar.

	Inside the strip, it may happen that the FG coordinates $(\rho,\varphi)$ break down.
	However, away from that strip, $z_{L_{\pm,1}}$ will cover (infinitely many copies of) a fundamental domain.
	Moreover, since $M_{\pm,1}$ is in fact an \emph{isometry}, it is enough
	to know the metric in one fundamental domain, i.e.\ we can extend the
	metric to the whole of $\HH$ by the identity
	$z_{L_{\pm,1}}(\rho,\varphi+2\pi) \sim M_{\pm,1} \cdot z_{L_{+,1}}(\rho,\varphi)$.
	The underlying topological space of the geometry described by
	$L_{\pm,1}(\varphi)$ looks therefore the same as for the standard cusp,
	namely $\HH / \{ z \sim z \pm 1 \}$.

    \item Isometry Group

      In contrast to the case of standard cusps, but as in the case of 
      exotic funnnels, in this case the isometry 
      group is neither compact nor connected. Indeed,   
      \be
      \Isom(ds^2(L_{\pm,n})) = \Stab (L_{\pm,n}) = \RR \times \ZZ_n, 
      \ee
      the non-compact continuous factor $\RR$ being generated by the vector field 
      $\xi_v$ \eqref{eq:xi} corresponding to the Virasoro stabiliser $v(\varphi)$ given in
      \eqref{eq:vpmn}. The fact that this $v(\varphi)$ has $n$ double zeros implies 
      that asymptotically $\xi_v$ has $n$ fixed points on the boundary at
      $\rho\to\infty$ (in contrast to the previously discussed case of exotic funnels, 
      where asymptotically either the tangential or the radial component of the Killing vector 
      vanishes at certain points). 
      The 
      additional invariance of $L_{\pm,n}$ (and hence of the metric) under 
      $\varphi \to \varphi + \frac{2\pi}{n}$ accounts for the second factor $\ZZ_n$.

	Similar to the exotic funnels, the exotic cusps can be seen as a deformation of the (degenerate) disc geometries.
	Indeed, there exists a family of metrics $ds^2(L_{\pm,n}^{(t)})$
obtained from a deformed prepotential \eqref{eq:Ft}, namely 
	\begin{equation}
	  F_{\pm,n}^{(t)} = \pm \frac{t\varphi}{2\pi} + \tan\left( \frac{n\varphi}{2} \right) \quad , \quad L_{\pm,n}^{(t)} = \frac12 \Sch(F_{\pm,n}^{(t)})
	  \label{eq:F_exotic_par_family}
	\end{equation}
all, as indicated by the notation, with monodromy in the conjugaacy class
$\sigma = \pm$ respectively 
(a similar but not identical one-parameter family was given in \cite{Balog_Feher_Palla}). 
	It is clear that for $t \to 0$ one has $F_{\pm,n}^{(t)} \to F_{0,n}$ and
	accordingly the family of metrics $ds^2(L_{\pm,n}^{(t)})$ converges to
	$ds^2(L_{0,n})$.
	Again, this deformation breaks part of the isometry group $\SL^{(n)}(2,\RR)$ in the sense that this time after the deformation only the parabolic direction is preseved as an isometry. 
	As for the exotic funnels, it suffices to study the case $n=1$.

	From \eqref{eq:F_exotic_par_family} and the corresponding expressions
	\eqref{eq:psins} 
	for $\psi_{1,2}$ and \eqref{eq:stab_L_disc} respectively one finds for $t \to 0$:  
	\begin{equation}
	  \begin{split}
	    &v^{ell,t}_{\pm,1}(\varphi) = 1 \mp \frac{t}{2\pi} \left( 1 + \cos(\varphi) - \varphi \sin(\varphi) \right) + \OO(t^2) \\
	    &v^{hyp,t}_{\pm,1}(\varphi) = \sin(\varphi) \pm \frac{t}{\pi} \cos^2\left( \frac{\varphi}{2} \right)\left( \varphi - \sin(\varphi) \right) + \OO(t^2)  \\
	    &v^{par,t}_{\pm,1}(\varphi) = 1 + \cos(\varphi) \mp \frac{2t}{\pi} \cos^4\left(\frac{\varphi}{2} \right) + \OO(t^2)
	  \end{split}
	\end{equation}
	which shows that only $v^{par,t}_{\pm,1}$ remains periodic. 
 This also
agrees with the stabiliser obtained from the $t\to 0$ expansion of the $t$-deformed
stabilising vector field \eqref{eq:vpmn} 
	\begin{equation}
	  v^{(t)}_{\pm,1} = \frac{\cos^2\left( \frac{\varphi}{2} \right)}{\frac12 + \frac{t}{2\pi} \cos^2\left(\frac{\varphi}{2}\right)} = 1 + \cos(\varphi) \mp \frac{2t}{\pi} \cos^4\left( \frac{\varphi}{2} \right) + \OO(t^2) 
	\end{equation}
	This shows that the deformation of $F_{0,1}$ to $F^{(t)}_{\pm,1}$
	breaks the elliptic and hyperbolic isometries of $ds^2(L_{0,1})$ while
	the parabolic isometry remains an isometry even after the deformation.
      \end{enumerate}

      \section{Virasoro Coadjoint Orbits as Moduli Spaces of Hyperbolic Structures}\label{sec:moduli_spaces}

      We have seen that there is a remarkably close connection between
      Virasoro coadjoint orbits and spaces of hyperbolic metrics on the
      cylinder $S$. In this Section, we will try to make this more precise
      and show that Virasoro coadjoint orbits define moduli spaces of
      possibly singular hyperbolic metrics on the disc, the punctured
      disc or the annulus. We start with an intuitive description of the
      moduli spaces in terms of hyperbolic metrics on $S$ and then define
      them more abstractly and conveniently in terms of hyperbolic
      structures later on.

      \subsection{Moduli Spaces of Hyperbolic Metrics: the (Geo-)Metric Perspective}
      \label{sub:mod1}

      Recall that by a hyperbolic metric on a surface $S$ 
      we mean a metric $g$ of constant negative curvature, 
      $R(g)=-2$. We have seen that hyperbolic metrics on the (topological) cylinder $S$ 
      are universally such that $S$ exhibits an asymptotic (ideal) boundary 
      $\del_{\infty}S$, and we can impose the (asymptotic) boundary condition
      that a hyperbolic metric $g$ approaches some 
      standard reference metric $g_0$ (e.g.\ the standard Poincar\'e metric) near
      $\del_\infty S$. We will loosely write this as $g\to g_0$, and we can thus
      formally consider the moduli space of all such metrics on $S$ modulo
      (orientation preserving) diffeomorphisms which are trivial on the boundary. We
      denote this space by $\M(S)$, 
      \begin{equation}
	\M(S) = \frac{\{ g \colon R(g) = -2 \;,\; g\to g_0\}
      }{\Diff_0^+(S)}
      \label{eq:M}
    \end{equation}
    As it stands, this moduli space is not particularly 
    well-defined, not only because we have
    not specified the rate at which $g\to g_0$, but also because we have not 
    specified the smoothness conditions on the metric $g$ in the interior 
    (or the kinds of singularities that we allow). With hindsight, the 
    definition of $\M(S)$ can be amended and enriched appropriately, but this is not 
    particularly enlightning and therefore not how we will proceed. 

    Instead, we will explore two options to define $\M(S)$
    directly. The first, based on the approach we have explored thus
    far, is to define $\M(S)$ as the space $\M_{FG}(S)$ of hyperbolic
    metrics that we have found explicitly in the Fefferman-Graham (FG)
    gauge (which in particular fixes the $\Diff^+_0(S)$-symmetry). The
    second, to be explored subsequently, is to define $\M(S)$ 
    in terms of \textit{hyperbolic structures} on $S$.

    In order to describe $\M(S)$ more concretely, in 
    Section \ref{sec:gauge_theory} we adopted the strategy to locally choose a gauge for the 
    $\Diff^+_0(S)$-symmetry, i.e.\ to choose a local coordinate system. An
    extremely useful and convenient choice of gauge is the Fefferman-Graham (FG)
    gauge, in which $g_{\rho\rho}=1,g_{\rho\varphi}=0$. The FG gauge always
    exists  for
    sufficiently large $\rho$, and (as we showed in Section \ref{sub:diffbc}) 
    completely fixes the asymptotic $\Diff^+_0(S)$ symmetry.
    Thus we were led to consider the 
    space 
    \be
    \M_{FG}(S) = 
    \{ g_{\mu\nu}(\rho,\varphi) \colon R = -2\;,\; g_{\varphi\varphi}\to
    1\;,\; g_{\rho\rho}=1\;,\; g_{\rho\varphi}=0 \}
    \ee
    of hyperbolic metrics in the FG gauge. By explicitly solving the $R=-2$
    equation, one finds that a hyperbolic metric in the FG gauge with the 
    specified asymptotic behaviour is fully determined by a single (and 
    arbitrary) periodic function $L(\varphi)$, in terms of which the metric
    takes the form \eqref{eq:FG_gauge}
    \be
    ds^2(L) = d\rho^2 + \left( e^{\rho} - L(\varphi) e^{-\rho} \right)^2d\varphi^2
    \ee
    so that one has the remarkably simple result that the FG moduli space is
    simply the space of periodic functions $L(\varphi)$, 
    \be
    \M_{FG}(S) = \{ ds^2(L)\} = \{L\} 
    \ee
    Now recall from Section \ref{sub:diffbc} that $\Diff^+(S^1)$ acts on the metric
    $ds^2(L)$ by the Virasoro coadjoint action on $L$
    \begin{equation}
      \Diff^+(S^1)\ni f\colon ds^2(L) \mapsto ds^2(L^f) 
    \end{equation}
    This shows that the moduli space $\M_{FG}(S)$ can be naturally (i.e.\ in a 
    $\Diff^+(S^1)$-equivariant way) identified 
    with the (smooth) dual $\vir^*$ of the Virasoro algebra, 
    \be
    \M_{FG}(S)\cong \vir^*
    \ee
    and that it decomposes into Virasoro coadjoint orbits
    $\OO_L = \{L^f\colon f\in\Diff^+(S^1)\}$ under the above action of $\Diff^+(S^1)$.
    Our main interest will not be the total space $\M_{FG}(S)$ itself, but rather 
    the individual Virasoro coadjoint orbits, regarded as moduli spaces of
    hyperbolic metrics in their own right. In order to better understand these, 
    we now study the asymptotics of the metrics $ds^2(L)$ 
    (for which the FG gauge is certainly available and suitable) in more
    detail. 

    To that end, let us now return to the $\Diff^+(S^1)$ action and its significance 
    for the moduli problem at hand.  
    As shown in 
    Section \ref{sec:bulk_extension}, within each orbit $\OO_L$, the geometries
    corresponding to points $L$ and $L^f$, say, are related by a bulk
    diffeomorphism $\tilde f$ which restricts to $f$ on the asymptotic (ideal) boundary. 
    However, one is nevertheless justified to treat them as distinct points in the 
    moduli space of hyperbolic metrics on the cylinder $S$, precisely because they
    are related by a diffeomorphism that is non-trivial on the boundary, 
    $\tilde{f}\notin \Diff^+_0(S)$, and which should therefore not be treated as 
    a gauge symmetry of the problem: rather, the residual $\tilde{f}$- or
    $\Diff^+(S^1)$-symmetry should be treated as a global symmetry that relates
    asymptotically inequivalent geometries. 

    This difference in the asymptotic geometry can thus be detected by looking 
    at the geometry close to 
    the ideal boundary $\del_{\infty}S$.
    For example, considering the curves defined by the condition $\rho = cst$ in
    the FG gauge, their extrinsic curvature defined by the metric $ds^2(L^f)$ is
    given by \cite{SSS,MSY}
    \begin{equation}
      k_{\rho}(\varphi) = \frac{e^{2\rho} + L^{f}(\varphi)}{e^{2\rho}-
      L^{f}(\varphi)} \sim_{\rho\to\infty} 1 + 2 e^{-2\rho} L^f(\varphi) + \ldots
    \end{equation}

    The geometry near the ideal boundary can therefore be characterised precisely 
    by $L^f(\varphi)$. Thus two geometries $L$ and $L^f$ are asymptotically equal, 
    iff $L^f=L$ i.e.\ iff $f$ is in the Virasoro stabiliser of $L$, $f\in \Stab(L)$, which, 
    as we have seen in Section \ref{sec:Isom}, correspond to isometries of
    $ds^2(L)$. Thus from this geometric perspective we recover that the moduli
    space of metrics of the form $ds^2(L^f)$ for a given $L$ should be identified
    with the corresponding Virasoro coadjoint orbit through $L$, 
    \be
    \{g\in\M_{FG} (S)\colon k \to 1+ 2 e^{-2\rho}L^f, f\in \Diff^+(S^1)\}
    \cong \OO_L 
    \label{eq:mls}
    \ee
    Combining this with our insights how the geometries in distinct orbits differ
    from each other, and with the classification of Virasoro coadjoint orbits, 
    this leads us to our 

    \paragraph{Main Statement (preliminary version)}
    A Virasoro coadjoint orbit $\OO_{\sigma,n_0}$ of type $(\sigma,n_0)$ 
    describes a moduli space of possibly singular hyperbolic metrics on $S$,
    with prescribed behaviour of the geometry near the ideal boundary
    $\del_{\infty}S$, namely with extrinsic curvature determined by $L\in
    \OO_{\sigma,n_0}$. 

    Thus for example, for $(\sigma,n_0)=(0,n)$ (degenerate monodromy) one obtains
    a moduli space of hyperbolic metrics on the disc for $n=1$ or on 
    an $n$-fold branched cover of the disc for $n>1$; likewise for 
    elliptic monodromies $(\alpha,n_0)$ one obtains moduli spaces of hyperbolic metrics
    with conical singularities, etc.; each with their characteristic asymptotics 
    as visible e.g.\ via the extrinsic curvature.  

    From the description given here, however, 
    it is a priori far from evident that this kind of moduli problem 
    involving fixing the subleading behaviour of the extrinsic curvature of the
    boundary curve, as in \eqref{eq:mls},  
    is particuarly interesting or natural (even though from a
    physics perspective it arises somewhat naturally in the context of the
    quantisation of two-dimensional JT gravity \cite{SSS,MSY,MertensTuriaci:Review}). 

    Below, we will reformulate the statement more carefully in terms of hyperbolic
    structures and the naturally induced projective structures on the ideal boundary which
    capture the behaviour of the geometry near the ideal boundary, and we will then 
    see that this kind of moduli problem is indeed completely natural also from
a geometric perspective. 

    \subsection{Hyperbolic and Projective Structures}\label{sub:hsps}

    Let us recall the notion of geometric structures (as e.g.\ defined by Goldman
    \cite{Goldman,GoldmanHiggs}). 
    Given a space $M$, and a space $X$ admitting a transitive $G$-action
    for some Lie group $G$, 
    by a \emph{geometric structure} on $M$
    modelled by $(X,G)$ one understands an atlas of charts $\phi_i
    \colon U_i \subset M \to Z_i \subset X$, for $U_i, Z_i$ open sets
    of $M$ and $X$ respectively, whose transition functions $\phi_i
    \circ \phi_j^{-1} $ take values in $G$. I.e.\ for each connected 
    component $C$ of $U_i \cap U_j$ there exists a $g_{C,ij}\in G$ such that 
    $\phi_i\rvert_C = g_{C,ij}\circ \phi_j\rvert_C$.

    For example, a \emph{Euclidean} structure on $M$ is modelled on $(X=\RR^n,G=
    E_n)$ where $E_n =\Isom(\RR^n)$ denotes the Euclidean group of isometries of $\RR^n$.
    Likewise, a \emph{spherical} structure on $M$ is modelled on $(S^n, SO(n+1))$.

    In the following, we will be interested in two specific geometric structures,
    namely a \textit{hyperbolic structure} on a two-dimensional surface
    $\Sigma$, 
    and a \textit{projective structure} on the circle $S^1$, and the relation between
    them:

    \begin{enumerate}
  \item Hyperbolic Structure

    A hyperbolic structure on a two-dimensional oriented surface $\Sigma$ 
    is a geometric structure
    modelled on $(\HH,\PSL(2,\RR))$,  where $\PSL(2,\RR) = \Isom^+(\HH)$
    is the group of orientation-preserving isometries of the Poincar\'e
    upper-half plane $\HH$. Equivalently, one could use the Poincar\'e 
    disc model $(\DD,\PSU(1,1))$. 

    A hyperbolic structure on $\Sigma$ is equivalent to a hyperbolic metric 
    (constant scalar curvature $R=-2$) on $\Sigma$:
    \begin{itemize}
      \item 
	Given a hyperbolic structure on $\Sigma$, i.e.\ an atlas of charts $\{
(U_i, \phi_i) \}$ taking values in $\HH$, one can define a hyperbolic metric on
$U_i \subset \Sigma$ by pulling back the standard hyperbolic metric $ds^2_{\HH}$ on $\HH$ by the charts $\phi_i$. 
	Since the transition functions take values in $\Isom^+(\HH)$, one can patch
	this metric together to obtain a globally defined hyperbolic metric on $\Sigma$.
      \item Conversely, any hyperbolic metric is locally isometric to the the standard
	Poincar\'e metric $ds^2_{\HH}$ on $\HH$, and using these local isometries as
	charts one obtains a hyperbolic structure in the sense defined above. 
    \end{itemize}
    This definition can be extended to surfaces $\Sigma$ with boundary by 
    replacing $\HH$ in the definition by its closure $\bar{\HH}$ (or $\DD$
    by the closed Poincar\'e disc $\bar{\DD}=\{z\in\CC: |z|\leq 1\}$).

    It is important to realise that, even though everything in this definition
    is perfectly smooth, the transition functions entail an
    identification which may or may not lead to a singularity in the resulting 
    geometry. Indeed the prime example of a singular hyperbolic structure that can 
    be obtained in this way 
    (explained e.g.\ in Section 4.3.2 of \cite{GoldmanHiggs}) 
    is precisely a branched hyperbolic structure with 
    a singularity of cone angle $2\pi n, n>1$, just as we found for the covering 
    geometry of the disc.

    More generally, we will see below that the uniformisation maps $z_L\colon S \to \HH$ 
    that we have
    constructed for both singular and non-singular 
    hyperbolic metrics on the cylinder $\Sigma = S$ provide us 
    precisely with such hyperbolic charts that define a hyperbolic structure.

  \item Projective Structure 

    A projective structure on $S^1$ is a geometric structure modelled on
    $(\RP^1,\PSL(2,\RR))$. Thus it provides a local identification of $S^1$ 
    with the projective line $\RP^1$, defined modulo the projective equivalence
    given by the fractional linear (M\"obius) action of $\PSL(2,\RR)$ on $\RP^1$.

    As we will recall  below, projective structures on $S^1$ are in
    one-to-one correspondence with Hill potentials $L(\varphi)$ (see e.g.\
    \cite{Ovsienko}). 

  \item Projective Structures from Hyperbolic Structures

    In our construction of the uniformisation map $z_L$ for 
    hyperbolic metrics on the cylinder $S$ we had seen in \eqref{eq:imz0} that
    the ideal boundary $\del_\infty S$ at $\rho\to\infty$ is mapped to the ideal
    boundary $\{\Im(z_L)=0\}\subset \del\HH \cong \RP^1$ of $\HH$. Thus given 
    a hyperbolic structure on $S$, one can restrict the charts to the ideal boundary. 
    This restriction defines an atlas of the (ideal) boundary $\del_{\infty}S$ whose charts
    take values in $\del\HH \cong \RP^1 \cong \RR \cup \{ \infty \}$. 
    Moreover, since they originate from a hyperbolic structure, 
    the transition functions of this atlas take values in $\PSL(2,\RR)$.
    Therefore, a hyperbolic structure on $S$ defines by restriction a projective structure 
    on $\del_\infty S$.

    \end{enumerate}

    \subsection{Projective Structures from Hyperbolic Structures and Hill's Equation}
    \label{sub:proj}

    We saw in Section \ref{sec:M} that the global properties of the geometries that 
    we are studying are to a large extent encoded in the $\PSL(2,\RR)$ monodromy 
    of the uniformisation map. However, as already mentioned 
    at the end of Section \ref{sec:M}, 
    usually the global aspects of a geometry are encoded
    in the transition functions of local charts (both in the standard definition 
    of a manifold, and in the more restrictive sense of a geometric structure), 
    not in some kind of monodromy. We will now establish the link between the 
    two descriptions and explain, in particular, how we can 
    translate the $\PSL(2,\RR)$-monodromies of the prepotential $F_L$ and 
    the uniformisation map 
    $z_L$ into $\PSL(2,\RR)$-valued transition functions, thus 
    defining a projective structure on $S^1$ and a 
    hyperbolic structure on the cylinder $S$.

    We start with the projective structure on $S^1$ (and discuss the hyperbolic structure 
    in Section \ref{sub:hs}). 
    To that end we first need to introduce a suitable open covering
    $U_i$ of the circle $S^1$, which will then also provide us with a corresponding
    open covering $V_i = \RR \times U_i$ of the cylinder $S=\RR\times S^1$. 
    Let us concretely regard 
    $S^1$ as embedded in $\CC$ as the unit circle, covered by the two open sets
    \begin{equation}
      U_1 = \{ e^{i\varphi_1} \mid \varphi_1 \in (-\pi,\pi)  \} \quad , \quad U_2 =
      \{ e^{i\varphi_2} \mid \varphi_2 \in (0,2\pi)  \}
      \label{eq:U}
    \end{equation}
    with local coordinates $\varphi_i$. 
    Their overlap is given by $U_{12} = U_1 \cap U_2 = S^1 - \{ (-1,0), (1,0) \}$.
    This has two connected components, $U_{12} = C_+ \cup C_-$
    (in order to obtain a \textit{good} cover, 
    with connected intersections, we would need at least three open sets, but this
    is not convenient for discussing the kind of geometric structures on $S^1$ and $S$
    that we are interested in). On the upper arc $C_+$, one has $\varphi_2 = \varphi_1$,  
    and therefore the transition function here is trivial, and we will not need to 
    consider $C_+$ any further in the following. On the lower arc $C_-$, however,
    one has 
    \be
    \varphi_2\rvert_{C_-} = \varphi_1\rvert_{C_-} + 2\pi
    \ee
    We can now take a fresh look at the prepotential $F_L$ \eqref{eq:F}.
    Since $F_L$ is best thought of as a map $F_L: S^1 \to \del\HH \cong \RP^1$ \eqref{FL2}
    (with $\PSL(2,\RR)$-monodromy $M_L$), we can think of the restrictions 
    \be
    F_i = F_L\rvert_{U_i}
    \ee
    as defining local $\RP^1$-valued maps on $S^1$. Due to the non-trivial monodromy 
    $M_L$ of $F_L$, 
    \be
    F_L(\varphi + 2\pi) = M_L\cdot F_L(\varphi) 
    \ee
    on the non-trivial overlap $C_-$ these maps $F_i$ are related by 
    \begin{equation}
      F_2(\varphi_2) = F_1(\varphi_1 + 2\pi) = M_L\cdot F_1(\varphi_1)
    \end{equation}
    Hence $\{ (U_i, F_i) \}$ defines a projective structure on $S^1$. 

    In Section \ref{sec:properties_z}, we had noted that asymptotically the
    uniformisation map behaves as 
    \begin{equation}
      \lim_{\rho \to \infty} z_{L}(\rho,\varphi) =
      \frac{\psi_2(\varphi)}{\psi_1(\varphi)} = F_{L}(\varphi)
      \label{eq:imz1}
    \end{equation}
    Together with the result (to be established below) that the uniformisation
    map $z_L$ defines a hyperbolic structure on $S$, 
    this provides a very concrete realisation of the general 
    statement made in Section \ref{sub:hsps} that projective structures
    arise as boundary values of hyperbolic structures. 

    Returning to the projective structure, note that 
    there is a one-to-one correspondence between these projective structures 
    defined by $F_L$ and the Hill potentials $L(\varphi)$, provided e.g.\ by Hill's equation 
    \cite{Ovsienko}. Indeed, we had already seen that given any two Wronskian
    normalised solutions to the Hill's equation with potential $L(\varphi)$, 
    $F_L= \psi_2/\psi_1$ defines such a projective structure. The $\SL(2,\RR)$
    action on the $\psi_i$ induces a $\PSL(2,\RR)$ action on $F_L$ defining 
    an equivalent projective structure. Conversely, 
    given a projective structure on $S^1$, i.e.\ an atlas $\{ (U_i, F_i)
    \}$ with transition function $M \in \PSL(2,\RR)$, the $F_i$ patch together into
    a quasi-periodic map $F$ with monodromy $M$.
    Then, $L(\varphi) = \frac12 \Sch(F)$ defines the corresponding Hill's
    potential, and two $F$ related by $\PSL(2,\RR)$ lead to the same $L$. 

    There is also an action of  $\Diff^+(S^1)$ on projective structures. It 
    acts by precomposition, i.e.\ by pullback $F_L\to F_L\circ f$. If $f\in
    \Stab(L)\subset \Diff^+(S^1)$, i.e.\ if $L^f=L$, 
    then $F_L\circ f$ and $F_L$ define equivalent projective structures, 
    while for $f \notin \Stab(L)$, $F_L$ is mapped to an inequivalent distinct
    (but diffeomorphic) projective structure
    \be
\label{eq:FFf}
    F_{L^f} = F_L\circ f 
    \ee
    Hence the space of 
    projective structures can be identified with the space of Hill potentials, 
    thus with $\vir^*$, 
    \be
\label{eq:projvir}
    \mathrm{Proj}(S^1) \cong \vir^* 
    \ee
    and decomposes into Virasoro coadjoint orbits
    under the (coadjoint) action of $\Diff^+(S^1)$. In particular, 
    the orbits can be labelled by parameters $(\sigma,n_0)$, and 
    the projective structures of type $(\sigma,n_0)$ can be labelled
    by either the prepotentials $F_{\sigma,n_0}\circ f$ (modulo
    fractional linear transformations), or directly by the corresponding
    Virasoro quadratic differentials or Hill potentials $(L_{\sigma,n_0})^f$
    for $f\in\Diff^+(S^1)$.

    \subsection{Hyperbolic Structures from the Uniformisation Map and Gauge Theory}\label{sub:hs}

    In order to show how a hyperbolic structure on $S$ arises from the
uniformiation map $z_L$, we extend the open sets $U_i$ of the circle $S^1$
used above to open sets $V_i = \RR \times U_i$ which cover $S$, 
    with local coordinates $(\rho_i,\varphi_i)$. The overlap
    $V_{12} = V_1\cap V_2 = (\RR\times C_+)\cup (\RR\times C_-)$,  
    with trivial transition functions on $\RR\times C_+$ and 
    \be
    \quad (\rho_2,\varphi_2)\rvert_{\RR\times C_-} =
    (\rho_1,\varphi_1 + 2\pi)\rvert_{\RR\times C_-}
    \ee
    With this in hand,  let us now reconsider the uniformisation map 
    $z_L \colon S \to \HH$. Given $z_L$ and the above open sets $V_i$, by
    restriction of $z_L$ to the $V_i$ we obtain local $\HH$-valued maps 
    \be
    \phi_i = z_L\rvert_{V_i} 
    \ee
    The non-trivial monodromy of $z_L$, 
    \be
    z_L(\rho,\varphi+2\pi) = M_L\cdot z_L(\rho,\varphi)
    \ee
    now implies that on
    the non-trivial part $\RR\times C_-$ of the overlap $V_{12}$ 
    the local $\HH$-valued 
    maps $\phi_i$ are related by 
    \begin{equation}
      \phi_2(\rho_2,\varphi_2) = \phi_1(\rho_1,\varphi_1 + 2\pi) = M_L \cdot
      \phi_1(\rho_1,\varphi_1)
    \end{equation}
    and thus the transition function between the $\HH$-valued maps $\phi_i$ 
    is given by the monodromy $M_L\in\PSL(2,\RR)$ of the Hill
    equation defined by $L(\varphi)$. 

For constant $L_0$, i.e.\ the simple model disc, cone, funnel and cusp 
geometries already described in 
Section \ref{sec:examples} this is already the end of the story, i.e.\ these data
are already sufficient to define a hyperbolic structure on $S$. Indeed, 
we know that in these cases the FG coordinates are global coordinates on 
$S$ and thus the open sets $V_i$ defined above already provide a complete
open covering of $S$, with the hyperbolic charts $\phi_i$ and the
$\SL(2,\RR)$-valued constant transition function $M_{L_0}$. 

For other $L_1\neq L_0$ (either for $L_1= L_0^f$ in the orbit of a
constant $L_0$ or for $L_1$ in an exotic orbit with no constant
representative), a bit more work is required to extend the above
data to a hyperbolic structure on all of $S$. Indeed, as we have
discussed before, the FG coordinates $(\rho,\varphi)$, and hence
also the unformisation map $z_{L_1}(\rho,\varphi)$ may (and often will)
only be defined asymptotically. 
One thus needs to find a way 
to continue these geometries into the interior (bulk). The candidate
interior geometries are precisely the model geometries described
by constant $L_0$. 

For an $L_1$ of the form $L_1=L_0^f$, one could envisage using the change of 
coordinates provided by the asymptotic diffeomorphism $\tilde{f}$ to 
accomplish this continuation. After all, its key property is that it relates 
the geometry corresponding to $L_0^f$ to that corresponding to $L_0$, 
via $ds^2(L_0^f) = \tilde f^*ds^2(L_0)$ \eqref{eq:ftllf}, 
and the latter does extend to the entire bulk of $S$. This works,
and we will come back to this below (see the discussion around 
Figure \ref{fig:attaching_an_annulus}). However, for present 
purposes this is not the most useful and insightful way to proceed, 
in particular because this procedure does not work for the exotic 
orbits. 
Instead, we will adopt a gauge theory perspective which, as we will see 
immediately, will allow us to deal with both cases simultaneously.

Thus consider the hyperbolic geometry defined asymptotically 
by some $L_1$ through the gauge field \eqref{eq:A}
\be
A(L_1)
    = \mat{\frac{d\rho}{2}}{-L_1(\varphi)e^{-\rho}d\varphi}{e^{\rho}d\varphi}{-\frac{d\rho}{2}}
\ee
We would like
to extend this gauge field to the entire bulk interior by performing 
a gauge transformation to a suitable internal gauge field $A(L_0)$
that allows this. This gauge transformation is performed 
in some intermediate region 
(so as not to change the asymptotics, namely
the induced projective structure on $\del_\infty S$). As shown in 
Section \ref{sec:gbulk}, such a gauge transformation has the form
	\begin{equation}
	  A(L_1) = A(L_0)^{h_{L_0L_1}} \quad , \quad h_{L_0L_1} = g_{L_0}^{-1}g_{L_1} \in \PSL(2,\RR) 
	  \label{eq:glue_A}
	\end{equation}
with $h_{L_0L_1}$ an allowed (periodic) gauge transformation if and only if
$L_0$ and $L_1$ have the same monodromy, $M_{L_0}=M_{L_1}$. We are now able 
to show that one can indeed find a suitable interior completion for any $L_1$:

\begin{enumerate}
\item Standard Orbits: $L_1 = L_0^f$

As shown in \eqref{eq:mfm}, in this case the solution vectors $\Psi_{L_1}$ and
$\Psi_{L_0}$ (and hence $g_{L_1}$ and $g_{L_0}$) can be chosen to have the same 
momodromy and one can glue the external $A(L_0^f)$ to the internal $A(L_0)$ via 
the gauge transformation $h_{L_0L_1}$. 

\item Degenerate Orbits: $L_1 = L_{0,n}$

      Even though in this special case the uniformisation map $z_{L_{0,n}}$ provides a priori global
      coordinates for the corresponding geometry, it is instructive to see how we 
      can obtain the hyperbolic
      structure by extending $A(L_{0,n})$ from some asymptotic region into the
      bulk by using the gauge transformation \eqref{eq:glue_A}, and using 
the standard disc geometry based on $L_{0,1}$ as the internal geometry. 

      The novelty compared with the above case $L_1=L_0^f$ 
      is that we are confronted with a large gauge transformation.
      Recall that away from the origin, the uniformisation map $w_{L_{0,n}}$
\eqref{eq:w}
      defines an $n$-sheeted cover of the disc, cf. Section
      \ref{sec:degenerate_monodromy_geometry}.
      The asymptotic geometry is thus rather given by $n$ pairs
      $(S_k,A_k(L_{0,n}))$, where $S_k = \RR \times [2\pi k/n,2\pi (k+1)/n)$
        is a copy (the $k$th sheet) of $S$ and $A_k(L_{0,n}) =
        A(L_{0,n})\rvert_{S_k}$.
      For each sheet $S_k$, we extend the gauge field
      $A_k(L_{0,n})$ by $A(L_{0,1})$ (we can do this since, 
regarded as elements of $\PSL(2,\RR)$, 
      the monodromies 
\eqref{eq:hol0n} of $A_k(L_{0,n})$ are the same for all $n$).
      In this way, we can realise the hyperbolic structure for $(S,A(L_{0,n}))$
as an $n$-fold cover of $(S,A(L_{0,1}))$, precisely as anticipated and
described in Section \ref{sec:degenerate_monodromy_geometry}.

\item Exotic Orbits I: $L_1= L_{\sigma,1}$, $\sigma \in \{\ell,\pm\}$

\begin{enumerate}
\item $\sigma \in \{\ell,+\}$

For any $L_1 = L_{\sigma,1}\in\OO_{\sigma,1}$, the monodromy can be chosen to be 
equal to the monodromy of the model geometry $L_{\sigma,0}$, i.e.\ a standard 
funnel for $\sigma=\ell$ or a standard cusp for $\sigma=+$. Thus one can glue 
the external exotic geometry defined by $A(L_1)$ (consisting of a single sheet
for $n=1$) to the standard internal 
geometry defined by $A(L_0)$ via a gauge transformation. The main difference
compared with the first case $L_1=L_0^f$ 
is that in this case $h_{L_0L_1}$ is a large gauge transformation, i.e.\ has 
non-trivial winding number $n=1$ (cf. the discussion in Section \ref{sec:G}).

\item $\sigma = -$

At first sight, for the exotic cusp with $L_1= L_{-,1}$ with monodromy $M_-$ 
there appears to be no suitable interior geometry. However, as
noted way back in \eqref{eq:Aminus}, a suitable gauge field with holonomy
$M_-$ can simply be constructed by taking the standard gauge field $A(L_{+,0}=0)$
for the cusp 
with holonomy $M_+$ and reversing the orientation of  the circle. 
This clearly has the effect of replacing the holonomy $M_+$ by $M_- = (M_+)^{-1}$ 
and the flat $\SL(2,\RR)$ gauge field that accomplishes this is
	\begin{equation}
	  A(L_0) \equiv \mat{\frac{d\rho}{2}}{\phantom{-}0}{-e^{\rho}d\varphi}{-\frac{d\rho}{2}}
	\end{equation}
In this way one can glue the external gauge field $A(L_{-,1})$ to the interior
orientation-reversed cusp geometry described by this gauge field $A(L_{0})$ 
via a large gauge transformation.  
\end{enumerate}
 \item Exotic Orbits II: $L_1 = L_{\sigma,n}$, $\sigma \in \{\ell, \pm \}$, $n >
      1$

      For higher winding numbers, $n > 1$, we proceed as for the 
      degenerate
      orbits.
      The gauge field $A(L_{\sigma,n})$ in an asymptotic region is given by
      $n$ gauge fields $A_k(L_{\sigma,n}) =
      A(L_{\sigma,n})\rvert_{S_k}$ defined on $n$-copies $S_k$ of $S$.
      For each copy $S_k$, the $\PSL(2,\RR)$ monodromy / holonomy 
      of $A_k(L_{\sigma,n})$ is given by
      $M_{\sigma/n}$, and we
      therefore extend the gauge field on each copy separately by
      $A(L_{\sigma/n,1})$.
      The resulting gauge field $A(L_{\sigma,n})$ has 
      monodromy $(M_{\sigma/n})^n= M_\sigma$
\eqref{eq:MMn}, as required, and the corresponding geometry   
 then indeed defines an $n$-fold cover of
      $(S,A(L_{\sigma/n,1}))$, as anticipated in Section \ref{sec:F}.

\end{enumerate}
To summarise, by this construction 
we have defined a flat gauge field $A$ on $S$ in such a way that in the bulk interior it takes the
form $A = A(L_0)$, while close to the ideal boundary it takes the form $A =
A(L_1)$. This is schematically illustrated in Figure \ref{fig:A_gluing}.

	\begin{figure}[htb]
	  \centering
    \hspace{3cm}
	  \begin{tikzpicture}
	    \tkzDefPoint(0,0){C}

	    \tkzDefPoint(0,.5){A1}
	    \tkzDefPoint(0,1.7){A2}
	    \tkzDefPoint(0,2){A3}
	    \tkzDefPoint(0,2.3){A4}

	    \tkzDrawCircle[thick,black,dashed,pattern=north east lines,pattern color=gray](C,A3)
	    \tkzDrawCircle[thick,black,fill=white](C,A2)
	    \tkzDrawCircle[thick,black](C,A4)
	    \tkzDrawCircle[thick,black](C,A1)

	    \tkzLabelPoint(0,-0.6){\small $A(L_0)$}
	    \tkzLabelPoint(135:3.5){\small $A(L_1)$}
	    \tkzLabelPoint(10:4.5){\small $A(L_1) = A(L_0)^{h_{L_0L_1}}$}

	    \tkzDefPoint(9:2.7){X}
	    \tkzDefPoint(10:1.85){Y}

	    \tkzDefPoint(132:2.7){P}
	    \tkzDefPoint(130:2.2){Q}

	    \tkzDrawSegment[-latex,thick](X,Y)
	    \tkzDrawSegment[-latex,thick](P,Q)

	  \end{tikzpicture}
	  \caption{Pictorial representation of the gluing construction.
The gauge field $A(L_0)$ extends the gauge field $A(L_1)$ in the asymptotic annular region 
to the entire bulk interior via a gauge transformation. 
The gluing happens in the shaded region.
For visualisation purposes, $S$ is represented as an annulus.}
	  \label{fig:A_gluing}
	\end{figure}
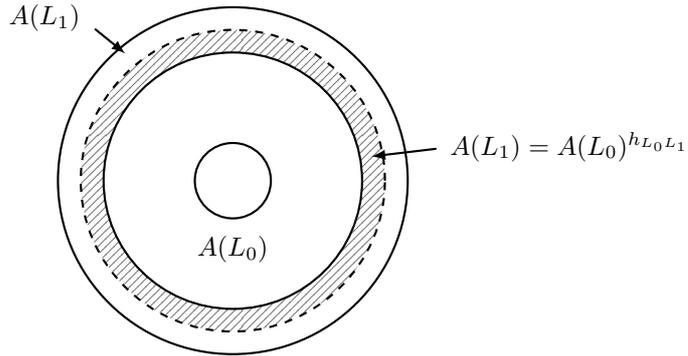

The gauge field $A$, in turn, defines a hyperbolic metric
which in the bulk looks like $ds^2(L_0)$
but close to the ideal boundary looks like $ds^2(L_1)$.
As a consequence of the fact that we now have a hyperbolic
metric on all of $S$ and not just asymptotically, 
this now also defines a hyperbolic structure on $S$. 
One can also see this 
explicitly from the transition function between the local charts 
provided by the uniformisation maps $z_{L_0}$ and $z_{L_1}$: one has
\be
z_{L_1} = g_{L_1}\cdot i = g_{L_1} (g_{L_0})^{-1} \cdot z_{L_0} = 
g_{L_0} h_{L_0L_1} (g_{L_0})^{-1} \cdot z_{L_0}
\ee
which is $\PSL(2,\RR)$-valued, as required, and conjugate to the 
gauge transformation $h_{L_0L_1}$. 

In the case $L_1=L_0^f$, this gauge theoretic gluing procedure is equivalent to 
the gluing of geometries via $\tilde{f}$ briefly mentioned above. Indeed, as 
already noted in \eqref{eq:z_via_gf_and_tf_g}, one has 
$z_{L_0^f} = g_{L_0^f} \cdot i = \tilde f^* z_{L_0}$, 
which implies directly that the metrics are glued according to 
$\tilde f^*ds^2(L_0) = ds^2(L_0^f)$ \eqref{eq:ftllf2}. This is illustrated in
Figure \ref{fig:attaching_an_annulus}.

	\begin{figure}[htb]
	  \centering
          \hspace{3cm}
	  \begin{tikzpicture}
	    \tkzDefPoint(0,0){C}

	    \tkzDefPoint(0,.5){A1F_0}
	    \tkzDefPoint(0,1.7){A2}
	    \tkzDefPoint(0,2){A3}
	    \tkzDefPoint(0,2.3){A4}

	    \tkzDrawCircle[thick,black,dashed,pattern=north east lines,pattern color=gray](C,A3)
	    \tkzDrawCircle[thick,black,fill=white](C,A2)
	    \tkzDrawCircle[thick,black](C,A4)
	    \tkzDrawCircle[thick,black](C,A1)

	    \tkzLabelPoint(0,-0.6){\small $ds^2(L_0)$}
	    \tkzLabelPoint(135:3.5){\small $ds^2(L_0^f)$}
	    \tkzLabelPoint(10:4.5){\small $\tilde f^*ds^2(L_0) = ds^2(L_0^f)$}

	    \tkzDefPoint(9:2.7){X}
	    \tkzDefPoint(10:1.85){Y}

	    \tkzDefPoint(132:2.7){P}
	    \tkzDefPoint(130:2.2){Q}

	    \tkzDrawSegment[-latex,thick](X,Y)
	    \tkzDrawSegment[-latex,thick](P,Q)

	  \end{tikzpicture}
	  \caption{Attaching an annulus to $(S,ds^2(L_0))$ using $\tilde f$. The gluing happens in the shaded region.
	  One then extends the metric to the annular region via pullback: $\tilde f^*ds^2(L_0) = ds^2(L_0^f)$. For visualisation purposes, $S$ is represented as an annulus.}
	  \label{fig:attaching_an_annulus}
	\end{figure}
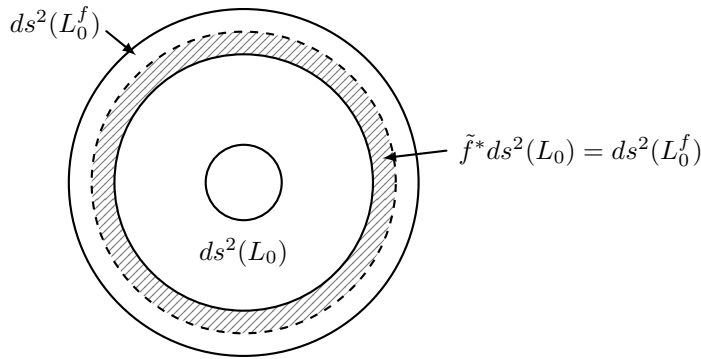

We close this Section with some more remarks on this construction:

\begin{itemize}
\item 
While we have so far motivated and described the above procedure from the point 
of view of extending
a given asymptotic geometry to the interior, it is also illuminating to 
turn this reasoning around. Namely, one can start with the interior geometry
$(S,A(L_0))$. By gluing an annulus to $S$ and extending $A(L_0)$ to $A(L_1)$ 
by \eqref{eq:glue_A} 
one can then effectively change the boundary condition, i.e.\ the induced projective 
structure. 
\item
In this way, the gluing construction allows us to describe a
hyperbolic structure on $S$ with any prescribed boundary condition
by starting with a suitable prototypical geometry $A(L_0)$ in the
interior.  An immediate and important consequence is that the only
singularities in the bulk can be those of the model geometries
$ds^2(L_0)$ and are thus either conical or cuspidal (depending on
whether the monodromy is elliptic or parabolic), and that exotic
funnels with hyperbolic monodromy still have the characteristic
property of possessing a periodic geodesic at the throat.
\item 
In particular, this gauge theoretic construction and this perspective
may shed some light on the possible physical significance and
interpretation of the exotic funnel and cusp geometries. We will come
back to this in Section \ref{sec:outlook}.
\end{itemize}

      \subsection{Moduli Spaces of Hyperbolic Structures: the Projective Perspective}
      \label{sub:modp}

      Combining the above observations, we define the moduli space of 
      hyperbolic structures $\M_{\sigma,n_0}(S)$ on $S$ 
      of type $(\sigma,n_0)$ to be the moduli space of hyperbolic structures on $S$
      which induce a projective structure of type $(\sigma,n_0)$ on the ideal 
      boundary $\del_\infty S$. 
      With this preparation, we can now rephrase the preliminary version of the 
      Main Statement from Section \ref{sub:mod1}: 

      \paragraph{Main Statement (revisited)}

There is a canonical isomorphism 
\begin{equation}
\label{eq:ms2}
\M_{\sigma,n_0}(S) \cong \OO_{\sigma,n_0}
\end{equation}
between the moduli space 
$\M_{\sigma,n_0}(S)$ of hyperbolic structures 
and the Virasoro 
coadjoint orbit $\OO_{\sigma,n_0}$ obtained by associating to any $L_{\sigma,n_0} 
\in\OO_{\sigma,n_0}$ the hyperbolic metric $ds^2(L_{\sigma,n_0})$.

Using the precise understanding of the individual $(\sigma,n_0)$ geometries 
acquired in Section \ref{sec:geometry}, the details of this assertion
for the individual Virasoro orbits are described in Table
\ref{tab:main_statement}.

\begin{table}[htb]
  \centering
  \begin{tabularx}{\textwidth}{c|Y|Y|c}
    $\M_{\sigma,n_0}$ & topology & singularity & Virasoro orbit \\ \hline
    $\M_{0,1}$ & unit disc $\DD$ & none (smooth) & $\OO_{0,1}\cong \Diff^{+}(S^1) / \PSL(2,\RR)$ \\ \hline
    $\M_{0,n}$ & $n$-fold branched covers $\DD^*_{0,n}$ of punctured disc $\DD^*$ & conical & $\OO_{0,n}\cong \Diff^{+}(S^1) /
    \PSL^{(n)}(2,\RR)$ \\ \hline
    $\M_{\alpha,0}$ & cone $\DD^*_{\alpha}$ with opening angle $2\pi\alpha$, $\alpha
    \in (0,1)$ & conical &
    $\OO_{\alpha,0}\cong \Diff^{+}(S^1) / S^1$ \\ \hline
    $\M_{\alpha,n}$ & excess cone $\DD^*_{1 + \alpha}$ with opening angle $2\pi(1 + \alpha)$, $\alpha \in (0,1)$ (and $n$-fold branched covers of $\DD^*_{1 + \alpha/n}$) & conical &
    $\OO_{\alpha,n}\cong \Diff^{+}(S^1) / S^1$ \\ \hline
    $\M_{\ell,0}$ & funnel with geodesic end of length $2\pi \ell$ & none
    (smooth) &
    $\OO_{\ell,0}\cong \Diff^{+}(S^1) / S^1$ \\ \hline
    $\M_{\ell,n}$ & exotic funnel $\mathcal F_{\ell,1}$ (and $n$-fold branched coverings of $\mathcal F_{\ell/n,1}$) & none
    (smooth)  &
    $\OO_{\ell,n}\cong \Diff^{+}(S^1) / \RR \times \ZZ_n$ \\ \hline
    $\M_{+,0}$ & cusp $\DD^*_0$ & cuspidal &
    $\OO_{+,0}\cong \Diff^{+}(S^1) / S^1$ \\ \hline
    $\M_{\pm,n}$ & exotic cusps $\DD^*_{\pm,1}$ (and their $n$-fold branched covers) & cuspidal
     &
    $\OO_{\pm,n}\cong \Diff^{+}(S^1) / \RR \times \ZZ_n$ 
  \end{tabularx}
  \caption{Overview of the Main Statement}
  \label{tab:main_statement}
\end{table}
      Note that in order to define a moduli space of a particular type $(\sigma,n_0)$, 
      it was not necessary to specify a priori what kind of singularity one is 
      perhaps willing to admit - instead  the structure of the resulting geometry 
      is an output rather than an input and determined by the boundary condition
      encoded in the boundary projective structure. This is in line with standard
      results in hyperbolic geometry like \cite{Nitsche}, where it is 
      proved that an isolated singularity of a hyperbolic metric is 
      either conical or a cusp. 

For the surfaces $S$ considered here 
      our results provide a somewhat more precise and stringent 
      relationship between singularities and boundary conditions: 
starting with
one of the standard interior geometries, the type of boundary condition that one
can obtain is determined  by the monodromy $\sigma$ and cannot be changed 
arbitrarily. 

      In this sense, the entries in Table \ref{tab:main_statement} 
can also be regarded as
      providing an answer to the question what sort of bulk hyperbolic
      structure on $S$ is required to induce a given projective
      structure on the boundary.

      Nevertheless, there remain many interesting questions which go beyond 
the scope of this article (and/or the abilities of the authors). We mention some 
of them in Section \ref{sec:outlook} below.

\subsection{Outlook}\label{sec:outlook}

Here we provide a quick outlook on some possible implications of
the work we have presented. On the physics side, we focus on the
possible significance and interpretation of the exotic cuspidal and
funnel geometries that we have already discussed extensively 
in this paper:

\begin{itemize}

\item First of all, as we have seen, these exotic geometries can
be obtained via large gauge transformations from the standard cusp
and funnel geometries. We had already seen that for the asymptotic
geometries in Section \ref{sec:G}, and we have shown in Section
\ref{sub:hs} how to extend this procedure to all of $S$.  Thus,
should one want to include these configurations in the JT path
integral, say, this can be done straightforwardly in the gauge
theoretic formulation of JT gravity.

\item
While we have worked exclusively with the (thermal) Euclidean
signature hyperbolic metrics, it would be of interest to extend
this to the Lorentzian case (we will come back to this
below).  In particular, one  would want to check if the exotic
orbits lead to the kink geometries of \cite{SchallerStrobl}, which 
are also obtained from large gauge transformations.

\item Next, we have a very clear understanding now in precisely
which way the exotic geometries differ from their standard counterparts:
they arise in a sector of the theory characterised by non-standard
and topologically inequivalent boundary conditions, labelled by a
winding number $n_0\neq 0$ (while the standard boundary condition
corresponds to $n_0=0$). Moreover, as shown in Section 
\ref{sec:geometry}, they can be regarded as 
deformations of the disc geometry that break the isometry group
of the disc to a hyperbolic or parabolic group respectively.

\item Thus, as for the standard geometries
\cite{MertensTuriaci:Defects} one may also be able to think of and
realise these exotic geometries in JT gravity, say, as arising from
the insertion of suitable defect operators which break the symmetry 
and create these new topological sectors.

\end{itemize}

On the mathematics side, we conclude this discussion with some (to the best 
of our knowledge) open questions:

	\begin{itemize}
\item 
In \cite{DuvalGuieu} the authors pose the question whether each Virasoro coadjoint orbit can be realised as a conformal class of Lorentzian metrics on the cylinder.
It is thus natural to ask to which extent the techniques developed in
this article can be carried over to the Lorentzian case and if they can give an
affirmative answer. For Lorentzian signature, a suitable analogue of FG coordinates
are Eddington-Finkelstein (EF) coordinates (employed in this context e.g.\ in 
\cite{Jensen}), in which
the metric takes the form
\begin{equation}
ds^2(L) = (r^2 - L(\varphi))d\varphi^2  + 2dr d\varphi
\end{equation}
As before, one can show that diffeomorphisms $\tilde f$ preserving the EF gauge
are paramet\-rised by diffeomorphisms $f$ of the circle such that 
$\tilde f^*ds^2(L) = ds^2(L^f)$, which thus appears to be a suitable starting
point.

	  \item The Main Statement establishes a \emph{diffeomorphism}
	    $\M_{\sigma,n}\cong \OO_{\sigma,n}$.
	    Since $\OO_{\sigma,n}$ is naturally symplectic, it is equally
natural to ask if one can strengthen the correspondence to a
	    \emph{symplectomorphism}.
	    For this, one would first need to find a natural symplectic form on
	    $\M_{\sigma,n}$ which is an interesting question in its own right.
            See \cite{DuvalGuieu,AlekseevMeinrenken1} for some relevant considerations.

	  \item Our discussion was based on real differential geometric
and gauge theoretic 
	    considerations. 
	    From a mathematical point of view, it would perhaps be interesting to see if
	    there exists a complex analytic analogue of the observations in this
paper which allows one to
	    interpret the Virasoro orbits $\OO_{\sigma,n}$ as Teichm\"uller spaces
	    for the punctured disc possibly with a cuspidal or conical singularity
	    at the puncture, and in what sense 
	    this would yield a generalisation of the well-known correspondence of the universal Teichm\"uller space and the Virasoro orbit $\OO_{0,1} \cong \Diff^{+}(S^1) / \PSL(2,\RR)$ \cite{NV}.

	\end{itemize}

      \subsection*{Acknowledgements}

      We are grateful to Anton Alekseev, Rea Dalipi, Valdo Tatitscheff and George Thompson
      for interesting and stimulating discussions. We are also grateful to Shahin
Sheikh-Jabbari for bringing \cite{SJY1,SJY2} to our attention. 

      The work of M.\ Blau is supported by the NCCR SwissMAP (The Mathematics
      of Physics) of the Swiss Science Foundation.  The work of D.\ R.\
      Youmans was supported by the Deutsche Forschungsgemeinschaft (DFG,
      German Research Foundation) under Germany’s Excellence Strategy EXC
      2181/1 - 390900948 (the Heidelberg STRUCTURES Excellence Cluster).

      \appendix

      \section{Hill's Equation and the Classification of Virasoro Coadjoint Orbits}
      \label{app:vir}

      In this appendix we briefly recall the necessary definitions and results of the study of Virasoro coadjoint orbits.
      For a more detailed account of the subject we refer the interested reader to the
      standard literature, e.g.\ \cite{Balog_Feher_Palla,Witten} or the nice review in
      \cite{Oblak}.  Our presentation of the theory differs somewhat 
from other accounts by making systematic use of the Hill prepotential $F_L$
\eqref{eq:F} and
its properties. 

      \subsection{Virasoro Coadjoint Orbits: Basic Definitions}

      Recall that the Virasoro algebra $\vir$ is given by the central extension of the
      algebra of vector fields on the circle, and can thus be represented by pairs 
      $(v(\varphi),s)$, with $v(\varphi + 2\pi)=v(\varphi)$, 
      $v(\varphi)\del_\varphi$ a vector field on the circle, and central element $s\in\RR$. 
      Its (smooth) dual $\vir^*$ can then naturally 
      be identified with pairs $(L(\varphi)d\varphi^2, t)$ of quadratic differentials
      on the circle and dual central elements $t \in \RR$, where $L(\varphi)$ is smooth
      and 
      periodic. The Virasoro algebra acts naturally on $\vir^*$ via the infinitesimal coadjoint action:
      \begin{equation}
	\delta_{(v(\varphi),s)} (L(\varphi)d\varphi^2, t) = (\delta_v L(\varphi) d\varphi^2, 0)
      \end{equation}
      where
      \begin{equation} 
	\label{eq:infcoad}
	\delta_{v} L(\varphi) = 2 v'(\varphi) L(\varphi) + v(\varphi) L'(\varphi) + \frac{t}{12} L'''(\varphi)
      \end{equation}
      This action integrates to the action of $\Diff^+(S^1)$ on $\vir^*$: for $f \in \Diff^+(S^1)$,
      \begin{equation}
	L^{f}(\varphi) = Ad^*_{f^{-1}}L(\varphi) = f'(\varphi)^2 L(f(\varphi)) + \frac{t}{12}\Sch(f)(\varphi)
	\label{eq:coad}
      \end{equation}
      with $\Sch(f)$ the Schwarzian derivative of $f$, 
      \begin{equation}
	\label{eq:sdef1}
	\Sch(f) = \frac{f'''}{f'} - \frac32 \left( \frac{f''}{f'} \right)^2 
	=\left( \frac{f''}{f'} \right)' - \frac12 \left( \frac{f''}{f'} \right)^2
      \end{equation}
      Throughout the paper, and as in the above equations, 
      we will represent $\Diff^+(S^1)$ by maps
      $e^{i f} \colon S^1 \to S^1$ parametrised by functions $f\colon \RR
      \to\RR$ which are quasi-periodic, $f(\varphi + 2\pi) = f(\varphi) + 2\pi$ and
      have positive derivative $f'(\varphi) > 0$ everywhere (and which therefore, strictly 
      speaking, represent the universal cover $\widetilde \Diff^+(S^1)$ of 
      $\Diff(S^1)$). 
      By abuse of notation, but for the sake of simplicity, 
      we will always write $f\in \Diff^+(S^1)$.
      For any smooth functions $f,g:\RR\to \RR$ one has the composition law
      \begin{equation}
	\label{Scomp}
	\Sch(f\circ g)(\varphi) = (g')^2\Sch(f)(g(\varphi)) + \Sch(g)(\varphi)
      \end{equation}
      which is the cocycle condition implied by \eqref{eq:coad} for $f,g$ (and hence
      $f\circ g$) quasi-periodic.

      Notice that the dual level $t$ is invariant under the coadjoint action, and that
      any $t\neq 0$ can be scaled to any other value by a scaling of $L$. 
      In order to match with formulas in the body of the paper, and with the normalisation of $L$
      as it appears naturally in Hill's equation to be discussed below, we set $t
      = 6$, and thus drop it from the notation. Thus we represent an element
      $(L(\varphi)d\varphi^2,t)$ of $\vir^*$ simply by the periodic function $L(\varphi)$. 

      The coadjoint orbits of the Virasoro group through $L(\varphi)$  are 
      thus of the form 
      \begin{equation}
	\OO_L = \{ L^{f}(\varphi) \mid f \in \Diff^+{S^1} \} \cong \Diff^+(S^1) / \Stab(L)
      \end{equation}
      Here $\Stab(L) \subset \Diff^{+}(S^1)$ denotes the stabiliser subgroup of $L$ under the       coadjoint action \eqref{eq:coad}. 
      From \eqref{eq:infcoad} one sees that, 
      infinitesimally, $\Stab(L)$ is generated by vector fields $v$ satisfying
      \begin{equation}
\label{eq:dvl0}
	\delta_v L = 2v'(\varphi)L(\varphi) + v(\varphi)L'(\varphi) + \frac{1}{2}v'''(\varphi) = 0
      \end{equation}

      \subsection{Hill's Equation and 
	\texorpdfstring{$\SL(2,\RR)$}{SL(2,R)}-Monodromy}\label{app:Hill_and_monodromy}

	In order to obtain a better understanding of the Virasoro coadjoint orbits and
	their classification, it turns out to be extremely useful to study
	an auxiliary problem, namely the properties of solutions to Hill's equation
	\begin{equation}
	  \psi''(\varphi) + L(\varphi) \psi(\varphi) = 0
	  \label{eq:Hill_eqn}
	\end{equation} 
	This is a second order differential equation that is naturally associated to a
	Virasoro coadjoint orbit through $L(\varphi)$. 
	Indeed, this equation is invariant under the $\Diff^+(S^1)$ transformation  
	\begin{equation}
	  \begin{split}
	    \psi(\varphi) &\mapsto \psi^f(\varphi) = \frac{\psi(f(\varphi))}{\sqrt{f'(\varphi)}}\\
	    L(\varphi) &\mapsto L^{f}(\varphi) = f'(\varphi)^2 L(f(\varphi)) + \frac12 \Sch(f)(\varphi)
	  \end{split}
	  \label{eq:Diff_action_on_Hill1}
	\end{equation}
	where $\Sch(f)$ is the Schwarzian derivative \eqref{eq:sdef1}, so that the second line is precisely the coadjoint 
	transformation \eqref{eq:coad} of $L(\varphi)$ (with $t=6$), while the first
	line says that $\psi(\varphi)$ transforms as a $(-1/2)$-density (``square-root of a
	vector field''). In particular, any $\Diff^+(S^1)$-invariant statement about 
	solutions of Hill's equation can be regarded as a statement about the coadjoint
	orbit $\OO_L$ through $L(\varphi)$.

	Thus let $\psi_1(\varphi),\psi_2(\varphi)$ be two linearly independent
	solutions of \eqref{eq:Hill_eqn}, which without loss of generality 
	we choose to have unit Wronskian, i.e.\
	\begin{equation}
	  \label{eq:wronski}
	  \psi_1(\varphi)\psi_2'(\varphi) - \psi_1'(\varphi)\psi_2(\varphi) = 1
	\end{equation} 
	and denote by $\Psi=\Psi_L = (\psi_1\; \psi_2)^t$ the corresponding nowhere
	vanishing solution vector. 

	Here are some statements about the solutions that follow immediately from these
	definitions:
	\begin{enumerate}
      \item 
	If $\hat{\Psi}_L$ is any other Wronskian-normalised solution vector to Hill's equation,
	corresponding to a different choice of basis of solutions
	$\hat{\psi}_{1,2}(\varphi)$, 
	then there is a constant $\SL(2,\RR)$ matrix $S$ such that 
	$\hat{\Psi}_L = S\Psi_L$.
      \item 
	Even though $L(\varphi)$ is periodic, in general $\Psi_L(\varphi)$ will not be periodic.
	However, $\Psi_L(\varphi + 2\pi)$ will again be a Wronskian-normalised solution
	vector and thus one has 
	\begin{equation}
	  \label{eq:psimono}
	  \Psi_L(\varphi + 2\pi) = M_{\Psi_L} \Psi_L(\varphi)
	\end{equation}
	for some constant $M_{\Psi_L}\in\SL(2,\RR)$, the monodromy matrix.
      \item \label{prop:Psi_change_of_basis} 
	Under a change of basis, the monodromy changes as
	\be
	\hat{\Psi}_L = S\Psi_L \quad\Rightarrow\quad M_{\hat{\Psi}_L} = SM_{\Psi_L} S^{-1}
	\label{eq:M_S_Psi_S_M_Psi_S}
	\ee
	In particular, the conjugacy class $[M_{\Psi_L}]$ of the monodromy matrix is            independent
	of the choice of basis and uniquely associated to $L$, $[M_{\Psi_L}] =
	[M_L]$. 
      \item \label{prop:M_diff_inv} The monodromy matrix is invariant under the $\Diff^+(S^1)$-action 
	\eqref{eq:Diff_action_on_Hill1}, i.e.\ for the $f$-transformed solution vector 
	\be
	(\Psi_L)^f = (\psi_1^f,\psi_2^f)^t = \Psi_{L^f} 
	\ee
	one has 
	\be
\label{eq:mfm}
	M_{\Psi_{L^f}} =M_{\Psi_{L}} 
	\ee
	This follows directly from the quasi-periodicity of $f$, $f(\varphi + 2\pi) =
	f(\varphi) + 2\pi$: by \eqref{eq:Diff_action_on_Hill1}, one has
	\begin{equation}
	  \psi_i^f(\varphi + 2\pi) = \frac{\psi_i(f(\varphi+2\pi))}{\sqrt{f'(\varphi +
	  2\pi)}} = \frac{\psi_i(f(\varphi)+ 2\pi)}{\sqrt{f'(\varphi)}} =
	  (M_{\Psi_L})_{ij} \psi^f_j(\varphi)
	  \label{eq:M_Psi_L_f = M_Psi_L}
	\end{equation}
	so that
	\begin{equation}
	  \Psi_{L^f}(\varphi + 2\pi) =
	  M_{\Psi_L}\Psi_{L^f}(\varphi)
	\end{equation}

	\end{enumerate}
	Together, Properties \ref{prop:Psi_change_of_basis} and \ref{prop:M_diff_inv} imply that the conjugacy class $[M_{\Psi_L}]$ is uniquely
	associated to the entire coadjoint orbit $\OO_L$ through $L$. 
The classification of $\SL(2,\RR)$ conjugacy classes will be recalled  in Appendix
\ref{app:virsl2r} below. 

	In addition, we note the following properties:
	\begin{enumerate}
	\addtocounter{enumi}{4}

      \item Since $\Psi_L$ is nowhere vanishing, 
	we can regard it as a quasi-periodic map from $\RR$ to
	$\RR^2 \setminus \{(0,0)\}$, and 
	we can therefore assign to it a winding number.
	This will be discussed in more detail in section \ref{app:F_winding} below. 
      \item An $\SL(2,\RR)$-matrix naturally associated to a Wronskian-normalised
	solution vector to Hill's equation is the Wronskian matrix 
	\begin{equation}
	  W_{\Psi_L}(\varphi) =
	  \left(\Psi_L(\varphi)
	  \Psi_L'(\varphi)\right)
	  = \mat{\psi_1(\varphi)}{\psi_1'(\varphi)}{\psi_2(\varphi)}{\psi_2'(\varphi)}
	  \in \SL(2,\RR) 
	  \label{eq:appWdef}
	\end{equation}
	Properties of the Wronskian matrix will be discussed in Section 
	\ref{sec:Hill_and_gauge_theory}.
	  \end{enumerate}

	  \subsection{\texorpdfstring{$\SL(2,\RR)$}{SL(2,R)} Conjugacy Classes}
	  \label{app:virsl2r}

	  In $\SL(2,\RR)$ one has 4 distinct types of conjugacy classes:

	  \begin{enumerate}
	\item Degenerate conjugacy classes

	  An element $g\in \SL(2,\RR)$ is called degenerate if it is conjugate (and
	  hence equal) to $\pm$ the identity matrix, 
	  \begin{equation}
	    g = \pm \mat{1}{0}{0}{1} = \pm \mathbb{I}
	  \end{equation}
	  and evidently the corresponding conjugacy classes each consist of a single element.

	\item Elliptic conjugacy classes

	  An element $g\in \SL(2,\RR)$ is called elliptic if 
	  \begin{equation}
	    \vert\Tr g\vert < 2
	  \end{equation}
	  An elliptic  element $g$ is conjugate to a matrix of the form $\pm M_\alpha$, where
	  \be
	  M_\alpha = 
	  \mat{\cos(\pi\alpha)}{-\sin(\pi\alpha)}{\sin(\pi\alpha)}{\phantom{-}\cos(\pi\alpha)} 
	  \quad,\quad \alpha\in (0,1)
	  \ee

	\item Hyperbolic conjugacy classes

	  An element $g\in \SL(2,\RR)$ is called hyperbolic if 
	  \begin{equation}
	    \vert\Tr g\vert > 2
	  \end{equation}
	  A hyperbolic element $g$ is conjugate to a matrix of the form $\pm M_\ell$ where
	  \begin{equation}
	    M_\ell = \mat{e^{-\pi\ell}}{0}{0}{e^{+\pi\ell}}
	    \quad,\quad \ell \in \RR_+ 
	  \end{equation}

	\item Parabolic conjugacy classes

	  An element $g\in \SL(2,\RR)$ is called parabolic if 
	  \begin{equation}
	    \vert\Tr g\vert = 2 \quad,\quad g \neq \pm \mat{1}{0}{0}{1}
	  \end{equation}
	  A parabolic element $g$ is conjugate to a matrix of the form $\pm M_\pm$
	  (independent signs), where
	  \be
	  M_\pm = \mat{\phantom{\pm}1}{\phantom{+}0}{\pm1}{\phantom{+}1}
	  \ee

	  \end{enumerate}
	  There is an analogous classification of $\PSL(2,\RR)$ conjugacy classes, which
	  simply amounts to identifying the conjugacy classes $[\pm M]$.

	  \subsection{Hill Prepotential}\label{app:F}

	  Another invariant of a Virasoro coadjoint orbit, not detected by the conjugacy class of
	  the monodromy discussed above, is a kind of winding 
	  number associated to solutions of Hill's equation. It can conveniently be
	  described in terms of the ratio 
	  \be
	  F_{\Psi_L}(\varphi) = \frac{\psi_2(\varphi)}{\psi_1(\varphi)}
	  \label{eq:F}
	  \ee
	  of two Wronskian-normalised solutions of Hill's equation, which also plays an
	  important role in the study of Hill's equation and Virasoro orbits and has a number
	  of interesting properties in its own right. We summarise  these here first, and
	  then discuss the winding number in Appendix \ref{app:F_winding}. 

	  \begin{enumerate}
	\item First of all we note that, as a ratio of two $(-1/2)$-densities, $F_{\Psi_L}(\varphi)$
	  transforms as a scalar (function) under $\Diff^+(S^1)$, 
	  \be
\label{eq:Ff}
	  F_{\Psi_L}^f(\varphi)  = F_{\Psi_L}(f(\varphi)) 
	  \ee
	\item Under a change of basis $\Psi_L \rightarrow S\Psi_L$ with $S \in \SL(2,\RR)$, 
	  $F_{\Psi_L}$ transforms with the fractional linear $\PSL(2,\RR)$-transformation 
	  \be
	  S = \begin{pmatrix}a & b \\ c & d \end{pmatrix} \quad\Rightarrow\quad 
	  F_{S\Psi_L} = \frac{c+ dF_{\Psi_L}}{a + b F_{\Psi_L}} \equiv S\cdot F_{\Psi_L}
\label{eq:fracF}
	  \ee
	  The unusual form of the fractional linear transformation is due to our choice
	  of definition $F=\psi_2/\psi_1$ rather than $\psi_1/\psi_2$; this definition is 
	  more convenient for other reasons. 
	\item
	  In particular, if $\Psi_L$ has monodromy matrix $M_{\Psi_L}$, then $F_{\psi_L}$
	  transforms as 
	  \be
	  \label{eq:F_monodromy}
	  F_{\Psi_L}(\varphi + 2\pi) = M_{\Psi_L}\cdot F_{\Psi_L}(\varphi) 
	  \ee
	\item Given $F_{\Psi_L}$, the Hill potential $L$ itself can be recovered from it
	  by calculating the Schwarzian of $F_{\Psi_L}$, 
	  \begin{equation}
	    \label{eq:LSchF}
	    \frac12 \Sch(F_{\Psi_L})(\varphi) = L(\varphi)
	  \end{equation}
	  In this sense, $F_L(\varphi)$ serves as 
	  a \textit{prepotential} for the Hill's potential $L(\varphi)$ and, for lack of a
	  better name, this is how we will refer to $F_{\Psi_L}$. 
	\item
	  Moreover, note that 
	  \begin{equation}
	    F_{\Psi_L}'(\varphi) = \frac{\psi_1(\varphi)\psi_2'(\varphi) - \psi_1'(\varphi)\psi_2(\varphi)}{\psi_1^2(\varphi)} = \frac{1}{\psi_1^2(\varphi)} > 0
	  \end{equation}
	  It follows that, given $F_{\Psi_L}$, one can recover $\psi_1$ and $\psi_2$ by
	  \begin{equation}
	    \psi_1(\varphi) = \frac{1}{\sqrt{F_{\Psi_L}'(\varphi)}} \quad , \quad
	    \psi_2(\varphi) =
	    \frac{F_{\Psi_L}(\varphi)}{\sqrt{F_{\Psi_L}'(\varphi)}}
	    \label{eq:psi_from_F}
	  \end{equation}
	  Conversely, given some monotone $F(\varphi)$, with these definitions the $\psi_{1,2}$ are
	  solutions of Hill's equation with $L = \tfrac12 \Sch(F)$ and unit Wronskian. 
	  Therefore, in order to construct (representatives of) Virasoro coadjoint orbits
	  it suffices to choose appropriate functions $F$ subject to the quasi-periodicity
	  condition \eqref{eq:F_monodromy}.

\item For any other point $L^f$ in the coadjoint orbit of $L$ one can choose the
prepotential to be simply $F^f_{\Psi_L} = F_{\Psi_L}\circ f$ \eqref{eq:Ff}. With
this choice, the construction \eqref{eq:psi_from_F} implies that the monodromy
of $\Psi_{L^f}$ is the same for all $L^f$ (and not just in the same conjugacy
class).
	  \end{enumerate}

	  \subsection{Hill's Equation and Winding Numbers}\label{app:F_winding}

	  Turning to the winding number of the prepotential 
	  $F_{\Psi_L}$, note first of all that the
	  nowhere zero solution vector $\Psi_L$ can be regarded as a quasi-periodic function
	  \be
	  \Psi_L\colon \RR \to 
	  \RR^2 \setminus \{(0,0)\}
	  \ee
	  (quasi-periodic referring to the possibly non-trivial monodromy).  Independently
	  of the monodromy it is thus possible to associate to $\Psi_L$ a winding number, 
	  which indicates how often the origin is circled as the argument 
	  changes from $\varphi$ to $\varphi + 2\pi$. One can also relate this to the
	  winding number of the corresponding prepotential $F_{\Psi_L} = \psi_1/\psi_2$
	  \eqref{eq:F}.
	  Since $\psi_1(\varphi)$ may have zeros, it is best to think of this prepotential 
	  as the homogeneous coordinate of a map 
	  \be
	  \label{FL2}
	  F_{\Psi_L}: S^1 \rightarrow \RR\cup\{\infty\} \cong \RP^1 \cong S^1 
	  \ee
         (with a monodromy)  from the circle $S^1$ to the projective line $\RP^1$.  
	  In order to better understand this winding number, and its relation to that 
	  of the solution vector $\Psi_L$ itself, let us note
	  the following:
	  \begin{enumerate}
	\item 
	  The prototypical example of a map $F:S^1\to\RP^1$ 
	  with winding number $n$ in this projective sense is 
	  \be
	  F_{0,n}(\varphi) = \tan\left( \frac{n\varphi}{2} \right)
	  \ee 
	  A lift of this map to a map $\Psi: \RR \to \RR^2\setminus \{(0,0)\}$ is 
	  provided by the Wronskian normalised solution vector
	  \be
	  \Psi_{0,n}(\varphi) = \vek{
	    \psi_1=\cos (n\varphi/2)/\sqrt{n/2} }{
	      \psi_2=\sin (n\varphi/2)/\sqrt{n/2} }
	      \ee
	      one finds for the constant value $L(\varphi) = L_{0,n}\equiv n^2/4$ (cf.\ the 
	      classification in Section \ref{app:classification_vir_orbits} below). This map
	      $\Psi$ appears to have ``winding number $n/2$'' and a monodromy
	      $(-1)^n\mathbb{I}$ in
	      $\SL(2,\RR)$, and it is thus clearly 
	      more convenient to think of it simply as a map with winding
	      number $n$ in $\PSL(2,\RR)$. 
	    \item
	      This winding number is not changed if one shifts $n$ by $\alpha \in (0,1)$,
	      i.e.\ also 
	      \be
	      F_{\alpha,n}(\varphi) = \tan\left( \frac{(\alpha+n)\varphi}{2} \right)
	      \ee 
	      has winding number $n$ in this sense (in addition to having $\PSL(2,\RR)$ 
	      monodromy $[M_\alpha]$).
	      As we will see below, these functions $F_{0,n}$ and $F_{\alpha,n}$ are associated 
	      with Virasoro orbits with degenerate or elliptic monodromy respectively
	      (with similar constructions for the other cases). 
	    \item
	      To understand the implications of the winding for the hyperbolic
	      geometry in Section \ref{sec:F}, it will be useful to know that the 
	      winding number of $\Psi_L$ is actually the same as the winding number of the Wronskian
	      $\SL(2,\RR)$ matrix $W_{\Psi_L}$ \eqref{eq:appWdef} around the compact 
	      $\SO(2)\subset\SL(2,\RR)$ subgroup. 
	      \end{enumerate}
	      The last assertion is actually a special case of a 
	      general statement about loops in $\SL(2,\RR)$ (or even $\GL(2,\RR)$), 
	      noted parenthetically and without proof in section 2.2 of \cite{PS}: ``A loop in 
	      $\SL(2,\RR)$ has a winding number which is the winding number of its
	      first column, which is a non-zero vector in $\RR^2$.'' Since the first 
	      column of the Wronskian matrix $W_{\Psi_L}$ is just the nowhere vanishing
	      vector $\Psi_L$, this implies the above statement. 

	      We close this section with a quick proof of the just cited
	      statement in \cite{PS}. Thus let $h\in\GL(2,\RR)$. The condition
	      $\det h \neq 0$ implies in particular that the two column vectors 
	      are non-zero. We can thus parametrise 
	      them by polar coordinates, e.g.\ as
	      \be
	      h=\mat{r_1\cos\alpha_1}{-r_2\sin\alpha_2}{r_1\sin\alpha_1}{r_2\cos\alpha_2}
	      \ee
	      Now $\det h = r_1 r_2 \cos(\alpha_2 - \alpha_1)$, and thus $\det h \neq 0$ 
	      implies that $\alpha_2 - \alpha_1 = \beta \in (-\pi/2,+\pi/2)$, and 
	      \be
	      h=\mat{r_1\cos\alpha_1}{-r_2\sin(\alpha_1+\beta)}{r_1\sin\alpha_1}{r_2\cos(\alpha_1+\beta}
	      \ee
	      Now consider a loop $h(t)$ in $\SL(2,\RR)$. Since $\beta(t) \in (-\pi/2,+\pi/2)$, it
	      cannot wind. Therefore $\alpha_1(t)$ and $\alpha_2(t) = \alpha_1(t) + \beta(t)$
	      have the same winding,  and the winding number of $h(t)$ around $\SO(2)$ is the
	      same as the winding number of the first column vector. Since the argument does
	      not actually use periodicity of $h(t)$, it also works
	      when $h(t)$ has a monodromy. 

\subsection{Hill's Equation and Infinitesimal Virasoro Stabilisers} \label{app:stab}

It follows from \eqref{eq:Diff_action_on_Hill1} that 
$\psi_{1,2}$ behave as $(-1/2)$-densities under diffeomorphisms. 
Therefore quadratic (or bilinear) expressions in $\psi_{1,2}$ transform as vector fields. 
Moreover, any such bilinear expression $v=\psi_i \psi_j$ 
actually solves the
Virasoro stabiliser equation $\delta_v L=0$ \eqref{eq:dvl0}, 
\be
\psi_i'' + L\psi_i = 0 \quad\Rightarrow\quad 2(\psi_i\psi_j)'L + 
(\psi_i\psi_j)L' + \frac12 (\psi_i\psi_j)'''=0 
\ee
as is readily verified using Hill's equation. 
However, due to the possibly
non-trivial monodromy of $\Psi_L$ discussed above, the bilinears are not guaranteed to be periodic
in $\varphi$ (and if not, they are not valid solutions of the stabiliser
equation). Of course, 
if $\Psi_L$ has degenerate monodromy, any bilinear is periodic, and there 
are thus three solutions (generating the Lie algebra $\mathfrak{sl}(2,\RR)$). 
An important obervation due to \cite{Kirillov} is that even when the monodromy is 
non-degenerate there is always
at least one non-trivial periodic bilinear solution. 
Indeed, using the explicit
expressions for the monodromy matrices obtained above, one can check explicitly 
that if 
$\Psi_L$ has monodromy $M_\alpha$, $M_\ell$ or $M_\pm$, then precisely the
linear combination 
\be
  v^{ell}(\varphi) = \frac12\left( \psi_1^2(\varphi) + \psi_2^2(\varphi) \right)
\;,\;
v^{hyp}(\varphi) = \psi_1(\varphi)\psi_2(\varphi) 
\;,\;
  v^{par}(\varphi) = \psi_1^2(\varphi)
\label{eq:appbil}
\end{equation}
is periodic respectively. These vector fields also generate corresponding elliptic /
hyperbolic / parabolic $\SL(2,\RR)$ transformations, 
e.g.\ via the fractional linear
tranformation \eqref{eq:fracF} on the prepotential $F_{\Psi_L}(\varphi)$,
arising from 
\be
F(\varphi + \epsilon v(\varphi)) \approx F(\varphi) + \epsilon v(\varphi)
F'(\varphi) \;\;.
\ee
Indeed, using the representation \eqref{eq:psi_from_F} of the $\psi_i$, 
one sees that the $F'(\varphi)$ cancels and that 
\be
\begin{aligned}
F(\varphi + \epsilon v^{ell}) &\approx F + (\epsilon/2)(1+F^2) \approx \mat{1}{-\epsilon/2}{\epsilon/2}{1} \cdot F \\
F(\varphi + \epsilon v^{hyp}) &\approx F + \epsilon F \approx \mat{1-\epsilon/2}{0}{0}{1+\epsilon/2} \cdot F \\
F(\varphi + \epsilon v^{par}) &\approx F + \epsilon  \approx \mat{1}{0}{\epsilon}{1} \cdot F 
\end{aligned}
\ee
which has precisely the form of an infinitesimal elliptic / hyperbolic /
parabolic $(\mathrm{P})\SL(2,\RR)$ transformation respectively.

	      \subsection{Classification of Virasoro Coadjoint
	      Orbits}\label{app:classification_vir_orbits}

	      It turns out that the two data discussed above, 
              a $\PSL(2,\RR)$ conjugacy class plus a winding
	      number,  together suffice to completely classify 
	      the Virasoro coadjoint orbits.
	      These data can also elegantly be regarded as parametrising the conjugacy classes
	      of $\uSL(2,\RR)$, the universal covering group of $\PSL(2,\RR)$
              \cite{Segal}, but
	      we will not pursue this perspective here.

	      Thus the classification of Virasoro orbits is in terms of pairs $(\sigma,n_0)$, 
	      where we write $\sigma=0$ for the single degenerate conjugacy class $[\pm\mathbb{I}]$, 
	      $\sigma=\alpha\in (0,1)$ for the elliptic conjugacy classes $[\pm M_\alpha]$, 
	      $\sigma=\ell \in \mathbb{R}_+$ for the hyperbolic conjugacy classes $[\pm M_\ell]$, 
	      and finally $\sigma=\epsilon \in \{+,-\}$  for the parabolic conjugacy classes $[\pm
	      M_\epsilon]$. Here and in the following, $n_0\in \NN_0$ 
	      while $n\in \NN$ is a non-zero positive integer. 

	      Here are then the resulting orbits, with a corresponding choice of
	      $F_{\sigma,n_0}$, 
	      and their stabilisers (whose generators can be determined from
\eqref{eq:appbil}). 
	      \begin{itemize}
		\item $(\sigma,n_0) = (0,n)$: Degenerate monodromy

		  These orbits have a monodromy in
		  the degenerate conjugacy class $[\pm\mathbb{I}]$. A convenient and simple
		  choice of prepotential is 
		  \begin{equation}
		    F_{0,n} = \tan\left( \frac{n\varphi}{2} \right)
		  \end{equation}
		  which has the desired degenerate monodromy (it is periodic) and winding number
		  $n$. It leads to the constant representative 
		  \begin{equation}
		    L_{0,n} = \frac12\Sch\left( F_{0,n} \right) = \frac{n^2}{4}
		  \end{equation}
		  The stabiliser of $L_{0,n}$ is generated by the vector fields
		  $\{\del_\varphi,\cos(n\varphi)\del_{\varphi},\sin(n\varphi)\del_{\varphi}\}$
		  which generate an $n$-fold cover of $\PSL(2,\RR)$
		  \begin{equation}
		    \Stab(L_{0,n}) = \PSL^{(n)}(2,\RR)
		  \end{equation}

		\item $(\sigma,n_0)=(\alpha,n_0)$: Elliptic monodromy

		  These orbits have a monodromy in the elliptic conjugacy class $[\pm
		  M_{\alpha}]$ with $\alpha \in (0,1)$. 
		  A simple choice of prepotential is 
		  \begin{equation}
		    F_{\alpha,n_0} = \tan\left(\frac{(\alpha + n_0)\varphi}{2}\right) 
		  \end{equation}
		  which indeed has monodromy $M_\alpha$ and winding number
		  $n_0$, leading to the constant representative 
		  \begin{equation}
		    L_{\alpha,n_0} = \frac12\Sch(F_{\alpha,n_0}) =  \frac{(\alpha + n_0)^2}{4}
		  \end{equation}
		  In this case, the stabiliser of $L_{\alpha,n_0}$ is only one-dimensional
		  and is generated by the vector field $\del_{\varphi}$ which integrates to 
		  \begin{equation}
		    \Stab(L_{\alpha,n_0}) = S^1
		  \end{equation}

		\item $(\sigma,n_0) = (\ell,0)$: Hyperbolic monodromy -- standard orbits

		  These orbits are characterised by a monodromy in the hyperbolic conjugacy
		  class $[\pm M_{\ell}]$. A suitable prepotential is 
		  \begin{equation}
		    F_{\ell,0}
		    = e^{\ell\varphi}
		  \end{equation}
		  with $\ell \in \RR_+$, leading once again to a constant representative, namely
		  \begin{equation}
		    L_{\ell,0} = \frac12\Sch(F_{\ell,0}) =  -\frac{\ell^2}{4}
		  \end{equation}
		  The stabiliser of $L_{\ell,0}$ is again one-dimensional.
		  It is generated by the vector field $\del_{\varphi}$ which integrates to
		  \begin{equation}
		    \Stab(L_{\ell,0}) = S^1
		  \end{equation}

		\item $(\sigma,n_0) = (\ell,n)$: Hyperbolic monodromy -- exotic orbits

		  These orbits are again characterised by a monodromy in the hyperbolic
		  conjugacy class $[\pm M_{\ell}]$ and a non-zero winding number. 
		  They are ``exotic'' in the sense that they do not admit any constant
		  representative. Indeed, the list up to this point has exhausted all constant
		  values of $L(\varphi)=L_0$ except $L_0=0$ (which will appear below for
		  $(\sigma,n_0)=(+,0)$), and thus the remaining orbits cannot have any constant 
		  representative. A convenient and simple choice of prepotential is
		  \begin{equation}
		    F_{\ell,n} = e^{\ell \varphi}
		    \tan\left( \frac{n \varphi}{2} \right)
		  \end{equation} 
		  which obviously has the same hyperbolic monodromy as $F_{\ell,0}$ but in
		  addition has winding number $n$. This simple choice of prepotential leads to 
		  a rather complicated expression for the representative $L_{\ell,n}$  of 
		  the orbit, namely
		  \begin{equation}
		    \label{eq:Lln}
		    L_{\ell,n}(\varphi) = \frac12 \Sch(F_{\ell,n}) = -\frac{\ell^2}{4}
		    -\frac12 \frac{n(n^2 + \ell^2)}{n + \ell\sin(n\varphi)} + \frac{3}{4}
		    \frac{n^2(n^2 - \ell^2)}{(n + \ell \sin(n\varphi))^2}
		  \end{equation}
		  (and it is unlikely that some more complicated choice of prepotential could 
		  lead to a significantly simpler expression for the representative). 

		  Note that in the limit $\ell \to 0$ one obtains the
		  representative $L_{0,n}$.
		  The exotic hyperbolic orbits may thus be seen as a deformation of the orbits
		  with degenerate monodromy.

		  The stabiliser of $L_{\ell,n}$ is generated by the vector field 
		  \begin{equation}
		    \label{eq:vln}
		    v(\varphi)\del_\varphi = \frac{\sin(n\varphi)}{n + \ell \sin(n \varphi)} \del_{\varphi}
		  \end{equation}
		  which has $2n$ simple zeros.
		  This gives a second interpretation of the winding number $n$ in this case.
		  The action of the vector field integrates to an action of $\RR$.
		  However, since $L_{\ell,n}(\varphi)$ is invariant under $\varphi \to
		  \varphi + \frac{2\pi}{n}$, the full stabiliser of
		  $L_{\ell,n}$ is given by the product 
		  \begin{equation}
		    \Stab(L_{\ell,n}) = \RR \times \ZZ_n
		  \end{equation}

		\item $(\sigma,n_0) = (+,0)$: Parabolic monodromy -- standard orbits

		  This orbit is characterised by a monodromy in the parabolic conjugacy class
		  $[\pm M_+]$. A simple prepotential with monodromy $M_+$, i.e.\ 
		  with $F_{+,0}(\varphi+2\pi) = F_{+,0}(\varphi) + 1$ is evidently 
		  \begin{equation}
		    F_{+,0} = \frac{\varphi}{2\pi}
		  \end{equation}
		  leading to 
		  \begin{equation}
		    L_{+,0} = \frac12\Sch(F_{+,0}) = 0
		  \end{equation}
		  The stabiliser of $L_{+,0}$ is one-dimensional and generated by the vector
		  field $\del_{\varphi}$ which again integrates to
		  \begin{equation}
		    \Stab(L_{+,0}) = S^1
		  \end{equation}

		\item $(\sigma,n_0) = (\pm,n)$: Parabolic monodromy -- exotic orbits

		  These orbits are characterised by a monodromy in the parabolic
		  conjugacy class $[\pm M_{\pm}]$, with non-zero winding.
		  Like the exotic hyperbolic orbits, they do not admit a constant
		  representative. 
		  A convenient choice of representative is found from the prepotential 
		  one obtains by adding winding to $F_{+,0}$ (and we can now allow for both
		  signs), namely 
		  \begin{equation}
		    F_{\pm,n} = \pm
		    \frac{\varphi}{2\pi} + \tan\left( \frac{n \varphi}{2} \right)
		  \end{equation}
		  leading to 
		  \begin{equation}
		    \label{eq:Lpmn}
		    L_{\pm,n}(\varphi) = \frac{1}{2}\Sch(F_{\pm,n}) = \frac{n^3}{8}\left( \frac{3\left(
		      \frac{n}{2}\pm\frac{1}{2\pi} \right)}{\left( \frac{n}{2} \pm
		      \frac{1}{2\pi}\cos^2\left( \frac{n \varphi}{2} \right) \right)^2} -
		      \frac{2}{\frac{n}{2} \pm \frac{1}{2\pi}\cos^2\left( \frac{n \varphi}{2} \right)} \right)
		    \end{equation}
		    Note that on these orbits there exists a curve
		    $L^{(t)}_{\pm,n}$, $t>0$, obtained from the deformed prepotential
		    \begin{equation}
		      F^{(t)}_{\pm,n} = \pm \frac{t
		      \varphi}{2\pi} + \tan\left( \frac{n \varphi}{2} \right) \label{eq:Ft}
                    \end{equation}
		    Indeed for all $t>0$ the monodromy of $F^{(t)}_{\pm,n}$ is conjugate to $M_\pm$, 
		    \begin{equation}
		      M_{\pm}^{(t)} = \mat{1}{0}{\pm t}{1} \sim 
		      \mat{1}{0}{\pm 1}{1}  = M_{\pm} 
		    \end{equation}
		    and its winding number is $n$. Thus the corresponding 
		    $L_{\pm,n}^{(t)}$ lie in the same orbit
		    for all $t>0$. 
		    In the limit $t\to 0$ one then obtains the standard degenerate
		    representative $L_{0,n}=n^2/4$.
		    Thus, similarly to the exotic hyperbolic orbits, the exotic parabolic
		    orbits can also be seen as a deformation of orbits with degenerate monodromy.
		    However, there is a qualitative difference since here the deformation is
		    not through a family of orbits, as
		    it was in the exotic hyperbolic case, but
		    in terms of just two distinct orbits (and it is thus not completely 
		    clear if $t\to 0$
		    is really in some sense ``close'' to $t=0$ or not). 

		    The stabiliser of $L_{\pm,n}$ is generated by the vector field
		    \begin{equation}
		      \label{eq:vpmn}
		      v(\varphi) = \frac{\cos^2\left( \frac{n\varphi}{2}
		    \right)}{\frac{n}{2} \pm \frac{1}{2\pi}\cos^2\left( \frac{n \varphi}{2}
		  \right)}\del_{\varphi}
		\end{equation}
		which has $n$ double-zeros.
		Therefore, the winding number $n$ admits a second interesting
		interpretation in this case.

		The action of the vector field integrates to an action of $\RR$.
		However, since $L_{\pm,n}$ is invariant under $\varphi \to \varphi +
		\frac{2\pi}{n}$, the full stabiliser is given by
		the product 
		\begin{equation}
		  \Stab(L_{\pm,n}) = \RR \times \ZZ_n
		\end{equation}

	    \end{itemize}

	    This is the complete list. In particular, there are no orbits corresponding
	    to $(\sigma,n_0)=(0,0)$ and $(\sigma,n_0)=(-,0)$.

	    For convenience, we summarise the classification of Virasoro coadjoint orbits in
	    Table \ref{tab:classification_Vir_orbits}, where we adopt the convention
	    $\alpha \in (0,1)$, $\ell \in \RR_+$ and $n \in \NN$.

	    \begin{table}[htb]
	      \centering
	      \bgroup 
	      \def\arraystretch{2}
	      \begin{tabular}{c|c|c|c|c}
		monodromy & $(\sigma,n_0)$ & $F_{\sigma,n_0}$  & $L_{\sigma,n_0}$ &
		$\Stab(L_{\sigma,n_0})$ \\ \hline

		degenerate & $(0,n)$ & $\tan\left( \frac{n\varphi}{2} \right)$ & $\frac{n^2}{4}$ & $\PSL^{(n)}(2,\RR)$ \\ \hline

		elliptic & $(\alpha,0)$ & $\tan\left( \frac{\alpha\varphi}{2} \right)$ & $\frac{\alpha^2}{4}$ & $S^1$ \\ \hline

		elliptic & $(\alpha,n)$ & $\tan\left( \frac{(\alpha+n)\varphi}{2} \right)$ & $\frac{(\alpha+n)^2}{4}$ & $S^1$ \\ \hline

		hyperbolic & $(\ell,0)$ & $e^{\ell\varphi}$ & $-\frac{\ell^2}{4}$ & $S^1$ \\ \hline

		hyperbolic & $(\ell,n)$ & $e^{\ell\varphi}\tan\left( \frac{n\varphi}{2}
		\right)$ & \eqref{eq:Lln} & $\RR \times \ZZ_n$ \\ \hline 

		parabolic & $(+,0)$ & $\frac{\varphi}{2\pi}$ & $0$ & $S^1$ \\ \hline

		parabolic & $(\pm,n)$ & $\pm\frac{\varphi}{2\pi} + \tan\left( \frac{n
		\varphi}{2} \right)$ & \eqref{eq:Lpmn}  & $\RR \times \ZZ_n$
	      \end{tabular}
	      \egroup
	      \vspace{.2cm}
	      \caption{Summary of the classification of Virasoro coadjoint orbits with $\alpha \in (0,1)$, $\ell \in \RR_+$ and $n \in \NN$.}
	      \label{tab:classification_Vir_orbits}
	    \end{table}

	    \pagebreak

	    \section{Bulk Extension of
	      \texorpdfstring{$\Diff^+(S^1)$}{Diff(S1)}: Explicit Examples}
	      \label{app:explicit_calc}

	      We explicitly compute the bulk extension $\tilde f$ of a given
	      diffeomorphism $f\in \Diff^+(S^1)$ on the boundary for constant $L_0$.
	      The explicit construction also provides insight into the nature and
		validity of FG coordinates for a general point $L_0^f$.

	      As outlined in Section \ref{sec:bulk_extension}, $\tilde f$ is constructed by
	      comparing the images of the FG-cylinder $S$ under the uniformisation maps
	      $z_{L_0}$ and $z_{L_0^f}$
	      \begin{equation}
		z_{L_0^f}(\rho,\varphi) = z_{L_0}(\rho_0,\varphi_0)
	      \end{equation}
	      where $(\rho_0,\varphi_0)$ are the coordinates of the reference point $L_0$.
	      This defines $\tilde f = (\rho_0,\varphi_0)$ as the coordinate
	      transformation $(\rho_0(\rho,\varphi),\varphi_0(\rho,\varphi))$.

	    \subsection{Degenerate Monodromy: Covering Geometries of the Disc}\label{sub:covering}

	    Let us consider a general point $L^{f}_{0,n} = \frac{n^2}{4}f'^2  +
	    \frac{1}{2}\Sch(f)$ in the Virasoro orbit $\OO_{L_{0,n}}$ and coordinates
	    $(\rho,\varphi)$ such that 
	    \begin{equation}
	      ds^2(L_{0,n}^f) = d\rho^2 + \left( \frac{n}{2} e^{\rho} -
	      \frac{2}{n}
	      \left(\frac{n^2}{4}f'^2(\varphi)  +
	      \frac{1}{2}\Sch(f)(\varphi)  \right)e^{-\rho} \right)^2d\varphi^2
	    \end{equation}
	    A basis of the associated Hill problem is given by 
	    \begin{equation}
	      \psi^f_1(\varphi) = \frac{1}{\sqrt{n/2}} \frac{\cos(n f(\varphi) /
	      2)}{\sqrt{f'(\varphi)}} \quad ,\quad 
	      \psi^f_2(\varphi) = \frac{1}{\sqrt{n/2}} \frac{\sin(n f(\varphi) / 2)}{\sqrt{f'(\varphi)}}.
	    \end{equation}
	    With those, one computes the Poincar\'e disc coordinate
	    \begin{equation}
	      w_{L_{0,n}^f}(\rho,\varphi) = e^{in f(\varphi)} \left( n^2
	      e^{2\rho} + \left( in f'(\varphi) - \frac{f''(\varphi)}{f'(\varphi)}
	      \right)^2 \right) = e^{in f(\varphi)}u(\rho,\varphi)
	    \end{equation}
	    where 
	    \begin{equation}
	      u(\rho,\varphi) = \frac{n^2 e^{2\rho} + \left( in f'(\varphi) -
	      \frac{f''(\varphi)}{f'(\varphi)} \right)^2}{n^2 e^{2\rho} + \left( in f'(\varphi) + \frac{f''(\varphi)}{f'(\varphi)} \right)^2}.
	    \end{equation}
	    By comparing $w_{L_{0,n}}(\rho_0,\varphi_0)$ and
	    $w_{L_{0,n}^f}(\rho,\varphi)$, we can define $(\rho_0,\varphi_0)$ as functions of $(\rho,\varphi)$ and as such the diffeomorphism $\tilde f(\rho,\varphi) = (\rho_0(\rho,\varphi),\varphi_0(\rho,\varphi))$.

	    Let $u(\rho,\varphi) = x(\rho,\varphi) + i y(\rho,\varphi)$, with
	    \begin{equation}
	      x(\rho,\varphi) = \frac{ n^2 \left( e^{2\rho} - f'(\varphi )^2 \right) +
	      \left(\frac{f''(\varphi )}{f'(\varphi )}\right)^2}{n^2\left(  f'(\varphi
		)+ e^{\rho}\right)^2+\left(\frac{f''(\varphi )}{f'(\varphi )}\right)^2}
		\quad ,\quad y(\rho,\varphi) = \frac{-2 n  f''(\varphi )}{n^2\left( f'(\varphi )+ e^{\rho}\right)^2+\left(\frac{f''(\varphi )}{f'(\varphi )}\right)^2}
	      \end{equation}
	      Then
	      \begin{equation}
		w(\rho,\varphi) = e^{i n f(\varphi)} u(\rho,\varphi) = \tanh\left(
		\frac{\rho_0(\rho,\varphi)}{2} \right) e^{in\tilde \varphi(\rho,\varphi)} = w(\rho_0,\varphi_0).
	      \end{equation}
	      It follows immediately that 
	      \begin{equation}
		\tanh\left(\frac{\rho_0(\rho,\varphi)}{2}\right) = \lvert u \rvert,\qquad
		e^{in \varphi_0(\rho,\varphi)} = e^{in f(\varphi) + i \arg u},
	      \end{equation}
	      which implies 
	      \begin{equation}
		\cosh(\rho_0(\rho,\varphi)) = \frac{1 + \lvert u \rvert^2}{1- \lvert u \rvert^2}
	      \end{equation}
	      Finally, with $e^{i\arg u} = \sqrt{u / \bar u}$ we arrive at
	      \begin{equation}
		\begin{split} 
		  \cosh\left(\rho_0(\rho,\varphi)\right) &= \frac{n^2(e^{2\rho} +
		  f'^2(\varphi)) + \left( \frac{f''(\varphi)}{f'(\varphi)} \right)^2}{2n^2 e^{\rho} f'(\varphi)} \\
		  e^{in\varphi_0(\rho,\varphi)} &= e^{in f(\varphi)}\ \sqrt{\frac{n^2
		    e^{2\rho} + \left( in f'(\varphi) - \frac{f''(\varphi)}{f'(\varphi)}
		  \right)^2}{n^2 e^{2\rho} + \left( in f'(\varphi) + \frac{f''(\varphi)}{f'(\varphi)} \right)^2}}
		\end{split}
		\label{eq:finite_trafo_deg}
	      \end{equation}
	      This generalises results of \cite{Choi_Larsen}, in which the case $n = 1$ was studied.

	      Finally, notice that  
	      \begin{equation}
		\lim_{\rho\to\infty} \rho_0(\rho,\varphi) = \infty \quad , \quad
		\lim_{\rho\to\infty} \varphi_0(\rho,\varphi) = f(\varphi)
	      \end{equation}
	      which shows that $\tilde f$ indeed extends $f$.

	      \subsection{Elliptic Monodromy: Conical Geometries}\label{sub:cone}

	      In this case we start with $L_{\alpha,n_0} = \frac{(\alpha+n_0)^2}{4}$ with 
	      $\alpha\in(0,1)$ and $n_0\in\NN_0$. 
	      Note that in the body of the paper it was occasionally useful to distinguish
	      the case $n_0=0$ from $n_0=n\in\NN$. However, here
	      the discussion of both cases can be treated simultaneously and for simplicity 
	      we omit this distinction in this section and 
	      set $\beta = \alpha + n_0$ in the following. 

	      The construction of $\tilde f$ in for elliptic monodromy is analogous to
	      the case of degenerate monodromy discussed in the previous section.
	      In particular, the form of $\tilde f$ is the same
	      \begin{equation}
		\begin{split} 
		  \cosh\left(\rho_0(\rho,\varphi)\right) &= \frac{\beta^2(e^{2\rho} +
		  f'^2(\varphi)) + \left( \frac{f''(\varphi)}{f'(\varphi)}
		\right)^2}{2\beta^2 e^{\rho} f'(\varphi)} \\
		e^{i\beta\varphi_0(\rho,\varphi)} &= e^{i \beta f(\varphi)}\
		\sqrt{\frac{\beta^2
		  e^{2\rho} + \left( i \beta f'(\varphi) - \frac{f''(\varphi)}{f'(\varphi)}
		\right)^2}{\beta^2 e^{2\rho} + \left( i \beta f'(\varphi) + \frac{f''(\varphi)}{f'(\varphi)} \right)^2}}
	      \end{split}
	      \label{eq:finite_trafo_ell}
	    \end{equation}
	    As before   
	    \begin{equation}
	      \lim_{\rho\to\infty} \rho_0(\rho,\varphi) = \infty \quad , \quad
	      \lim_{\rho\to\infty} \varphi_0(\rho,\varphi) = f(\varphi)
	    \end{equation}
	    which shows that $\tilde f$ extends $f$.

	    Notice that the $(\rho_0,\varphi_0)$-coordinates provide a globally
	    defined coordinate system, while the $(\rho,\varphi)$-coordinates might suffer from various coordinate singularities.

	    Moreover, it turns out that for non-trivial $f$, i.e.\ $f \neq id$, the
	    $(\rho,\varphi)$-coordinates do not cover the full space.
	    For example, consider the point $\rho_0 = 0$.
	    By direct inspection of \eqref{eq:finite_trafo_ell}, it follows that  
	    \begin{equation}
	      \cosh\left(\rho_0(\rho,\varphi)\right) = \frac{\beta^2(e^{2\rho} +
	      f'^2(\varphi)) + \left( \frac{f''(\varphi)}{f'(\varphi)} \right)^2}{2\beta^2 e^{\rho} f'(\varphi)} = 1
	    \end{equation}
	    This in turn would imply that 
	    \begin{equation}
	      0 = \beta^2(e^{2\rho} + f'^2(\varphi)) + \left(
	      \frac{f''(\varphi)}{f'(\varphi)} \right)^2 - 2\beta^2 e^{\rho} f'(\varphi) =
	      \beta^2 (e^{\rho} - f'(\varphi))^2 + \left( \frac{f''(\varphi)}{f'(\varphi)} \right)^2
	    \end{equation}
	    which cannot be satisfied for general non-trivial $f \in \Diff^+(S^1)$.
	    In particular, the conical singularity at $\rho_0 = 0$ is not covered by the
	    $(\rho,\varphi)$-coordinates. 
	    This shows in particular that the FG coordinates
	    $(\rho,\varphi)$ cannot be global and consequently that $\tilde f$ is only a
	    \emph{local} diffeomorphism.
	    Note that the same argument, with $\beta \to
	    n\in\NN$, applies to the
	    coordinate transformation
	    \eqref{eq:finite_trafo_deg} in the degenerate case discussed above.

	    In Section \ref{sec:bulk_extension} we claimed that $\tilde f$ integrates the
	    vector fields $\xi_v(L_{\beta})$.
	    We now show this explicitly.
	    Let us first remark that in the shifted coordinate $\rho \to \rho +
	    \log(\beta/2)$, the vector fields $\xi_v(L_{\beta})$ defined in \eqref{eq:xi} take the form
	    \begin{equation}
	      \begin{split} 
		\xi_v(L_{\beta}) &= \xi^{\rho}_v(L_{\beta}) \del_{\rho} +
		\xi_v^{\varphi}(L_{\beta})\del_{\varphi} \\
		&= -v'(\varphi) \del_{\rho} + \left(  v(\varphi) - \frac{v''(\varphi)}{2}
		\frac{1}{\frac{\beta^2}{4}\left( (e^{2\rho} - 1 \right)} \right).
		\label{eq:xi_ell}
	      \end{split}
	    \end{equation}
	    Consider an infinitesimal diffeomorphism $f(\varphi) = \varphi = \varepsilon v(\varphi) + \OO(\varepsilon^2)$.
	    Expanding, \eqref{eq:finite_trafo_ell} in $\varepsilon$, we find 
	    \begin{equation}
	      \begin{split} 
		\cosh(\rho_0(\rho,\varphi)) &= \frac{1}{2e^{\rho}}\left( e^{2\rho} + 1 + 2\varepsilon v'(\varphi) \right)\left( 1 - \varepsilon v'(\varphi) \right) + \OO(\varepsilon^2) \\
		&= \frac12\left( e^{\rho} + e^{-\rho} + 2 \varepsilon v'(\varphi) e^{-\rho} \right)(1 - \varepsilon v'(\varphi)) + \OO(\varepsilon^2) \\
		&= \cosh(\rho) - \varepsilon v'(\varphi) \sinh(\rho) + \OO(\varepsilon^2) \\
		&= \cosh(\rho - \varepsilon v'(\varphi)) + \OO(\varepsilon^2),
	      \end{split}
	    \end{equation}
	    so that 
	    \begin{equation}
	      \rho_0(\rho,\varphi) - \rho = - \varepsilon v'(\varphi) +
	      \OO(\varepsilon^2) = \varepsilon \xi^{\rho}_v(L_{\beta}) + \OO(\varepsilon^2).
	    \end{equation}
	    Likewise,
	    \begin{equation}
	      \begin{split} 
		\varphi_0(\rho,\varphi) &= f(\varphi) + \frac{1}{2 i \beta} \log\left(
		\frac{\beta^2 e^{2\rho} + \left( i\beta f'(\varphi) -
		\frac{f''(\varphi)}{f'(\varphi)} \right)^2}{\beta^2 e^{2\rho} +\left(
		i\beta f'(\varphi) + \frac{f''(\varphi)}{f'(\varphi)} \right)^2} \right) \\
		&= \varphi + \varepsilon v(\varphi) + \frac{1}{2i\beta} \log\left(
		\frac{\beta^2 e^{2\rho} - \beta^2 + 2i\beta\varepsilon(i\beta v'(\varphi) -
	      v''(\varphi))}{\beta^2 e^{2\rho} - \beta^2 + 2 i \beta \varepsilon (i\beta v'(\varphi) + v''(\varphi))} \right) + \OO(\varepsilon^2) \\
	      &= \varphi + \varepsilon v(\varphi) + \frac{1}{2i\beta} \log\left( 1 -
	      \frac{\varepsilon i \beta v''(\varphi) }{\frac{\beta^2}{4}\left( e^{2\rho} - 1 \right)} \right) + \OO(\varepsilon^2)\\
	      &= \varphi + \varepsilon \left( v(\varphi) + \frac{v''(\varphi)}{2}
	      \frac{1}{\frac{\beta^2}{4}\left( e^{2\rho} - 1 \right)} \right) + \OO(\varepsilon^2).
	    \end{split}
	    \label{eq:f_tilde_ell}
	  \end{equation}
	  Hence,
	  \begin{equation}
	    \begin{split} 
	      \varphi_0(\rho,\varphi) - \varphi &= \varepsilon \left( v(\varphi) +
	      \frac{v''(\varphi)}{2} \frac{1}{\frac{\beta^2}{4}\left( e^{2\rho} - 1 \right)} \right) + \OO(\varepsilon^2) \\
	      &= \varepsilon \xi^{\varphi}_v(L_{\beta}) + \OO(\varepsilon^2),
	    \end{split}
	  \end{equation}
	  in accordance with \eqref{eq:xi_ell}.

	  \subsection{Hyperbolic Monodromy: Annular Geometries}\label{sub:funnel}

	  In the following we will repeat the preceding discussion in the case of Virasoro
	  orbits with hyperbolic monodromy that admit a constant representative.

	  Consider a general point $L^{f}_{\ell,0} =  -\frac{\ell^2}{4} f'^2  +
	  \frac{1}{2}\Sch(f)$ in the Virasoro orbit $\OO_{L_{\ell,0}}$ and coordinates
	  $(\rho,\varphi)$ in which, after shifting $\rho \to \rho + \log(\ell/2)$, the metric takes the form
	  \begin{equation}
	    ds^2(L_{\ell,0}^f) = d\rho^2 + \left(  \frac{\ell}{2}e^{\rho} - \left(
	    -\frac{\ell^2}{4}f'^2(\varphi)
	    + \frac{1}{2}\Sch(f)(\varphi) \right) \frac{2}{\ell}e^{-\rho} \right)^2\varphi^2
	  \end{equation}
	  A basis of the associated Hill problem is given by 
	  \begin{equation}
	    \psi^f_1 = \frac{1}{\sqrt{\ell}} \frac{e^{-\ell
	    f(\varphi)/2}}{\sqrt{f'(\varphi)}}\quad , \quad 
	    \psi^f_2 = \frac{1}{\sqrt{\ell}} \frac{e^{\ell f(\varphi)/2}}{\sqrt{f'(\varphi)}}
	    \label{eq:psif_hyp}
	  \end{equation}
	  In this case, substituting \eqref{eq:psif_hyp} into \eqref{eq:zhyp} leads to
	  \begin{equation}
	    z_{L_{\ell,0}^f}(\rho,\varphi) = e^{\ell f(\varphi)} \frac{ e^{\rho} + i \left(
	    f'(\varphi) - \frac{1}{\ell} \frac{f''(\varphi)}{f'(\varphi)}\right)}{
	      e^{\rho} - i \left( f'(\varphi) + \frac{1}{\ell}
	      \frac{f''(\varphi)}{f'(\varphi)}\right)} = e^{\ell f(\varphi)} u(\rho,\varphi)
	      \label{eq:zhypf}
	    \end{equation}
	    where 
	    \begin{equation}
	      u(\rho,\varphi) = \frac{ e^{\rho} + i \left( f'(\varphi) - \frac{1}{\ell} \frac{f''(\varphi)}{f'(\varphi)}\right)}{ e^{\rho} - i \left( f'(\varphi) + \frac{1}{\ell} \frac{f''(\varphi)}{f'(\varphi)}\right)}
	    \end{equation}
	    The comparison of \eqref{eq:z_tilde_hyp} with \eqref{eq:zhypf} defines us the
	    diffeomorphism $\tilde f$ in the form $\tilde f(\rho,\varphi) =
	    (\rho_0(\rho,\varphi), \varphi_0(\rho,\varphi))$.

	    Setting $z(\rho_0,\varphi_0) = z(\rho,\varphi)$ defines $(\rho_0,\varphi_0)$ as functions of $(\rho,\varphi)$.
	    In detail, consider 
	    \begin{equation}
	      z_{L^f_{\ell,0}}(\rho,\varphi) = e^{\ell f(\varphi) + \log\lvert
	      u(\rho,\varphi) \rvert}\ e^{i \arg u(\rho,\varphi)}=
	      e^{\ell\varphi_0(\rho,\varphi)} \frac{e^{\rho_0(\rho,\varphi)} +
	    i}{e^{\rho_0(\rho,\varphi)} - i} = z_{L_{\ell,0}}(\rho_0,\varphi_0)
	  \end{equation}
	  Notice that  
	  \begin{equation}
	    \left\lvert \frac{e^{\rho_0(\rho,\varphi)} + i}{e^{\rho_0(\rho,\varphi)} - i}\right\rvert^2 = 1
	  \end{equation}
	  i.e.\  
	  \begin{equation}
	    \frac{e^{\rho_0(\rho,\varphi)} + i}{e^{\rho_0(\rho,\varphi)} - i} =
	    e^{i\arg u(\rho,\varphi)}  \quad , \quad e^{\ell \varphi_0(\rho,\varphi)} = e^{\ell f(\varphi) }\lvert u(\rho,\varphi) \rvert
	    \label{eq:tilde_rho_hyp}
	  \end{equation} 
	  From \eqref{eq:tilde_rho_hyp}, we obtain 
	  \begin{equation}
	    e^{\rho_0(\rho,\varphi)} = i \frac{e^{i\arg u(\rho,\varphi)} + 1}{e^{i\arg u(\rho,\varphi)} - 1} = \cot\left( \frac{\arg u(\rho,\varphi)}{2} \right)
	  \end{equation} 
	  which implies 
	  \begin{equation}
	    \sinh(\rho_0(\rho,\varphi))) = \cot\left( \arg u(\rho,\varphi) \right).
	  \end{equation}
	  Let $u(\rho,\varphi) = x(\rho,\varphi) + i y(\rho,\varphi)$, with 
	  \begin{equation}
	    x(\rho,\varphi) = \frac{ e^{2\rho} - f'^2(\varphi) + \frac{1}{\ell^2} \left(
	    \frac{f''(\varphi)}{f'(\varphi)} \right)^2}{ e^{2\rho} + \left( f'(\varphi) +
	    \frac{1}{\ell} \frac{f''(\varphi)}{f'(\varphi)}\right)^2} \quad , \quad y(\rho,\varphi) = \frac{2 e^{\rho}f'(\varphi)}{ e^{2\rho} + \left( f'(\varphi) + \frac{1}{\ell} \frac{f''(\varphi)}{f'(\varphi)}\right)^2}
	  \end{equation}
	  Then 
	  \begin{equation}
	    \cot\left( \arg u(\rho,\varphi) \right) = \frac{x(\rho,\varphi)}{y(\rho,\varphi)} 
	  \end{equation}
	  Finally, we obtain 
	  \begin{equation}
	    \begin{split} 
	      \sinh\rho_0(\rho,\varphi) &= \frac{\ell^2 \left( e^{2\rho} - f'^2(\varphi) \right) + \left( \frac{f''(\varphi)}{f'(\varphi)} \right)^2}{2 \ell^2 e^{\rho}f'(\varphi)}   \\
	      e^{\ell\varphi_0(\rho,\varphi)} &=e^{\ell f(\varphi)}\ \sqrt{\frac{\ell^2 e^{2\rho} + \left( \ell f'(\varphi) - \frac{f''(\varphi)}{f'(\varphi)} \right)^2}{\ell^2 e^{2\rho} + \left( \ell f'(\varphi) + \frac{f''(\varphi)}{f'(\varphi)} \right)^2}}
	    \end{split}
	    \label{eq:tilde_f_hyp}
	  \end{equation}
	  It is again straightforward to see that 
	  \begin{equation}
	    \lim_{\rho\to\infty} \rho_0(\rho,\varphi) = \infty \quad , \quad
	    \lim_{\rho\to\infty} \varphi_0(\rho,\varphi) = f(\varphi)
	  \end{equation}
	  which again shows that $\tilde f$ extends $f$.

	  For completeness, let us show that $\tilde f$ integrates the vector fields
	  $\xi_{v}(L_{\ell,0})$.
	  For this, recall first that we have studied the shifted the coordinate $\rho \to \rho + \log(\ell/2)$.
	  Under this change of coordinates, the vector fields $\xi_v(L_{\ell,0})$ defined in \eqref{eq:xi} become 
	  \begin{equation}
	    \begin{split} 
	      \xi_v(L_{\ell,0}) &= \xi_v^{\rho}(L_{\ell,0}) \del_{\rho} +
	      \xi_v^{\varphi}(L_{\ell,0}) \del_{\varphi} \\
	      &= -v'(\varphi) \del_{\rho} + \left( v(\varphi) - \frac{v''(\varphi)}{2} \frac{1}{\frac{\ell^2}{4}\left( e^{2\rho} + 1 \right)} \right)\del_{\varphi}.
	    \end{split}
	    \label{eq:xi_hyp}
	  \end{equation}

	  Now, consider again an infinitesimal diffeomorphism $f(\varphi) = \varphi + \varepsilon v(\varphi) + \OO(\varepsilon^2)$.
	  Expanding \eqref{eq:tilde_f_hyp} in $\varepsilon$, we find
	  \begin{equation}
	    \begin{split} 
	      \sinh(\rho_0(\rho,\varphi)) 
	      &= \frac{e^{-\rho}}{2\ell^2}(1 - \varepsilon v'(\varphi)) \left( \ell^2 \left( e^{2\rho} - (1 + 2\varepsilon v'(\varphi)) \right) \right) + \OO(\varepsilon^2) \\
	      &= \frac{e^{-\rho}}{2}\left( e^{2\rho} - 1 - 2\varepsilon v'(\varphi) - \varepsilon v'(\varphi) e^{2\rho} + \varepsilon v'(\varphi) \right) + \OO(\varepsilon^2) \\
	      &= \sinh(\rho) - \varepsilon \cosh(\rho) v'(\varphi) + \OO(\varepsilon^2) \\
	      &= \sinh(\rho - \varepsilon v'(\varphi)) + \OO(\varepsilon^2).
	    \end{split} 
	  \end{equation}
	  from which we conclude
	  \begin{equation}
	    \rho_0(\rho,\varphi) - \rho = -\varepsilon v'(\varphi) +
	    \OO(\varepsilon^2) = \varepsilon \xi^{\rho}_v(L_{\ell,0}) + \OO(\varepsilon^2).
	  \end{equation}
	  Likewise, 
	  \begin{equation}
	    \begin{split} 
	      \varphi_0(\rho,\varphi) &= f(\varphi) + \frac{1}{2\ell} \log\left( \frac{\ell^2 e^{2\rho} + \left( \ell f'(\varphi) - \frac{f''(\varphi)}{f'(\varphi)} \right)^2}{\ell^2 e^{2\rho} + \left( \ell f'(\varphi) + \frac{f''(\varphi)}{f'(\varphi)} \right)^2} \right) \\
	      &= \varphi + \varepsilon v'(\varphi) + \frac{1}{2\ell} \log\left( \ell^2 \left( e^{2\rho} + 1 + 2\varepsilon (v'(\varphi) - \ell^{-1}v''(\varphi)) \right) \right) \\
	      &\phantom{= \varphi ~ } - \frac{1}{2\ell} \log\left( \ell^2 \left( e^{2\rho} + 1 + 2\varepsilon (v'(\varphi) + \ell^{-1}v''(\varphi)) \right) \right) + \OO(\varepsilon^2)\\ 
	      &= \varphi + \varepsilon \left( v(\varphi) - \frac{2 v''(\varphi)}{\ell^2\left( e^{2\rho} + 1 \right)} \right) + \OO(\varepsilon^2)
	    \end{split}
	  \end{equation}
	  so that
	  \begin{equation}
	    \begin{split} 
	      \varphi_0(\rho,\varphi) - \varphi &= \varepsilon \left( v(\varphi) - \frac{v''(\varphi)}{2}\frac{1}{\frac{\ell^2}{4}\left( e^{2\rho} + 1 \right)} \right) + \OO(\varepsilon^2) \\
	      &= \varepsilon \xi_v^{\varphi}(L_{\ell,0}) + \OO(\varepsilon^2)
	    \end{split}
	  \end{equation}
	  in accordance with \eqref{eq:xi_hyp}.

	  \subsection{Parabolic Monodromy: Cuspidal Geometries}\label{sub:cusp}
	  Finally, let us study the case of parabolic monodromy.

	  Consider a general point $L_{+,0}^{f} = \frac12 \Sch(f) \in \OO_{L_{+,0}}$ in
	  the Virasoro orbit passing through $L_{+,0} = 0$.
	  Let $(\rho,\varphi)$ be coordinates for which the
	  metric takes the form
	  \begin{equation}
	    ds^2(L_{+,0}^f) = d\rho^2 + \left( e^{\rho} - \frac12 \Sch(f) \right)^2 d\varphi^2
	  \end{equation}
	  A basis for the associated Hill problem is given by 
	  \begin{equation}
	    \psi_1^f = \sqrt{\frac{2\pi}{f'(\varphi)}},\qquad \psi_2^f = \frac{f(\varphi)}{\sqrt{2\pi f'(\varphi)}}. 
	    \label{eq:psif_par}
	  \end{equation}
	  The corresponding upper half plane coordinate is found from \eqref{eq:z} by substituting \eqref{eq:psif_par}:
	  \begin{equation}
	    z_{L_{+,0}^f}(\rho,\varphi) = \frac{1}{2\pi}  \left( f(\varphi) - \frac{2f''(\varphi) }{4e^{2\rho} + \left( \frac{f''(\varphi) }{f'(\varphi) } \right)^2} + i \frac{4 e^{\rho} f'(\varphi) }{4e^{2\rho} + \left( \frac{f''(\varphi) }{f'(\varphi) } \right)^2}\right).
	    \label{eq:zpar}
	  \end{equation}
	  The comparison between \eqref{eq:z_tilde_par} and \eqref{eq:zpar} yields the diffeomorphism 
	  \begin{equation}
	    \tilde f(\rho,\varphi) = (\rho_0(\rho,\varphi),\varphi_0(\rho,\varphi))
	  \end{equation}
	  where
	  \begin{equation}
	    \begin{split} 
	      e^{-\rho_0(\rho,\varphi)} = \frac{4 e^{\rho} f'(\varphi) }{4e^{2\rho} +
	      \left( \frac{f''(\varphi) }{f'(\varphi) } \right)^2}\quad ,\quad \varphi_0(\rho,\varphi) = f(\varphi)  - \frac{2f''(\varphi) }{4e^{2\rho} + \left( \frac{f''(\varphi) }{f'(\varphi) } \right)^2}
	    \end{split}
	    \label{eq:tilde_f_par}
	  \end{equation}
	  This recovers results of \cite{GNW}.

	  As before, we find 
	  \begin{equation}
	    \lim_{\rho \to \infty} \rho_0(\rho,\varphi) = \infty \quad , \quad
	    \lim_{\rho \to \infty} \varphi_0(\rho,\varphi) = f(\varphi)
	  \end{equation}
	  which shows that $\tilde f$ indeed extends $f$.

	  Notice that again the $(\rho_0,\varphi_0)$ provide a global
	  coordinate system.
	  For any non-trivial $f$, however, the $(\rho,\varphi)$ coordinates do not cover the cuspidal singularity.
	  Indeed, for $f \neq id$, by inspection of \eqref{eq:tilde_f_par}, 
	  \begin{equation}
	    e^{\rho_0} = \frac{4e^{2\rho} + \left( \frac{f''(\varphi)}{f'(\varphi)} \right)^2}{4e^{\rho}f'(\varphi)} > 0
	    \label{eq:par_FG_non_global}
	  \end{equation}
	  is strictly greater than zero, and hence there exists no values of
$(\rho,\varphi)$ such that $\rho_0(\rho,\varphi) = -\infty$. 
	  This means that the coordinate patch defined by $(\rho,\varphi)$ cannot cover
	  the cuspidal singularity, while the coordinate patch defined by
	  $(\rho_0,\varphi_0)$ clearly does.

	  \section{The Range of \texorpdfstring{$z_{L_{+,1}}$}{z\_\{L\_\{+,1\}\}}}\label{app:exotic_cusp}

	  As we have seen in Section \ref{subsub:exc}, the asymptotics of $z_{L_{+,1}}$ covers a full neighbourhood of the ideal boundary $\del \HH$.
	  In the following we want to study the actual range of $z_{L_{+,1}}$ in detail.
	  To this end, let us consider the family of constant-$\rho$ curves $\gamma_{\rho}(\varphi) = z(\rho=\text{fixed},\varphi)$.
	  In order to understand the qualitative behaviour of those curves, we study their extremal horizontal and vertical extension.
	  At these points, the tangent vector
	  \begin{equation}
	    \dot \gamma_{\rho}(\varphi) = \del_{\varphi} z(\rho,\varphi) = \frac{e^{2\rho} - L(\varphi)}{(e^{\rho} \psi_1(\varphi) + i \psi_1'(\varphi))^2}
	    \label{eq:gammadot}
	  \end{equation}
	  becomes either purely imaginary or purely real respectively.
	  It is clear that the latter happens when either $\psi_1 = 0$ or $\psi_1' = 0$.
	  From the definition of $F_{+,1}(\varphi)$ \eqref{eq:Fpmn} together with the corresponding $\psi_{1,2}$ (cf.\ \eqref{eq:psi_from_F}) one quickly finds that this can happen only at three points, namely
	  \begin{equation}
	    \psi_2(0) = \psi_1'(0) = \psi_1(\pm \pi)= 0 
	  \end{equation}
	  The condition of \eqref{eq:gammadot} to be purely imaginary, on the other hand, is given by
	  \begin{equation}
	    e^{2\rho}\psi_1^2(\varphi) - \psi_1'^2(\varphi) = 0
	    \label{eq:phi_star}
	  \end{equation}
	  which implies 
	  \begin{equation}
	    e^{\rho} = \pm \frac{\psi_1'(\varphi)}{\psi_1(\varphi)} = \mp \frac12 \frac{F''(\varphi)}{F'(\varphi)}
	    \label{eq:phi_star2}
	  \end{equation}
	  Now 
	  \begin{equation}
	    \frac{F''(\varphi)}{F'(\varphi)} = \frac{\pi \tan\left( \frac{\varphi}{2} \right)}{\pi + \cos^2\left( \frac{\varphi}{2} \right)}
	  \end{equation}
	  maps $(0,\pi)$ isomorphically to $\RR_+$ and $(-\pi,0)$ isomorphically to $\RR_-$ so that \eqref{eq:phi_star2} has exactly two solutions (which are symmetric around 0), $\varphi_{\pm} = \pm \varphi_*$ ($\varphi_* \in [0,\pi)$). 
	    Notice in particular that for $\rho \to \infty$, $\varphi_{\pm} \to \pm \pi$.

	    Moreover, a direct computation shows that the maximum(s) and minimum lie at the points
	    \begin{equation}
	      \gamma_{\rho}(0) = i e^{-\rho} \left( \frac{1}{2\pi} + \frac12 \right) \quad , \quad \gamma_{\rho}(\pm \pi) = \pm\frac12 + 2 i e^{\rho}
	    \end{equation}
	    which approach $0$ and $i\infty$ exponentially fast in the limit $\rho \to \infty$.
	    At the same time, the real part of the rightmost and leftmost point are given by
	    \begin{equation}
	      \begin{split} 
		{\rm Re}(\gamma_{\rho})(\varphi_{\pm}) &= \frac{e^{2\rho}\psi_1(\varphi_{\pm})\psi_2(\varphi_{\pm}) + \psi_1'(\varphi_{\pm})\psi_2'(\varphi_{\pm})}{e^{2\rho}\psi_1^2(\varphi_{\pm}) + \psi_1'^2(\varphi_{\pm})} 
		\stackrel{\eqref{eq:phi_star}}{=} \frac12 \left( \frac{\psi_2(\varphi_{\pm})}{\psi_1(\varphi_{\pm})} + \frac{\psi_2'(\varphi_{\pm})}{\psi_1'(\varphi_{\pm})} \right) \\
		&= \frac12 \left( 2F(\varphi_{\pm}) + F'(\varphi_{\pm})\frac{\psi_1(\varphi_{\pm})}{\psi_1'(\varphi_{\pm})} \right) 
		\stackrel{\eqref{eq:phi_star2}}{=} \frac12 \left(\vphantom{\frac12} 2F(\varphi_{\pm}) \pm e^{-\rho} F'(\pm\varphi) \right)
	      \end{split}
	    \end{equation}
	    where we used $\psi_2(\varphi) = F(\varphi)\psi_1(\varphi)$ in the second to last step.
	    Now recall that $F'(\varphi)$ is everywhere positive and $\varphi_{+} = -\varphi_{-} > 0$.
	    Moreover, since $\varphi_{\pm} \to \pm \pi$ as $\rho \to \infty$, it follows that ${\rm Re}(\gamma_{\rho})(\varphi_{\pm})\to \pm\infty$.

\begin{figure}[htb]
	      \centering
	      \begin{subfigure}
		\centering
		\begin{tikzpicture}
		  \coordinate (A) at (-1,0);
		  \coordinate (B) at (1,0);
		  \coordinate (L) at (-3,0);
		  \coordinate (R) at (3,0);

		  \draw[thick] (L) -- (R);
		  \draw[thick,dashed] (A) -- (-1,4);
		  \draw[thick,dashed] (B) -- (1,4);

		  \node[point] (Min) at (0,.3) {};
		  \node[point] (VL) at (-1,3) {};
		  \node[point] (VR) at (1,3) {};
		  \node[point] (HL) at (-2,2) {};
		  \node[point] (HR) at (2,2) {};

		  \draw (VL) to[out=180,in=90] (HL);
		  \draw (HL) to[out=-90,in=180] (Min);
		  \draw (Min) to[out=0,in=-90] (HR);
		  \draw (HR) to[out=90,in=0] (VR);

		  \node[below] at (A) {$\hspace{-10pt}-\frac12$};
		  \node[below] at (B) {$\frac12$};

		  \draw[-latex] (Min) -- ($(Min) + (0.7,0)$);
		  \draw[-latex] (VL) -- ($(VL) + (-0.7,0)$);
		  \draw[-latex] (VR) -- ($(VR) + (-0.7,0)$);
		  \draw[-latex] (HL) -- ($(HL) + (0,-0.7)$);
		  \draw[-latex] (HR) -- ($(HR) + (0,0.7)$);

		  \node[above] at (Min) {$\dot\gamma_{\rho}(0)$};
		  \node[above left] at (VL) {$\dot\gamma_{\rho}(-\pi)$};
		  \node[above right] at (VR) {$\dot\gamma_{\rho}(\pi)$};
		  \node[left] at (HL) {$\dot\gamma_{\rho}(-\varphi_*)$};
		  \node[right] at (HR) {$\dot\gamma_{\rho}(\varphi_*)$};

		\end{tikzpicture}
	      \end{subfigure}%
	      \begin{subfigure}
		\centering
		\begin{tikzpicture}
		  \coordinate (A) at (-1,0);
		  \coordinate (B) at (1,0);
		  \coordinate (L) at (-3,0);
		  \coordinate (R) at (3,0);

		  \draw[thick] (L) -- (R);
		  \draw[thick,dashed] (A) -- (-1,4);
		  \draw[thick,dashed] (B) -- (1,4);

		  \node (Min1) at (0,.4) {};
		  \node[point] (VL1) at (-1,2.7) {};
		  \node[point] (VR1) at (1,2.7) {};
		  \node (HL1) at (-1.7,2) {};
		  \node (HR1) at (1.7,2) {};

		  \node (Min2) at (0,.5) {};
		  \node[point] (VL2) at (-1,2.4) {};
		  \node[point] (VR2) at (1,2.4) {};
		  \node (HL2) at (-1.4,1.8) {};
		  \node (HR2) at (1.4,1.8) {};

		  \node (Min3) at (0,.3) {};
		  \node[point] (VL3) at (-1,3) {};
		  \node[point] (VR3) at (1,3) {};
		  \node (HL3) at (-2,2) {};
		  \node (HR3) at (2,2) {};

		  \draw (VL1.center) to[out=180,in=90] (HL1.center);
		  \draw (HL1.center) to[out=-90,in=180] (Min1.center);
		  \draw (Min1.center) to[out=0,in=-90] (HR1.center);
		  \draw (HR1.center) to[out=90,in=0] (VR1.center);

		  \draw (VL2.center) to[out=180,in=90] (HL2.center);
		  \draw (HL2.center) to[out=-90,in=180] (Min2.center);
		  \draw (Min2.center) to[out=0,in=-90] (HR2.center);
		  \draw (HR2.center) to[out=90,in=0] (VR2.center);

		  \draw (VL3.center) to[out=180,in=90] (HL3.center);
		  \draw (HL3.center) to[out=-90,in=180] (Min3.center);
		  \draw (Min3.center) to[out=0,in=-90] (HR3.center);
		  \draw (HR3.center) to[out=90,in=0] (VR3.center);

		  \node[below] at (A) {$\hspace{-10pt}-\frac12$};
		  \node[below] at (B) {$\frac12$};

		\end{tikzpicture}
	      \end{subfigure}
	      \caption{Left: Schematic picture of a $\gamma_{\rho}(\varphi)$ for fixed $\rho$ large enough. Right: Schematic picture of the family $\gamma_{\rho}$.}
	      \label{fig:cst_rho_curves_par}
	    \end{figure}
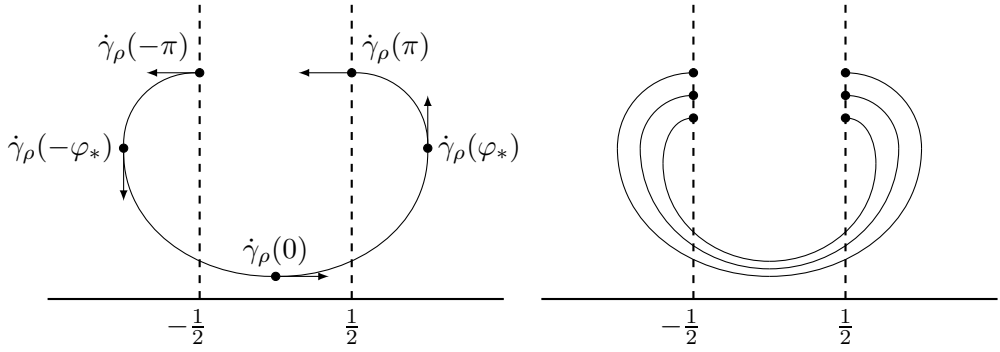

	    We display a sketch of the curve $\gamma_{\rho}(\varphi)$ in Figure \ref{fig:cst_rho_curves_par}.
	    We want to stress that the above analysis is only valid for $\rho$ large enough. 

	    To summarise the above discussion, the family $\gamma_{\rho}$ approaches the ideal boundary as $\rho\to\infty$ thereby sweeping out almost all of $\HH$.

\begin{thebibliography}{00}
\addcontentsline{toc}{section}{References}
\frenchspacing
\small
\addtolength{\itemsep}{-4pt}


  \bibitem{Kirillov} A.\ A.\ Kirillov, \textit{Orbits of the group of diffeomorphisms of
    a circle and local Lie superalgebras}, Funktsional.\ Anal.\ i
    Prilozhen.\ 15, no. 2 (1981) 75-76.

\bibitem{Segal} G.\ Segal, \textit{Unitary Representations of some Infinite
Dimensional Groups}, Commun.\ Math.\ Phys.\ 80 (1981) 301-342.

  \bibitem{Witten} E.\ Witten, \textit{Coadjoint orbits of the Virasoro group}, 
    Commun. Math. Phys. 114 (1988) 1-53.

\bibitem{Balog_Feher_Palla} J.\ Balog, L.\ Feh\'er, and L.\ Palla, \textit{Coadjoint
  orbits of the Virasoro algebra and the global Liouville equation}, 
  Int.\ J.\ Mod.\ Phys.\ A 13(02) (1998) 315-362.
  \href{https://arxiv.org/abs/hep-th/9703045}{arXiv:hep-th/9703045}.

\bibitem{AS1} A.\ Alekseev and S.\ Shatashvili, \textit{Path Integral
Quantization of the Coadjoint Orbits of the Virasoro Group and 2D Gravity},
Nucl.\ Phys.\ B323 (1989) 719–733.

\bibitem{RR1} B.\ Rai and V.\ Rodgers, \textit{From Coadjoint Orbits to
Scale Invariant WZNW Type Actions and 2-D Quantum Gravity Action}, Nucl.\ Phys.\ 
B341 (1990) 119–133.

\bibitem{MertensTuriaci:Review} T.\ Mertens, G.\ Turiaci, 
\textit{Solvable Models of Quantum Black Holes: A Review on Jackiw-Teitelboim
Gravity}, Living Rev.\ Rel.\ 26, 4 (2023).
\href{https://arxiv.org/abs/2210.10846}{arXiv:2210.10846}.

\bibitem{AP} A.\ Almheiri, J.\ Polchiniski, \textit{Models of AdS${}_2$ Backreaction and
Holography}, JHEP 11 (2015) 014. \href{https://arxiv.org/abs/1402.6334}{arXiv:1402.6334 [hep-th]}.

\bibitem{Jensen} K.\ Jensen, \textit{Chaos in AdS${}_2$ holography}, 
Phys.\ Rev.\ Lett/\ 117 (2016) 11,
111601. \href{https://arxiv.org/abs/1605.06098}{arXiv:1605.06098 [hep-th]}

\bibitem{MSY} J.\ Maldacena, D.\ Stanford, Z.\ Yang, \textit{Conformal
symmetry and its breaking in two dimensional Nearly Anti-de-Sitter space}, 
PTEP 2016 (2016) 12, 12C104.
\href{https://arxiv.org/abs/1606.01857}{arXiv:1606.01857 [hep-th]}

\bibitem{EMV} J.\ Engels\"oy, T.\ Mertens, H.\ Verlinde, 
\textit{An Investigation of AdS${}_2$ Backreaction and Holography}, 
JHEP 07 (2016) 139. \href{https://arxiv.org/abs/1606.03438}{arXiv:1606.03438
[hep-th]}


  \bibitem{GNW} M.\ Gautam, P.\ Nayak, and S.\ R.\ Wadia,
    \textit{Coadjoint orbit action of Virasoro group and two-dimensional quantum gravity
    dual to SYK/tensor models}, JHEP 11 (2017) 046.
    \href{https://arxiv.org/abs/1702.04266}{arXiv:1702.04266}

\bibitem{Mertens:Origins} T.\ Mertens, \textit{The Schwarzian Theory - Origins}, 
JHEP 05 (2018) 036. \href{https://arxiv.org/abs/1801.09605}{arXiv:1801.09605
[hep-th]}

\bibitem{SJY1} M.\ M.\ Sheikh-Jabbari, H.\ Yavartanoo, 
\textit{On Quantization of AdS3 Gravity I: Semi-Classical Analysis}, 
JHEP 1407 (2014) 104.
\href{https://arxiv.org/abs/1404.4472v2}{arXiv:1404.4472v2}

\bibitem{SJY2} M.\ M.\ Sheikh-Jabbari, H.\ Yavartanoo, 
\textit{On 3d Bulk Geometry of Virasoro Coadjoint Orbits: Orbit invariant
charges and Virasoro hair on locally AdS3 geometries}, 
Eur.\ Phys.\ J.\ C76 (2016) 493.
\href{https://arxiv.org/abs/1603.05272}{arXiv:1603.05272}

\bibitem{Banados} M.\ Ba\~nados, \textit{Three-dimensional quantum geometry and
black holes}, 
AIP Conf.Proc.\ 484 (1999) 1, 147-169.
\href{https://www.arxiv.org/abs/hep-th/9901148}{hep-th/9901148}. 

\bibitem{Oblak} B.\ Oblak, \textit{BMS Particles in Three Dimensions}, 
Springer Theses (2018). 
\href{https://arxiv.org/abs/1610.08526}{arXiv:1610.08526 [hep-th]}

\bibitem{VV} E.\ Verlinde, H.\ Verlinde, \textit{Conformal Field Theory and
Geometric Quantization}, in \textit{Superstrings '89}, Proceedings of the
Trieste Spring School, World Scientific (1990).  

\bibitem{NV} S.\ Nag, A.\ Verjovsky, 
\textit{$\Diff(S^1)$ and the Teichm\"uller Spaces}, Commun.\ Math,\ Phy.\ 130
(1990) 123-138.

\bibitem{HR} D.\ Hong, S.\ Rajeev, \textit{Universal Teichm\"uller Space and
$\Diff S^1/S^1$}, Commun.\ Math.\ Phys.\ 135 (1991) 401-411. 

\bibitem{MertensTuriaci:Defects} T.\ Mertens, G.\ Turiaci, 
\textit{Defects in Jackiw-Teitelboim Quantum Gravity}, 
JHEP 08 (2019) 127. \href{https://arxiv.org/abs/1904.05228}{arXiv:1904.05228}.

\bibitem{FK} T.\ Fukuyama, K.\ Kamimura, \textit{Gauge Theory of Two-Dimensional
Gravity}, Phys.\ Lett.\ B160 (1985) 259-262.

\bibitem{IT} K.\ Isler, C.\ Trugenberger, \textit{A Gauge Theory of
Two-dimensional Quantum Gravity}, Phys.\ Rev.\ Lett.\ 63 (1989) 834.

\bibitem{Nitsche} J.\ Nitsche, \textit{\"Uber die isolierten Singularit\"aten
der L\"osungen von $\Delta u = e^u$}, Math.Z.68 (1957) 316-324

\bibitem{FengShiXu} Yu Feng, Yiquian Shi, Bin Xu, \textit{Isolated singularities
of conformal hyperbolic metrics},
\href{https://arxiv.org/abs/1711.01018}{arXiv:1711/01018 [math.DG]}

\bibitem{Choi_Larsen} S.\ Choi and F.\ Larsen, \textit{AdS2 Holography and Effective
  QFT}, JHEP 11 (2023) 151. \href{https://arxiv.org/abs/2302.13917}{arXiv:2302.13917}.

\bibitem{SSS} P.\ Saad, S.\ Shenker, D.\ Stanford, \textit{JT gravity as a
matrix integral}.
\href{https://arxiv.org/abs/1903.11115}{arXiv:1903.11115}


\bibitem{Goldman} W.\ Goldman, \textit{Geometric structures on manifolds}, 
AMS Graduate Studies in Mathematicss 227 (2022).
\href{https://math.umd.edu/~wmg/gstom.pdf}{gstom.pdf}

\bibitem{GoldmanHiggs} W.\ Goldman, \textit{Higgs Bundles and
Geometric Structures on Surfaces}, in Oscar Garcia-Prada, Jean
Pierre Bourguignon, and Simon Salamon (eds), \textit{The Many Facets of
Geometry: A Tribute to Nigel Hitchin}, Oxford Academic (2010), 129-163; 
\href{https://arxiv.org/abs/0805.1793}{arXiv:0805.1793 [math.DG]}

\bibitem{SchallerStrobl} P.\ Schaller, T.\ Strobl, \textit{Diffeomorphisms versus
non-Abelian gauge transformations: an example of 1+1 dimensional gravity}, 
Phys.\ Lett.\ B337 (1994) 266-270.
\href{https://arxiv.org/abs/hep-th/9401110}{arXiv:hep-th/9401110}

\bibitem{Dunajski_Gavrea} M.\ Dunajski, and N.\ Gavrea, \textit{Elizabethan
vortices}, Nonlinearity 36, no. 8 (2023) 4169.
\href{https://arxiv.org/abs/2301.06191}{arXiv:2301.06191}


\bibitem{Barnich_Troessaert_I} G.\ Barnich and C.\ Troessaert, \textit{Symmetries 
of asymptotically flat 4 dimensional spacetimes at null infinity revisited}, 
Phys.\ Rev.\ Lett.\ 105 (2010) 111103. \href{https://arxiv.org/abs/0909.2617}{arXiv:0909.2617}.





  \bibitem{Wise} D.\ K.\ Wise, \textit{MacDowell–Mansouri gravity and Cartan
geometry}, Class.\ Quantum Grav. 27, no. 15 (2010) 155010.
\href{https://arxiv.org/abs/gr-qc/0611154}{arXiv:gr-qc/0611154}

    
\bibitem{GRS} A.\ Gorskii, B.\ Roy, K.\ Selivanov, 
\textit{Large gauge transformations and special orbits of the Virasoro group}, 
JETP Letters 53(1) (1991), 64-68.

  \bibitem{Wolpert} S.\ A.\ Wolpert, \textit{Cusps and the family hyperbolic
metric}, Duke Math. J. 138 (3) (2007) 423-443.
    \href{https://arxiv.org/abs/math/0508470}{arXiv:math/0508470}


\bibitem{Ovsienko} 
V.\ Ovsienko,  S.\ Tabachnikov, 
\textit{Projective Differential Geometry Old and New: From the Schwarzian
Derivative to the Cohomology of Diffeomorphism Group},  Cambridge University
Press (2004).

\bibitem{DuvalGuieu} C.\ Duval, L.\ Guieu, 
\textit{The Virasoro Group and Lorentzian Surfaces: the hyperboloid of one
sheet}, J.\ Geom.\ Phys.\ 33 (2000) 103-127.

\bibitem{AlekseevMeinrenken1} A.\ Alekseev, E.\ Meinrenken, 
\textit{Symplectic geometry of Teichm\"uller spaces for surfaces with ideal
boundary}, \href{https://arxiv.org/abs/2401.03029}{arXiv:2401.03029 [math.DG]}

\bibitem{PS} A.\ Pressley, G.\ Segal, \textit{Loop Groups}, Oxford University
Press (1988).  

\end{thebibliography}
\end{document}